\pdfoutput=1
\documentclass[preprint,11pt]{elsarticle1}
\usepackage{epsfig,amssymb,graphics,color,latexsym,psfrag,amsmath,pifont,mathtools,subfig,rotating,amscd}

     \usepackage[margin=2.5cm]{geometry}

\mathtoolsset{showmanualtags}

\newcommand{\bbN}{\mathbb{N}}
\newcommand{\tc}{\textcolor}

\newcommand{\cO}{{\cal O}}
\newcommand{\bA}{\mathbf{A}}
\newcommand{\bB}{\mathbf{B}}
\newcommand{\bC}{\mathbf{C}}
\newcommand{\bF}{\mathbf{F}}
\newcommand{\bG}{\mathbf{G}}
\newcommand{\bH}{\mathbf{H}}

\newcommand{\bP}{\mathbf{P}}

\newcommand{\bS}{\mathbf{S}}

\newcommand{\bX}{\mathbf{X}}
\newcommand{\bY}{\mathbf{Y}}
\newcommand{\bZ}{\mathbf{Z}}
\newcommand{\bI}{\mathbf{I}}

\newcommand{\bK}{\mathbf{K}}

\newcommand{\bSos}{\mathbf{S}_{\rm{os}}}
\newcommand{\bSsq}{\mathbf{S}_{\rm{sq}}}
\newcommand{\bKos}{\mathbf{K}_{\rm{os}}}
\newcommand{\bKsq}{\mathbf{K}_{\rm{sq}}}
\newcommand{\ds}{\displaystyle}
\newcommand{\bphi}{\mbox{\boldmath$\phi$}}

\newcommand{\btau}{\mbox{\boldmath$\tau$}}

\newcommand{\bxi}{\mbox{\boldmath$\xi$}}
\newcommand{\bnabla}{\mbox{\boldmath$\nabla$}}

\newcommand{\bGamma}{\mbox{\boldmath$\Gamma$}}
\newcommand{\eps}{\epsilon}
\newcommand{\bd}{{\bf d}}
\newcommand{\bx}{{\bf x}}
\newcommand{\bz}{{\bf z}}

\newcommand{\mrd}{\mathrm{d}}
\newcommand{\mre}{\mathrm{e}}
\newcommand{\mri}{\mathrm{i}}

\newcommand{\mrk}{\mathrm{k}}

\newcommand{\mrv}{\mathrm{v}}
\newcommand{\pt}{\partial_t}
\newcommand{\py}{\partial_y}
\newcommand{\pz}{\partial_z}
\newcommand{\pyy}{\partial_{yy}}

\newcommand{\We}{W\!e}
\newcommand{\bPi}{\bar{\Pi}}
\newcommand{\bSigma}{\bar{\Sigma}}

\newcommand{\bv}{{\bf v}}
\newcommand{\beq}{\begin{equation}}
\newcommand{\eeq}{\end{equation}}
\newcommand{\ba}{\begin{array}}
\newcommand{\ea}{\end{array}}
\newcommand{\non}{\nonumber}
\newcommand{\enma}[1]   {\ensuremath{#1}}
\newcommand{\diag}{\enma{\mathrm{diag}}}
\newcommand{\trace}{\enma{\mathrm{trace}}}
\newcommand{\matbegin}{
        \left[
}
\newcommand{\matend}{
        \right]
}

\newcommand{\tbo}[2]{
  \matbegin \begin{array}{c}
       #1 \\ #2
       \end{array} \matend }
\newcommand{\thbo}[3]{
  \matbegin \begin{array}{c}
       #1 \\ #2 \\ #3
       \end{array} \matend }
\newcommand{\tbt}[4]{
  \matbegin \begin{array}{cc}
       #1 & #2 \\ #3 & #4
       \end{array} \matend }
\newcommand{\obt}[2]{
  \matbegin \begin{array}{cc}
       #1 & #2
       \end{array} \matend }
\newcommand{\obth}[3]{
  \matbegin \begin{array}{ccc}
       #1 & #2 & #3
       \end{array} \matend }
\newcommand{\fbo}[4]{
 \matbegin \begin{array}{c}
                #1 \\ #2 \\ #3 \\ #4
                \end{array}\matend}
\newcommand{\obf}[4]{
 \matbegin \begin{array}{cccc}
                #1 & #2 & #3 & #4
                \end{array}\matend}
\newcommand{\btab}{\begin{tabular}}
\newcommand{\etab}{\end{tabular}}

\definecolor{bgblue}{rgb}{0.04,0.39,0.53}

 \biboptions{sort&compress}

\journal{J.\ Non-Newtonian Fluid Mech.}

\begin{document}

\begin{frontmatter}



\title{{\bf Nonmodal amplification of stochastic disturbances \\in strongly elastic channel flows}}


\author[mj]{Mihailo R.\ Jovanovi\'c}
\ead{mihailo@umn.edu}
\ead[url]{http://www.umn.edu/$\sim$mihailo}
\address[mj]{Department of Electrical and Computer Engineering, University of
Minnesota, Minneapolis, MN 55455, USA}
\author[sk]{Satish Kumar}
\ead{kumar030@umn.edu}
\ead[url]{http://www.cems.umn.edu/research/kumar/kumar.htm}
\address[sk]{Department of Chemical Engineering and Materials Science, University of Minnesota, Minneapolis, MN 55455, USA}

    \begin{abstract}
Nonmodal amplification of stochastic disturbances in elasticity-dominated channel flows of Oldroyd-B fluids is analyzed in this work. For streamwise-constant flows with high elasticity numbers $\mu$ and finite Weissenberg numbers $\We$, we show that the linearized dynamics can be decomposed into slow and fast subsystems, and establish analytically that the steady-state variances of velocity and polymer stress fluctuations scale as $\cO (\We^2)$ and $\cO (\We^4)$, respectively. This demonstrates that large velocity variance can be sustained even in weakly inertial stochastically driven channel flows of viscoelastic fluids. We further show that the wall-normal and spanwise forces have the strongest impact on the flow fluctuations, and that the influence of these forces is largest on the fluctuations in streamwise velocity and the streamwise component of the polymer stress tensor. The underlying physical mechanism involves polymer stretching that introduces a lift-up of flow fluctuations similar to vortex tilting in inertia-dominated flows. The validity of our analytical results is confirmed in stochastic simulations. The phenomenon examined here provides a possible route for the early stages of a bypass transition to elastic turbulence and might be exploited to enhance mixing in microfluidic devices.
    \end{abstract}

\begin{keyword}
     Elastic turbulence \sep frequency responses \sep inertialess flows \sep polymer stretching \sep singular perturbations \sep variance amplification \sep viscoelastic fluids.

\end{keyword}

\end{frontmatter}

% \linenumbers

\section{Introduction}
    \label{sec.intro}

\subsection{Background}

The classical approach to transition to turbulence examines the linearized equations for exponentially growing normal modes. The existence of these unstable modes implies exponential growth of infinitesimal perturbations to the laminar flow, and the corresponding eigenfunctions identify flow patterns that are expected to dominate early stages of transition. This approach agrees with experiments in many flows (e.g., those driven by thermal and centrifugal forces~\citep{Trefethen1993}) but it comes up short in matching experimental observations in wall-bounded shear flows (flows in channels, pipes, and boundary layers). The failure of hydrodynamic stability analysis in describing {\em the early stages of transition\/} is attributed in part to the nonnormal nature of the linearized equations, which may manifest itself by transient growth of perturbations~\citep{Butler1992,redhen93}, protrusion of pseudospectra to the unstable regions~\citep{Trefethen1993,treemb05}, and large receptivity to ambient disturbances~\citep{Farrell1993,Bamieh2001,jovbamJFM05}. Even in stable regimes -- owing to nonnormality -- perturbations that grow transiently before decaying due to viscosity can be configured, irregularities in laboratory design can lead to instability, and disturbances (such as free-stream turbulence or surface imperfections) can be amplified by orders of magnitude. These conclusions can be reached by performing transient growth, pseudospectra, or variance amplification analyses~\citep{Grossmann2000,schhen01,Schmid2007}. All of these methods demonstrate the importance of streamwise-elongated flow patterns of high and low streamwise velocity (streaks) in transitional wall-bounded shear flows of Newtonian fluids; this is at odds with modal stability results, but in agreement with experiments~\citep{matalf01} and direct numerical simulations~\cite{jacdur01} conducted in noisy environments. We note that in order to understand {\em the later stages of transition\/}, consideration of nonlinear interactions between streamwise-varying fluctuations and the streaks is required~\cite{hamkimwal95,wal97,wedker04}.

Transition to turbulence in viscoelastic fluids is important from both fundamental and technological perspectives~\citep{Larson1992}. The observation that transition can occur even when the effects of fluid elasticity dominate those of inertia -- which is a primary cause of transition in Newtonian fluids -- is particularly intriguing~\citep{Larson2000,Groisman2000,Groisman2004,
arrthodiogol06,Berti2008,thoshe09}. Improved understanding of transition mechanisms in viscoelastic fluids has broad applications, ranging from deeper insight into order-disorder transitions in spatially extended nonlinear dynamical systems to enhanced mixing in microfluidic devices through the addition of polymers~\citep{Groisman2001,Groisman2004}. The phenomenon of `elastic turbulence' occurs in the absence of inertial effects~\citep{Larson2000}, and it has been observed experimentally in shear flows with curved streamlines~\cite{Larson2000,Groisman2000,Groisman2004,bursegste06,bursegste07,junste10}. The transition in curvilinear flows is triggered by a purely elastic instability  that originates from the interactions between polymer stress fluctuations and the velocity gradients in the base flow~\cite{larshamul90,Larson1992,Shaqfeh1996}. Currently, it is not known whether fluid elasticity can promote transition in parallel shear flows with negligible inertial forces.

In spite of the linear stability of weakly inertial parallel shear flows of viscoelastic fluids, small fluctuations around the laminar base state can achieve significant transient growth. Early efforts used simulations of two-dimensional (2D) channel flows to probe their transient responses in both linear and nonlinear regimes~\citep{Sureshkumar1999,Keunings2002}. A new family of linearly stable transiently growing 2D stress modes was identified for the Oldroyd-B constitutive model~\cite{Kupferman2005}; these modes were obtained in Couette flow by setting the velocity and pressure fluctuations to zero and they do not couple back to the momentum equation. More recently, a similar result was shown for the three-dimensional (3D) upper convected Maxwell model (a special case of the Oldroyd-B model) with linear base velocity~\cite{Renardy2009}. In~\cite{Schumacher2006}, an exact solution to the Oldroyd-B model was constructed which displays non-monotonic transient responses in strongly elastic 2D Couette flow with arbitrarily low, but non-zero, inertia. Even in channel flows without inertia, the streamwise-independent velocity and stress fluctuations can exhibit transient growth that scales unfavorably with elasticity~\cite{jovkumPOF10}. Several explicit scaling relationships were established, and computations were used to identify the spatial structure of the initial conditions (in the polymer stress components) that grow the most with time.

Amplification of stochastic disturbances in channel flows of viscoelastic fluids was recently examined using linear systems theory~\citep{hodjovkumJFM08}. For the Oldroyd-B model, computations reported in~\cite{hodjovkumJFM08} demonstrated that streamwise-constant velocity fluctuations can experience considerable amplification even in the weakly inertial/strongly elastic regime. As in Newtonian fluids, this amplification is fundamentally nonmodal in nature: it cannot be described using the normal mode decomposition of classical hydrodynamic stability analysis~\citep{Grossmann2000,schhen01,Schmid2007}. Rather, it arises from an energy exchange involving the fluctuations in the streamwise/wall-normal polymer stress and the wall-normal gradient of the streamwise velocity~\citep{hodjovkumJFM09}.

Despite this recent progress, analytical results that describe amplification of stochastic disturbances in strongly elastic channel flows of viscoelastic fluids are lacking. Such results are important because of the physical insight they yield, and as a means to validate numerical simulations. The purpose of the present work is to address this issue.

\subsection{Preview of key results}

The key parameters that characterize channel flows of viscoelastic fluids are:
    the viscosity ratio, $\beta = \eta_s/(\eta_s + \eta_p)$, where $\eta_s$ and $\eta_p$ are the solvent and polymer viscosities;
    the Weissenberg number, $\We = \lambda U_o/L$, which represents the product of the polymer relaxation time $\lambda$ and the typical velocity gradient $U_o/L$;
    and the elasticity number, $\mu = \We/Re$, which quantifies the ratio of the polymer relaxation time $\lambda$ to the viscous diffusion time $\rho L^2 / (\eta_s + \eta_p)$.
Here, $Re = \rho U_o L/(\eta_s + \eta_p)$ is the Reynolds number, which represents the ratio of inertial to viscous forces, $U_o$ is the largest base velocity, $L$ is the channel half-height, and $\rho$ is the fluid density. By modeling ambient disturbances to streamwise-constant channel flows of Oldroyd-B fluids (with spanwise wavenumber $k_z$) as an additive white Gaussian forcing with zero mean and unit variance, we develop an explicit scaling of the variance (or energy) amplification of velocity fluctuations with the Weissenberg number $\We$,
    \beq
    E_{\mrv} (k_z;\We,\beta,\mu)
    \; = \;
    f (k_z;\beta,\mu)
    \; + \;
    \We^2 \, g (k_z;\beta,\mu).
    \label{eq.Ev-intro1}
    \eeq
Here, $f$ and $g$ denote $\We$-independent functions where $g$ accounts for the amplification from wall-normal and spanwise forces to the fluctuations in streamwise velocity, while $f$ accounts for the amplification from all other forcing to all other velocity components. It is worth noting that $E_{\mrv}$ quantifies the ensemble-average energy density (associated with the velocity field) of the statistical steady-state~\citep{Farrell1993}, and it is determined by integrating the power spectral density over all temporal frequencies~\citep{farioa94}.

Furthermore, considering flows with $\mu \gg 1$, we apply singular perturbation techniques to establish that the steady-state velocity variance scales as
    \beq
    E_{\mrv} (k_z;\We,\beta,\mu)
    \; = \;
    \mu \tilde{f}_0 (k_z)/\beta
    \; + \;
    \tilde{f}_1 (k_z)
    \,
    (1 - \beta)/\beta^2
    \; + \;
    \We^2 \, \tilde{g}_0 (k_z) \, (1 - \beta)^2/\beta
    \; + \;
    \cO(1/\mu).
    \non
    \eeq
Our analysis demonstrates that, in flows with high elasticity numbers, the linear $\mu$-scaling of the function $f$ in~(\ref{eq.Ev-intro1}) arises from the corresponding power spectral density becoming almost uniformly distributed over the temporal frequency band whose width is proportional to $\mu$. We also show that, from a physical point of view, no important viscoelastic effects take place in the contribution of the function $\tilde{f}_0$ to the variance amplification.

The last expression should be compared to the expression for the variance amplification in Newtonian fluids~\citep{Bamieh2001},
    \beq
    E_N (k_z;Re)
    \; = \;
    f_N (k_z)
    \; + \;
    Re^2 \, g_N (k_z).
    \label{eq.Ev-Newt}
    \eeq
At low $Re$ the $k_z$-dependence of $E_N$ is governed by $f_N (k_z)$,
    $
    E_N (k_z;Re)
    \approx
    f_N (k_z),
    $
and at high $Re$ it is governed by $g_N (k_z)$, $E_N (k_z) \approx Re^2 \, g_N (k_z)$. In this paper, we show that $\tilde{f}_0 (k_z) = f_N (k_z)$ which implies that the $k_z$-dependence of $\tilde{f}_0$ is characterized by viscous dissipation~\citep{Bamieh2001}. This clearly indicates that, at the level of velocity fluctuation dynamics, the behavior of Newtonian fluids with low $Re$ and the behavior of Oldroyd-B fluids with low $\We$ is dominated by diffusion. On the other hand, the $g$-functions in the expressions for $E_{\mrv}$ and $E_N$ exhibit peaks at $k_z = \cO(1)$; the values of $k_z$ where these peaks take place identify the spanwise length scales of the most energetic response of velocity fluctuations to stochastic forcing in Newtonian fluids with high $Re$, and in Oldroyd-B fluids with high $\We$.

We note that $g_N (k_z)$ and $\tilde{g}_0 (k_z)$ arise from fundamentally different physical mechanisms: in inertia-dominated flows of Newtonian fluids, vortex tilting is the main driving force for amplification; in elasticity-dominated flows of viscoelastic fluids, it is polymer stretching, which gives rise to an energy transfer from the base flow to fluctuations. In streamwise-constant channel flows of Newtonian fluids, the linearized dynamics of the wall-normal vorticity, $\eta$, are governed by~\cite{Butler1992}
    \beq
    \pt \eta
    \; = \,
    - Re \, U' (y) \, \pz v
    \; + \;
    \Delta \eta,
    \label{eq.eta-Newt}
    \eeq
where $v$ denotes the wall-normal velocity fluctuations, $\Delta$ is a Laplacian, and $- U'(y)$ is the base flow vorticity (in the spanwise direction $z$). The first term on the right-hand side of~(\ref{eq.eta-Newt}) represents the linearized vortex-tilting term which acts as a source in the vorticity equation. From a physical point of view, the spanwise vorticity of the base flow, i.e.\ $- U'(y)$, gets tilted in the wall-normal direction $y$ by the spanwise changes in $v$ which leads to the amplification of the wall-normal vorticity (and thereby streamwise velocity, $\eta = \pz u$)~\cite{Butler1992}. In this paper, we show that the linearized wall-normal vorticity equation in inertialess streamwise-constant flows of Oldroyd-B fluids assumes the following form
    \begin{subequations}
    \label{eq.eta-SOB}
    \begin{align}
    \label{eq.eta-SOB1}
    \pt \Delta \eta
    & \; = \,
    - \We
    \left( 1/\beta - 1\right)
    \left( U'(y) \, \Delta \, \pz \, + \, 2 \, U''(y) \, \partial_{yz} \right) \vartheta
    \; - \;
    \left( 1/\beta \right) \Delta \eta
    \\[0.1cm]
    &
    \; = \,
    - \We
    \left( 1/\beta - 1\right)
    \left(
    \partial_{y z}
    \left( U'(y) \tau_{22} \right)
    \, + \,
    \partial_{z z}
    \left( U'(y) \tau_{23} \right)
    \right)
    \; - \;
    \left( 1/\beta \right) \Delta \eta,
    \label{eq.eta-SOB2}
    \end{align}
    \end{subequations}
where $\vartheta$ in~(\ref{eq.eta-SOB1}) is obtained by filtering high temporal frequencies in the wall-normal velocity $v$; see Section~\ref{sec.main-Ev} for details. The terms $U'(y) \tau_{22}$ and $U'(y) \tau_{23}$ in~(\ref{eq.eta-SOB2}) represent stretching of the corresponding stress fluctuations by a background shear; gradients of these quantities provide a source in the vorticity equation even in the absence of inertia. Thus, base-shear stretching of stress fluctuations along with their spanwise variations gives rise to the amplification of $\eta$ (and consequently $u$) in inertialess flows of viscoelastic fluids. As in streamwise-constant inertial flows of Newtonian fluids, this amplification disappears either in the absence of spanwise variations in flow fluctuations, i.e.\ $\partial_z (\cdot) = 0$, or in the absence of the background shear, i.e.\ $U' = 0$.

\begin{figure}
    \centering
    {
    \subfloat[]
    {\includegraphics[width=0.65\textwidth]{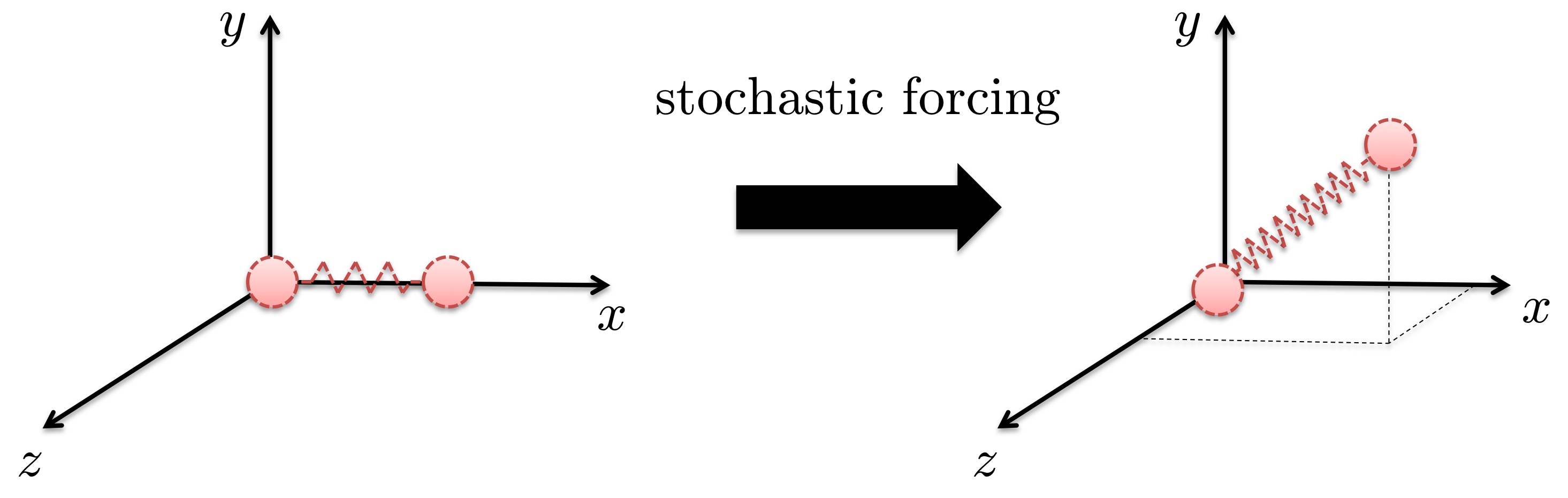}
    \label{fig.spring-xy-yz}}
    \\
    \subfloat[]
    {\includegraphics[width=0.45\textwidth]{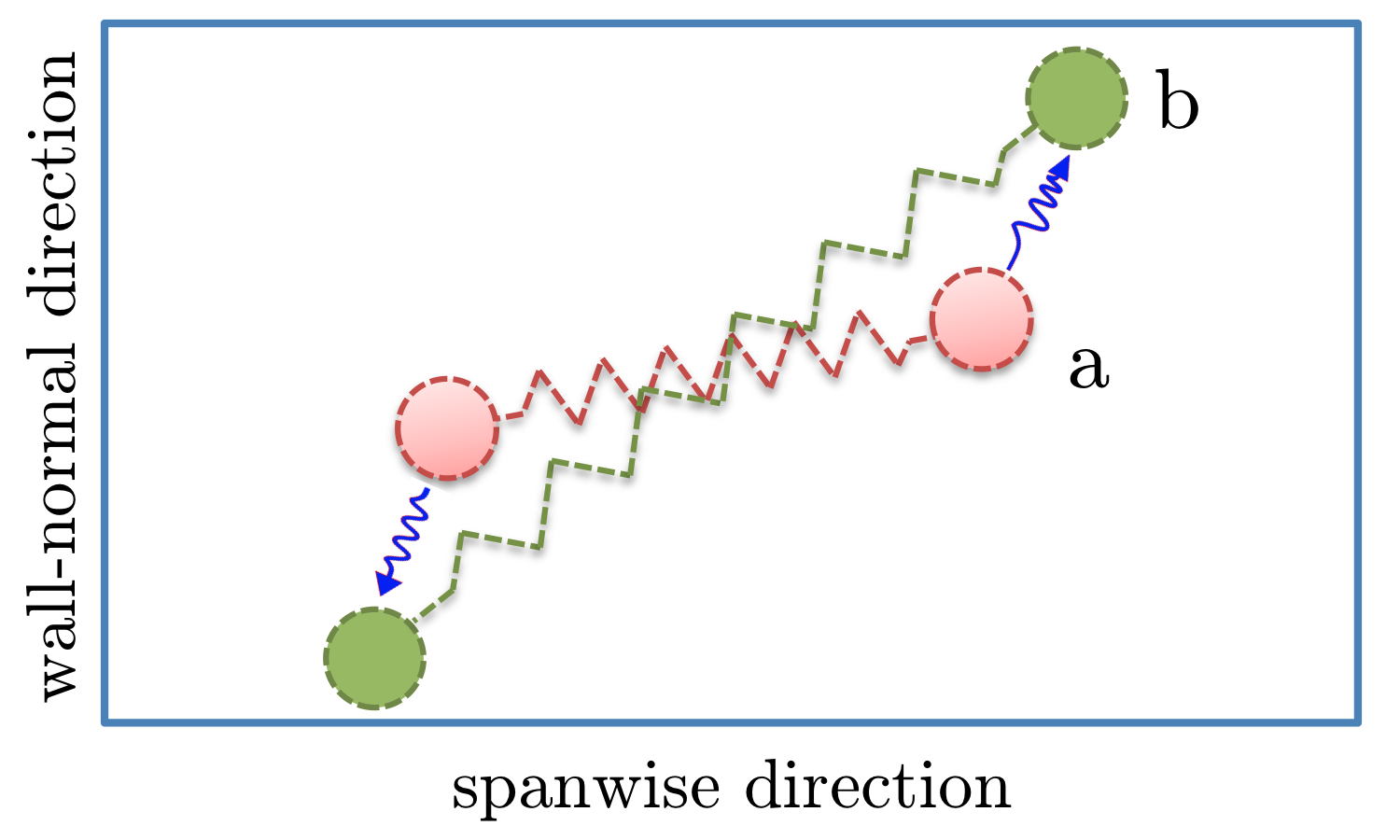}
    \label{fig.polymer-stretching}}
    \hspace*{0.5cm}
    \subfloat[]
    {\includegraphics[width=0.45\textwidth]{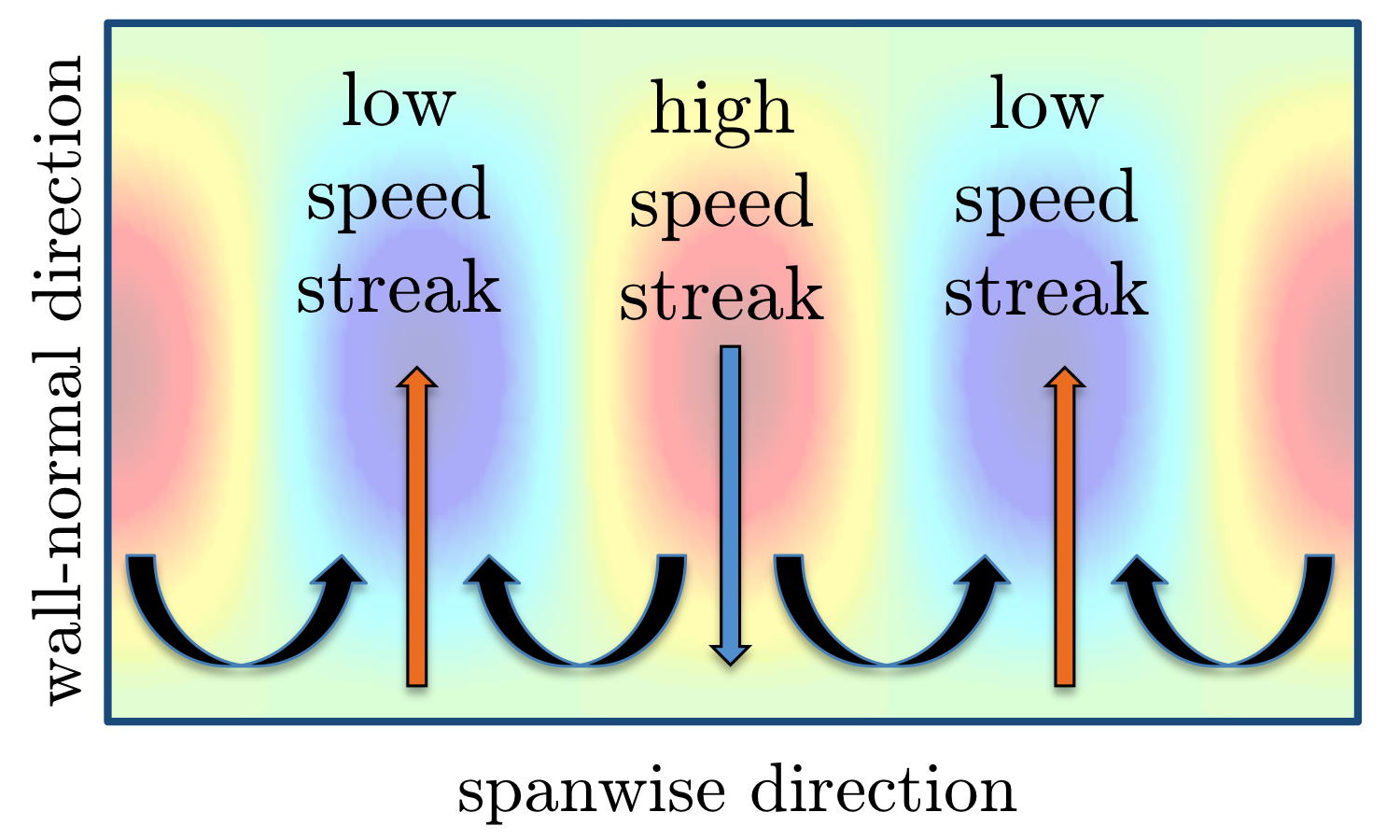}
    \label{fig.lift-up}}
    }
    \caption{
    (a) The steady-state configuration of an elastic dumbbell is perturbed out of the ($x,y$)-plane by stochastic forcing.
    (b) A projection of a perturbed dumbbell in the ($y,z$)-plane. Dumbbell stretching in the wall-normal and spanwise directions creates fluctuations in polymer stress components $\tau_{22}$, $\tau_{23}$, and $\tau_{33}$.
    (c) Streamwise vortices, generated by the gradients in $\tau_{22}$, $\tau_{23}$, and $\tau_{33}$, induce streamwise streaks through the lift-up mechanism. A spanwise momentum exchange is enabled by displacement of fluid particles in the wall-normal direction.
    }
    \label{fig.mechanism}
    \end{figure}

Additional insight into the above mechanism can be gained by considering the momentum conservation equation in planes perpendicular to the direction of the base flow. For streamwise-independent inertialess flows, there is a static-in-time relationship between the ($y,z$)-gradients in $\tau_{22}$, $\tau_{23}$, and $\tau_{33}$ and the wall-normal ($v$) and spanwise ($w$) velocity fluctuations
    \beq
    \ba{rcl}
    0
    & \!\! = \!\!\! &
    -\py \, p
    \, + \,
    (1-\beta)
    \left(
    \partial_y \tau_{22}
    \, + \,
    \partial_z \tau_{23}
    \right)
    \, + \,
    \beta \, \Delta v
    \, + \,
    d_2,
    \\[0.1cm]
    0
    & \!\! = \!\!\!  &
    -\pz \, p
    \, + \,
    (1-\beta)
    \left(
    \partial_y \tau_{23}
    \, + \,
    \partial_z \tau_{33}
    \right)
    \, + \,
    \beta \, \Delta w
    \, + \,
    d_3.
    \ea
    \non
    \eeq
Spatial variations in $v$ and $w$ induced by these stress gradients result in streamwise vorticity fluctuations (i.e., the streamwise `rolls'); these redistribute momentum in the ($y,z$)-plane and promote amplification of streamwise velocity fluctuations. As in the Newtonian case, this momentum exchange involves lifting of the low speed fluid away from the wall and movement of the high speed fluid towards the wall, and it is responsible for creation of low and high speed streaks that alternate in the spanwise direction. From a microscopic point of view, the end-to-end vectors of the elastic dumbbells that underlie the Oldroyd-B model are oriented in the streamwise direction in the base flow (the only non-zero diagonal element of the base polymer stress tensor is the streamwise component)~\cite{Bird1987,Larson1999}. Stochastic forcing $(d_2, d_3)$ perturbs the end-to-end vector and generates fluctuations in $\tau_{22}$, $\tau_{23}$, and $\tau_{33}$ (which are zero in the base flow). These fluctuations then lead to energy amplification through the mechanism described above; see Figure~\ref{fig.mechanism} for additional illustration.

Our second key result is an explicit formula for the steady-state variance maintained in the components of the polymer stress tensor by streamwise-constant stochastic forcing
    \beq
    E_{\tau} (k_z;\We,\beta,\mu)
    \; = \;
    a (k_z;\beta,\mu)
    \; + \;
    \We^2
    \,
    b (k_z;\beta,\mu)
    \; + \;
    \We^4
    \,
    c (k_z;\beta,\mu).
    \non
    \eeq
Here, $a$, $b$, and $c$ represent $\We$-independent functions which in flows with high $\mu$ also become elasticity-number-independent,
    \beq
    E_{\tau} (k_z;\We,\beta,\mu)
    \; = \;
    a_0 (k_z;\beta)
    \; + \;
    \We^2
    \,
    b_0 (k_z;\beta)
    \; + \;
    \We^4
    \,
    c_0 (k_z;\beta)
    \; + \;
    \cO (1/\mu).
    \non
    \eeq
We note that the $c$-function, which primarily originates from the polymer stretching, quantifies the amplification from the wall-normal and spanwise forces to the fluctuations in the streamwise component of the polymer stress tensor, $\tau_{11}$. Therefore, in high-$\We$ regimes the wall-normal and spanwise disturbances have the strongest influence, and the impact of these forces is largest on the streamwise velocity and polymer stress fluctuations. Furthermore, we demonstrate that, in flows with high elasticity numbers, the analysis of inertialess (or creeping) flows of Oldroyd-B fluids correctly predicts all important properties of the functions $a$, $b$, $c$, and $g$. On the other hand, the inertialess model provides a poor approximation at high temporal frequencies of the power spectral densities responsible for the generation of the function $f$ in~(\ref{eq.Ev-intro1}). In fact, we show that the problem of determining this function in inertialess flows becomes ill-posed. This ill-posedness arises from the absence of the inertial terms in the momentum equation and it cannot be alleviated by the addition of diffusion to the constitutive equations.

The above analytical results are obtained as a consequence of our discovery that the linearized dynamics can be decomposed into slow and fast subsystems. This observation is used to cast the equations into a standard singularly perturbed form for which existing methodology~\citep{kokkharei99} can be applied. The decomposition of the linearized dynamics at high $\mu$ is not obvious {\em a priori}, and it takes advantage of the intrinsic time scale ($\lambda$) in the Oldroyd-B constitutive equation. In addition, it facilitates derivation of the explicit analytical expressions for the steady-state variances of velocity and polymer stress fluctuations given above. Our success with uncovering the hitherto unknown dependence of the energy amplification on the Weissenberg and elasticity numbers points to the scaling and modeling steps as prerequisites for applying standard singular perturbation techniques.

The organization of the rest of the paper is laid out next. In Section~\ref{sec.2d3c}, we describe the streamwise-constant linearized model with forcing. In Section~\ref{sec.Hij}, we provide explicit scaling of the frequency responses from different forcing to different velocity and stress components with the Weissenberg number. In Section~\ref{sec.main}, we provide analytical expressions for the variance amplification and discuss physical mechanisms leading to amplification from forcing to flow fluctuation components. We also determine the spanwise length scales of flow structures that contribute most to the steady-state variance and show that the most energetic velocity fluctuations assume the form of high and low speed streaks. In Section~\ref{sec.sim}, we use stochastic simulations of the linearized dynamics to verify our analytical developments. The major contributions are summarized in Section~\ref{sec.concl} and the mathematical developments are relegated to the appendices. These developments make heavy use of singular perturbation techniques for stochastically forced linear systems and they provide important physical insight about the dynamics of strongly elastic fluids through transformation of the linearized equations into slow and fast subsystems.

\section{The streamwise-constant linearized model with forcing}
    \label{sec.2d3c}

\begin{figure}
    \begin{center}
    \includegraphics[width=0.5\textwidth]{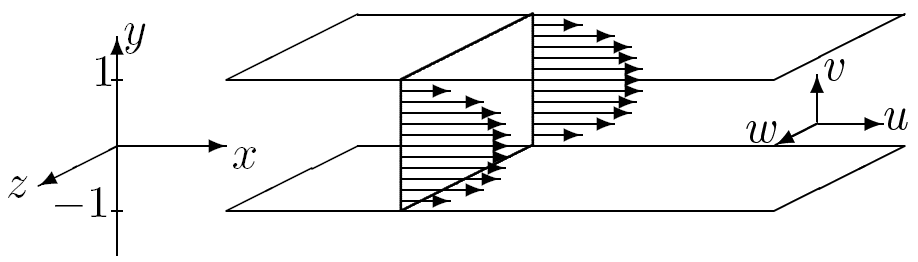}
    \end{center}
    \caption{Schematic of channel flow. In this paper, we study the linearized model for streamwise-constant three-dimensional fluctuations, which means that the dynamics evolve in the ($y,z$)-plane, but fluctuations in all three spatial directions are considered.}
    \label{fig.flow}
    \end{figure}

We consider incompressible channel flows of \mbox{Oldroyd-B} fluids with $\eps = 1/\mu \ll 1$; see Figure~\ref{fig.flow} for geometry. The equations governing the dynamics (up to first order) of velocity ($\bv = \left[\,u\,\,\,v\,\,\,w\,\right]^{T}$), pressure ($p$), and polymer stress tensor ($\btau$) fluctuations around base flow ($\overline{\bv},\overline{\btau}$) are brought to a non-dimensional form by scaling time with $\lambda$, length with $L$, velocity with $U_o$, polymer stresses with $\eta_p U_o/L$, pressure with $(\eta_s + \eta_p)U_o/L$, and forcing per unit mass with $(\eta_s + \eta_p) U_o/\rho L^2$
    \beq
    \ba{rcl}
    \!\!\!
    \eps \dot{\bv}
    & \!\!\!\! = \!\!\!\! &
    - \eps \, \We
    \left(
    \bnabla_\bv \overline{\bv}
    +
    \bnabla_{\overline{\bv}} \, \bv
    \right)
    \, - \,
    \bnabla p
    \, + \,
    (1 - \beta) \bnabla \! \cdot \! \btau
    +
    \beta \bnabla^2 \bv
    \, + \,
    \bd,
    \\[0.1cm]
    \!\!\!
    0
    &\!\!\!\! = \!\!\!\!&
    \bnabla \! \cdot \! \bv,
    \\[0.1cm]
    \!\!\!
    \dot{\btau}
    &\!\!\! = \!\!\!&
    \bnabla \bv
    +
    \left( \bnabla \bv \right)^{T}
    -
    \btau
    +
    \We
    \bigl(
    \btau \! \cdot \! \bnabla \overline{\bv}
    +
    \overline{\btau} \! \cdot \! \bnabla \bv
    +
    ( \overline{\btau} \! \cdot \! \bnabla \bv )^{T}
    +
    ( \btau\! \cdot \! \bnabla \overline{\bv})^{T}
    -
    \bnabla_\bv \overline{\btau}
    -
    \bnabla_{\overline{\bv}} \, \btau
    \bigr).
    \ea
    \label{eq.flow-lin}
    \eeq
Here, a dot signifies a partial derivative with respect to time $t$, $\bnabla$ is the gradient, \mbox{$\bnabla_\bv = \bv \cdot \bnabla$}, and $u$, $v$, and $w$ are the velocity fluctuations in the streamwise ($x$), wall-normal ($y$), and spanwise ($z$) directions, respectively. The linearized momentum equation is driven by the body force fluctuation vector $\bd$, which is purely harmonic in the horizontal directions, and stochastic in the wall-normal direction and in time,
    \beq
    \bd (x,y,z,t)
    \; = \;
    \Re
    \left(
    \bd (k_x,y,k_z,t)
    \,
    \mre^{\mri (k_x x \, + \, k_z z)}
    \right),
    \non
    \eeq
where the same notation is used to represent the field $\bd (x,y,z,t)$ and its Fourier transform in the horizontal directions $\bd(k_x,y,k_z,t)$; the difference between the two should be clear from the context. This spatio-temporal forcing will in turn yield velocity and polymer stress fluctuations of the same nature. We assume that $\bd(k_x,y,k_z,t)$ is a temporally stationary white Gaussian process with zero mean and unit variance; see~\cite{Farrell1993,Bamieh2001,jovbamJFM05} for additional details.

We study the linearized model for streamwise-constant three-dimensional fluctuations, which means that the dynamics evolve in the ($y,z$)-plane, but fluctuations in all three spatial directions are considered. This model is analyzed since the largest velocity variance in stochastically forced channel flows of viscoelastic fluids is maintained by streamwise-constant fluctuations~\citep{hodjovkumJFM08}. The linearized equations can be brought to an evolution form by removing pressure from the equations and by expressing $\bv$ in terms of the streamwise velocity and the ($y,z$)-plane streamfunction fluctuations, $\{ u = u$, $v = \pz \psi$, $w = - \py \psi \}$. By denoting
    \[
    \phi_1 \, = \, \psi,
    ~~~
    \phi_3 \, = \, u,
    \]
and by rearranging the polymer stress tensor components into
    \[
    \bphi_2 \, = \, \left[\,\tau_{22}\,\,\,\tau_{23}\,\,\,\tau_{33}\,\right]^T,
    ~~~
    \bphi_4
    \, = \,
    \left[\,\tau_{12}\,\,\,\tau_{13}\,\right]^T,
    ~~~
    \phi_5
    \, = \,
    \tau_{11},
    \]
system~(\ref{eq.flow-lin}) with fluctuations that are constant in the streamwise direction ($\partial_x (\cdot) \equiv 0$) and purely harmonic in the spanwise direction can be converted to
    \begin{subequations}
    \label{eq.lnse-2d3c}
    \begin{align}
    \label{eq.lnse-2d3c-v}
    \eps \, \dot{\phi}_1
    & \;=\;
    \beta \, \bS_{11} \, \phi_1
    \;+\;
    (1 \,-\, \beta) \, \bS_{12} \, \bphi_2
    \;+\;
    \bF_2 \, d_2
    \,+\,
    \bF_3 \, d_3,
    \\
    \label{eq.lnse-2d3c-tauv}
    \dot{\bphi}_2
    & \;=\;
    - \, \bphi_2
    \;+\;
    \bS_{21} \, \phi_1,
    \\
    \label{eq.lnse-2d3c-eta}
    \eps \, \dot{\phi}_3
    & \;=\;
    \beta \, \bS_{33} \, \phi_3
    \;+\;
    \eps \, \We \, \bS_{31} \, \phi_1
    \;+\;
    (1 \,-\, \beta) \, \bS_{34} \, \bphi_4
    \;+\;
    \bF_1 \, d_1,
    \\
    \label{eq.lnse-2d3c-taueta}
    \dot{\bphi}_4
    & \;=\;
    - \, \bphi_4
    \;+\;
    \We
    \left(
    \bS_{41} \, \phi_1
    \,+\,
    \bS_{42} \, \bphi_2
    \right)
    \;+\;
    \bS_{43} \, \phi_3,
    \\
    \label{eq.lnse-2d3c-tau11}
    \dot{\phi}_5
    & \;=\;
    - \, \phi_5
    \;+\;
    \We^2 \, \bS_{51} \, \phi_1
    \;+\;
    \We
    \left(
    \bS_{53} \, \phi_3
    \,+\,
    \bS_{54} \, \bphi_4
    \right),
    \\[0.0cm]
    \label{eq.lnse-2d3c-out}
    \left[
    \ba{c}
    {u} \\[0.cm]
    {v} \\[0.cm]
    {w}
    \ea
    \right]
    & \;=\;
    \left[
    \begin{array}{cc}
    0 & \bG_{u} \\[0.cm]
    \bG_{v} & 0 \\[0.cm]
    \bG_{w} & 0 \end{array}
    \right]
    \left[
    \ba{c}
    {\phi_1} \\[0.cm]
    {\phi_3}
    \ea
    \right].
    \end{align}
    \end{subequations}

Equations (\ref{eq.lnse-2d3c-v})-(\ref{eq.lnse-2d3c-tau11}) represent a system of partial differential equations (PDEs) in the wall-normal direction and in time driven by the body forcing, $\bd (y,k_z,t) = \left[\,d_1 (y,k_z,t)\,\,\,d_2 (y,k_z,t)\,\,\,d_3 (y,k_z,t)\,\right]^{T}$, and parameterized by the spanwise wavenumber, $k_z$, the Weissenberg number, $\We$, the elasticity number, $1/\eps$, and the viscosity ratio, $\beta$. The operators $\bF_j$ and $\bG_r$ are given by
    \[
    \ba{c}
    \bF_1 \, = \, \bI,
    ~~~
    \bF_2 \, = \, \mri k_z \Delta^{-1},
    ~~~
    \bF_3 \, = \, - \Delta^{-1} \py,
    \\[0.1cm]
    \bG_u \, = \, \bI,
    ~~~
    \bG_v \, = \, \mri k_z,
    ~~~
    \bG_w \, = \, - \py,
    \ea
    \]
and they, respectively, determine the way the forcing enters into the evolution model, and the way the velocity fluctuations depend on $\phi_1$ and $\phi_3$. On the other hand, the $\bS$-operators determine internal properties of the streamwise-constant evolution model (e.g., modal stability)
    \beq
    \ba{l}
    \bS_{11}
    \, = \,
    \Delta^{-1} \Delta^{2},
    ~~~
    \bS_{33}
    \, = \,
    \Delta,
    ~~~
    \bS_{31}
    \, = \,
    - \,\mri k_z U'(y),
    \\[0.1cm]
    \bS_{12}
    \, = \,
    \Delta^{-1}
    \left[
    \ba{ccc}
    \mri k_z \py
    &
    -
    \left(
    \pyy \, + \, k_z^2
    \right)
    &
    - \mri k_z \py
    \ea
    \right],
    ~~~
    \bS_{34}
    \, = \,
    \left[
    \ba{cc}
    \py
    &
    \mri k_z
    \ea
    \right],
    \\[0.1cm]
    \bS_{21}
    \, = \,
    \left[
    \ba{ccc}
    2 \mri k_z \py
    &
    -
    \left(
    \pyy \, + \, k_z^2
    \right)
    &
    -2 \mri k_z \py
    \ea
    \right]^T,
    ~~~
    \bS_{43}
    \, = \,
    \bS_{34}^T,
    \\[0.1cm]
    \bS_{41}
    \, = \,
    \left[
    \ba{c}
    \mri k_z \left( U'(y) \py  - U''(y) \right)
    \\
    - U'(y) \pyy
    \ea
    \right],
    ~~~
    \bS_{42}
    \, = \,
    \left[
    \ba{ccc}
    U'(y) & 0 & 0
    \\
    0 & U'(y) & 0
    \ea
    \right],
    \\[0.25cm]
    \bS_{51}
    \, = \,
    - 4 \mri k_z U'(y) U''(y),
    ~~~
    \bS_{53}
    \, = \,
    2 U'(y) \py,
    ~~~
    \bS_{54}
    \, = \,
    \left[
    \ba{cc}
    2 U'(y)
    &
    0
    \ea
    \right].
    \ea
    \non
    \eeq
Here, $\bI$ is the identity operator, $\Delta = \partial_{yy} - k_z^2 $ is a Laplacian with homogeneous Dirichlet boundary conditions, $\Delta^{-1}$ is the inverse of the Laplacian, $\Delta^2 = \partial_{yyyy} - 2 k_z^2 \partial_{yy} + k_z^4$ with homogeneous Cauchy (both Dirichlet and Neumann) boundary conditions, $\mri = \sqrt{-1}$,
    $
    U(y) = y
    $
in Couette flow,
    $
    U(y) = 1 - y^2
    $
in Poiseuille flow, and $U'(y) = \mrd U(y)/ \mrd y$. We note that operators $\bS_{11}$ and $\bS_{33}$, respectively, stand for the Orr-Sommerfeld and Squire operators in the streamwise-constant model of Newtonian fluids with $Re = 1$~\citep{schhen01}, and that $\bS_{31}$ denotes the vortex-tilting
term~\citep{Butler1992}. A comparison of the evolution model~(\ref{eq.lnse-2d3c}) and the linearized momentum, continuity, and constitutive equations~(\ref{eq.flow-lin}) reveals that, from a physical point of view,
$\bS_{12}$ and $\bS_{34}$ account for gradients of polymer stress fluctuations (i.e., $\nabla \cdot \btau$), $\bS_{21}$ and $\bS_{43}$ produce gradients of velocity fluctuations (i.e., $\nabla \bv$), $\bS_{41}$ captures both transport and stretching of base polymer stress by velocity fluctuations (i.e., $\bv \cdot \nabla \overline{\btau}$ and $\overline{\btau} \cdot \nabla \bv$), and $\bS_{42}$ and $\bS_{54}$ represent stretching of polymer stress fluctuations by base shear (i.e., $\btau \cdot \nabla \overline{\bv}$). Furthermore, operators $\bS_{51}$ and $\bS_{53}$ in~(\ref{eq.lnse-2d3c-tau11}) quantify transport and stretching of base polymer stress by velocity fluctuations (i.e., $\bv \cdot \nabla \overline{\btau}$ and $\overline{\btau} \cdot \nabla \bv$), respectively.

\section{Dependence of frequency responses on the Weissenberg number}
    \label{sec.Hij}

In this section, we examine the Weissenberg-number dependence of the frequency responses from different forcing to different velocity and polymer stress components. Application of the temporal Fourier transform to~(\ref{eq.lnse-2d3c}) enables us to determine the elements of the frequency response operator, $\bH$, that relates $\bv$ to $\bd$, $\bv = \bH \, \bd$. We also determine the elements of the frequency response operator associated with the stress components. We show that the frequency responses from wall-normal and spanwise forces to the fluctuations in streamwise velocity, $u$, and the streamwise component of the polymer stress tensor, $\tau_{11}$, scale linearly and quadratically with $\We$, respectively. Furthermore, these two forces introduce a linear dependence of $\tau_{12}$ and $\tau_{13}$ on $\We$, and the presence of the streamwise forcing introduces a similar effect on $\tau_{11}$. On the other hand, the responses from all other forces to all other velocity and stress components are $\We$-independent.

Although the analysis of the frequency responses of velocity fluctuations in Section~\ref{sec.PSDv} is similar to that of~\cite{hodjovkumJFM09}, it is revisited here because of the different scalings employed; to the best of our knowledge, the analysis of the frequency responses of polymer stress fluctuations in Section~\ref{sec.PSDtau} has not been done before. The scalings used in this work are well-suited for uncovering the conditions under which strong elasticity amplifies disturbances, and the resulting expressions for variance amplification will be analyzed in detail in Section~\ref{sec.main}.

\subsection{Frequency responses of velocity fluctuations}
    \label{sec.PSDv}

\begin{figure}
    \centering
    {
    \setlength{\unitlength}{1.25cm}
    \begin{picture}(12.5,3.75)(0,0)
   % Block diagram for the linearized NS equations

    %first part - dv
    \put(0,2.375){\vector(1,0){0.75}}
    \put(0.375,2.575){\makebox(0,0)[b]{$ d_2 $}}
    \put(0.75,2){\framebox(1,0.75){$\bF_2$}}
    \put(1.75,2.375){\vector(1,0){0.6}}
    \put(2.5,2.375){\circle{0.3}}

    %second part - orr-sommerfeld
    \put(2.65,2.375){\vector(1,0){0.6}}
    \put(3.25,2){\framebox(1.75,0.75){$ (1 + \mri \omega) \bKos $}}
    \put(5,2.375){\line(1,0){0.375}}
    \put(5.375,2.375){\circle*{0.08}}
    \put(5.375,2.375){\vector(1,0){0.375}}
    %third part - coupling
    \put(5.75,2){\framebox(1,0.75){$ \tc{red}{\We} \, \bC_p$}}
    \put(6.75,2.375){\vector(1,0){0.6}}
    \put(7.5,2.375){\circle{0.3}}

    %fourth part - Squire
    \put(7.65,2.375){\vector(1,0){0.6}}
    \put(8.25,2){\framebox(1.75,0.75){$ (1 + \mri \omega) \bKsq $}}
    \put(10,2.375){\vector(1,0){0.75}}
    \put(10.375,2.575){\makebox(0,0)[b]{$ u $}}

    % Map to u
    \put(10.75,2){\framebox(1,0.75){$\bG_u $}}
    \put(11.75,2.375){\vector(1,0){0.75}}
    \put(12.125,2.575){\makebox(0,0)[b]{$ u $}}
  
    %part 5 - du to v
    \put(0,3.375){\vector(1,0){0.75}}
    \put(0.375,3.575){\makebox(0,0)[b]{$ d_1 $}}
    \put(0.75,3){\framebox(1,0.75){$\bF_1$}}
    \put(1.75,3.375){\line(1,0){5.75}}
    \put(7.5,3.375){\vector(0,-1){0.85}}

    %part 6 - dw to v
    \put(0,1.375){\vector(1,0){0.75}}
    \put(0.375,1.575){\makebox(0,0)[b]{$ d_3 $}}
    \put(0.75,1){\framebox(1,0.75){$\bF_3$}}
    \put(1.75,1.375){\line(1,0){0.75}}
    \put(2.5,1.375){\vector(0,1){0.85}}

    %part 7 - du to w
    \put(5.375,2.375){\line(0,-1){1}}
    \put(5.375,1.375){\circle*{0.08}}
    \put(5.375,1.375){\vector(1,0){5.375}}
    \put(10.75,1){\framebox(1,0.75){$\bG_v$}}
    \put(11.75,1.375){\vector(1,0){0.75}}
    \put(12.125,1.575){\makebox(0,0)[b]{$v$}}

    %part 8 - dw to w
    \put(5.375,1.375){\line(0,-1){1}}
    \put(5.375,0.375){\vector(1,0){5.375}}
    \put(10.75,0){\framebox(1,0.75){$ \bG_w$}}
    \put(11.75,0.375){\vector(1,0){0.75}}
    \put(12.125,0.575){\makebox(0,0)[b]{$w$}}
    
    %part 9 - v
    \put(5.375,2.575){\makebox(0,0)[b]{$\psi$}}
\end{picture}
    }
    \caption{Block diagram of the velocity dynamics in the streamwise-constant linearized model. The capital letters denote the $\We$-independent operators, and $\omega$ denotes the temporal frequency. The operators $\bF_j$ and $\bG_r$ describe the way the forcing enters in the evolution model~(\ref{eq.lnse-2d3c}), and the way the velocity fluctuations depend on $\psi$ and $u$; $\bKos$ and $\bKsq$ govern the internal dynamics of $\psi$ and $u$; and $\bC_p$ captures the coupling from $\psi$ to $u$ which accounts for both vortex tilting and polymer stretching. From this block diagram it follows that
    (i) $d_2$ and $d_3$ induce a linear scaling of $u$ with $\We$;
    and
    (ii) the responses from all other forces to all other velocity components are $\We$-independent.}
    \label{fig.Block}
    \end{figure}
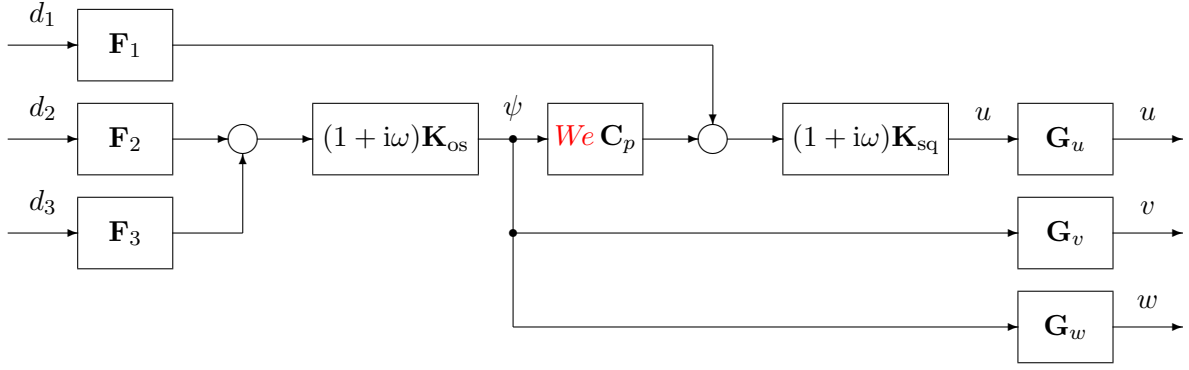

As shown in~\ref{sec.fr-v}, application of the temporal Fourier transform to~(\ref{eq.lnse-2d3c}) allows for elimination of the polymer stresses from the evolution model, which can be used to clarify the $\We$-dependence of the frequency responses from forcing to velocity components. The block diagram in Figure~\ref{fig.Block} provides a systems-level view of the velocity fluctuation dynamics in the streamwise-constant linearized model. The boxes represent different parts of the system and the circles denote summation of signals. Inputs into each box/circle are represented by lines with arrows directed toward the box/circle, and outputs of each box/circle are represented by lines with arrows leading away from the box/circle. The inputs specify the signals affecting subsystems, and the outputs designate the signals of interest or signals affecting other parts of the system~\citep{astmur08}.

All signals in Figure~\ref{fig.Block} are functions of the wall-normal coordinate $y$, the spanwise wavenumber $k_z$, and the temporal frequency $\omega$,  e.g.\ $u = u (y,k_z,\omega)$, with the following boundary conditions on $\psi$ and $u$,
    $
    \{
    \psi (\pm 1,k_z,\omega)
    =
    \py \psi (\pm 1,k_z,\omega)
    =
    u (\pm1,k_z,\omega)
    = 0
    \}.
    $
The capital letters in Figure~\ref{fig.Block} denote the Weissenberg-number-independent operators. These operators act in the wall-normal direction and some of them are parameterized by $k_z$ ($\bG_r$ and $\bF_j$ with $\{ r = u, v, w$; $j = 1, 2, 3 \}$), while the others depend on $k_z$, $\omega$, $\beta$, and $\eps$ ($\bKos$, $\bKsq$, and $\bC_p$). As discussed in Section~\ref{sec.2d3c}, the operators $\bF_j$ and $\bG_r$, respectively, describe the way the forcing enters into the evolution model, and the way the velocity fluctuations depend on the streamfunction and the streamwise velocity. The operator $\bC_p$ captures the coupling from the equation governing the dynamics of $\psi$ to the equation governing the dynamics of $u$, and it is defined as
    \beq
    \bC_p
    \; = \;
    \eps \, \bC_{p1}
    \, + \,
    \dfrac{1 \, - \, \beta}{(1 \,+\, \mri \omega)^2}
    \,
    \bC_{p2},
    \non
    \eeq
where $\bC_{p1} = - \mri k_z U'(y)$ denotes the vortex-tilting
term~\citep{Butler1992}, and
    \beq
    \bC_{p2}
    \, = \,
    \mri k_z
    \tilde{\bC}_{p2},
    ~~~
    \tilde{\bC}_{p2}
    \, = \,
    U'(y) \Delta \, + \, 2 U''(y) \py,
    \label{eq.Cp2}
    \eeq
denotes the term arising from polymer stretching~(see Section~\ref{sec.main} and~\ref{sec.fr-v}). Finally, $\bKos$ and $\bKsq$ govern the internal dynamics of $\psi$ and $u$, respectively. These two operators describe how the Orr-Sommerfeld and Squire operators (respectively, $\bSos = \Delta^{-1} \Delta^{2}$ and $\bSsq = \Delta$) in the streamwise-constant model of Newtonian fluids with $Re = 1$ are modified by elasticity,
    \beq
    \bK_{\mrk}
    =
    \left(
    \eps (\mri \omega)^2 \bI
    \, - \,
    ( \beta \bS_{\mrk} - \eps \bI)
    \mri \omega
    \, - \,
    \bS_{\mrk}
    \right)^{-1},
    ~~
    \mrk \, = \, \{ \mbox{os}, \, \mbox{sq} \}.
    \non
    \eeq

In the frequency domain, the forcing and velocity components are related by
    \[
    \ba{rcl}
    r (y,k_z,\omega;\We,\beta,\eps)
    & \!\! = \!\! &
    \ds{\sum_{j \, = \, 1}^{3}}
    \left[
    \bH_{r j} (k_z,\omega;\We,\beta,\eps) \, d_j (\cdot,k_z,\omega)
    \right] (y),
    ~~
    r \, = \, \{ u, v, w \},
    \ea
    \]
where $\bH_{r j}$ denotes the frequency response from $d_j$ to $r$. Each $\bH_{r j}$ represents an operator in $y$ parameterized by spatial and temporal frequencies ($k_z,\omega$) and key parameters associated with the constitutive equation ($\We,\beta,\eps$). The power spectral density maintained in $r$ by forcing evolution model~(\ref{eq.lnse-2d3c}) with white, unit variance, stationary stochastic process $d_j$ is determined by~\citep{farioa94}
    \beq
    \Pi_{r j} (k_z,\omega;\We,\beta,\eps)
    \, = \,
    \trace
    \left(
    \bH_{r j} (k_z,\omega;\We,\beta,\eps) \, \bH_{r j}^*(k_z,\omega;\We,\beta,\eps)
    \right),
    \non
    \eeq
where $\bH_{r j}^*$ is the adjoint of the operator $\bH_{r j}$. From a physical point of view, function $\Pi_{r j} (k_z,\omega)$ quantifies how the energy of  the velocity component $r$ arising from the forcing component $d_j$ is distributed over spanwise wavenumber, $k_z$, and temporal frequency, $\omega$. Furthermore, for a fixed value of $k_z$, the variance (energy) sustained in $r$ by $d_j$ is given by~\citep{Farrell1993}
    \beq
    E_{r j} (k_z;\We,\beta,\eps)
    \, = \,
    \dfrac{1}{2 \pi}
    \int_{-\infty}^{\infty}
    \Pi_{r j} (k_z,\omega;\We,\beta,\eps)
    \,
    \mrd \omega.
    \non
    \eeq

From the analysis presented in~\ref{sec.fr-v} (or, equivalently, from the block diagram in Figure~\ref{fig.Block}), it follows that the $\bH_{r j}$ are determined by
    \beq
    \ba{rcl}
    \bH_{u 1}(k_z,\omega;\We,\beta,\eps)
    & \!\! = \!\! &
    \bar{\bH}_{u 1}(k_z,\omega;\beta,\eps),
    \\[0.1cm]
    \bH_{u j}(k_z,\omega;\We,\beta,\eps)
    & \!\! = \!\! &
    \We
    \,
    \bar{\bH}_{u j}(k_z,\omega;\beta,\eps),
    ~~
    j = 2,3,
    \\[0.1cm]
    \bH_{r j}(k_z,\omega;\We,\beta,\eps)
    & \!\! = \!\! &
    \bar{\bH}_{r j}(k_z,\omega;\beta,\eps),
    ~~
    r = v,w;
    ~
    j = 2,3,
    \\[0.1cm]
    \bH_{r 1}(k_z,\omega;\We,\beta,\eps)
    & \!\! = \!\! &
    0,
    ~~
    r = v,w,
    \ea
    \label{eq.Hrj}
    \eeq
where the $\bar{\bH}_{r j}$ represent the $\We$-independent operators,
    \beq
    \ba{c}
    \bar{\bH}_{u 1}
    \, = \,
    (1 \, + \, \mri \omega)
    \,
    \bG_u \bKsq \bF_1
    \, = \,
    (1 \, + \, \mri \omega)
    \,
    \bKsq,
    \\[0.15cm]
    \bar{\bH}_{r j}
    \, = \,
    (1 \, + \, \mri \omega)
    \,
    \bG_r \bKos \bF_j,
    ~~
    r = v,w;
    ~
    j = 2,3,
    \\[0.15cm]
    \bar{\bH}_{u j}
    \, = \,
    \bG_u \bKsq
    \left(
    \eps (1 + \mri \omega)^2 \bC_{p1}
    \, + \,
    (1 - \beta) \bC_{p2}
    \right)
    \bKos \bF_j,
    ~~
    j = 2,3.
    \ea
    \non
    \eeq
Using the definitions of $\Pi_{r j}$ and the above expressions for $\bH_{r j}$, we obtain the following $\We$-scaling of the power spectral densities maintained in $r$ by stochastically forcing the linearized model with $d_j$
    \beq
    \begin{aligned}
    &
    \left[
    \begin{array}{ccc}
    \Pi_{u 1} (k_z,\omega;\We,\beta,\eps)
    &
    \Pi_{u 2} (k_z,\omega;\We,\beta,\eps)
    &
    \Pi_{u 3} (k_z,\omega;\We,\beta,\eps)
    \\[0.1cm]
    \Pi_{v 1} (k_z,\omega;\We,\beta,\eps)
    &
    \Pi_{v 2} (k_z,\omega;\We,\beta,\eps)
    &
    \Pi_{v 3} (k_z,\omega;\We,\beta,\eps)
    \\[0.1cm]
    \Pi_{w 1} (k_z,\omega;\We,\beta,\eps)
    &
    \Pi_{w 2} (k_z,\omega;\We,\beta,\eps)
    &
    \Pi_{w 3} (k_z,\omega;\We,\beta,\eps)
    \end{array}
    \right]
    \\
    =~
    &
    \left[ \begin{array} {ccc}
    {\bPi_{u 1} (k_z,\omega;\beta,\eps)}
    &
    {\bPi_{u 2} (k_z,\omega;\beta,\eps) \, \We^2}
    &
    {\bPi_{u 3} (k_z,\omega;\beta,\eps) \, \We^2}
    \\[0.1cm]
    {0}
    &
    {\bPi_{v 2} (k_z,\omega;\beta,\eps)}
    &
    {\bPi_{v 3} (k_z,\omega;\beta,\eps)}
    \\[0.1cm]
    {0}
    &
    {\bPi_{w 2} (k_z,\omega;\beta,\eps)}
    &
    {\bPi_{w 3} (k_z,\omega;\beta,\eps)}
    \end{array}
    \right],
    \end{aligned}
    \label{eq.hs-component}
    \end{equation}
where $\bPi_{r j}$ are the power spectral densities of the $\We$-independent operators $\bar{\bH}_{r j} (k_z,\omega;\beta,\eps)$. Moreover, the square-additive property of the power spectral density can be used to determine the aggregate effect of forces in all three spatial directions, $\bd$, on all three velocity components, $\bv$,
    \beq
    \Pi (k_z,\omega;\We,\beta,\eps)
    \, = \,
    \bPi_f (k_z,\omega;\beta,\eps)
    \, + \,
    \bPi_g (k_z,\omega;\beta,\eps) \, \We^2.
    \label{eq.Pi}
    \eeq
Here, $\Pi (k_z,\omega;\We,\beta,\eps)$ denotes the power spectral density of the frequency response operator $\bH(k_z,\omega;\We,\beta,\eps)$, $\bv \, = \, \bH \bd$, with
    \[
    \ba{c}
    \bPi_f
    \, = \,
    \bPi_{u 1} \, + \,
    \sum_{j \, = \, 2}^{3}
    \left(
    \bPi_{v j} \, + \, \bPi_{w j}
    \right),
    ~~
    \bPi_g
    \, = \,
    \bPi_{u 2} \, + \, \bPi_{u 3}.
    \ea
    \]

Similarly, the variance maintained in $\bv$ by $\bd$ is determined by
    \beq
    E_{\mrv} (k_z;\We,\beta,\eps)
    \; = \;
    f (k_z;\beta,\eps)
    \; + \;
    g (k_z;\beta,\eps) \, \We^2,
    \tag*{(${\rm E}_{\rm v}$)}
    \label{eq.Ev}
    \eeq
where
    $
    f
    \, = \,
    f_{u 1}
    \, + \,
    \sum_{j \, = \, 2}^{3}
    \left(
    f_{v j} \, + \, f_{w j}
    \right),
    $
    $
    g
    \, = \,
    g_{u 2} \, + \, g_{u 3},
    $
and, for example,
    \beq
    g_{u 2} (k_z;\beta,\eps)
    \, = \,
    \dfrac{1}{2 \pi}
    \int_{-\infty}^{\infty}
    \bPi_{u 2} (k_z,\omega;\beta,\eps)
    \,
    \mrd \omega.
    \non
    \eeq
Therefore, as can be seen from~(\ref{eq.hs-component}), variance amplification from wall-normal and spanwise forces to streamwise velocity is proportional to $\We^2$, while variance amplification for all other components of the frequency response operator $\bH$, $\bv = \bH \bd$, is Weissenberg-number independent.

\subsection{Frequency responses of polymer stress fluctuations}
    \label{sec.PSDtau}

\begin{figure}
    \centering
    {
    \includegraphics[width=0.99\textwidth]{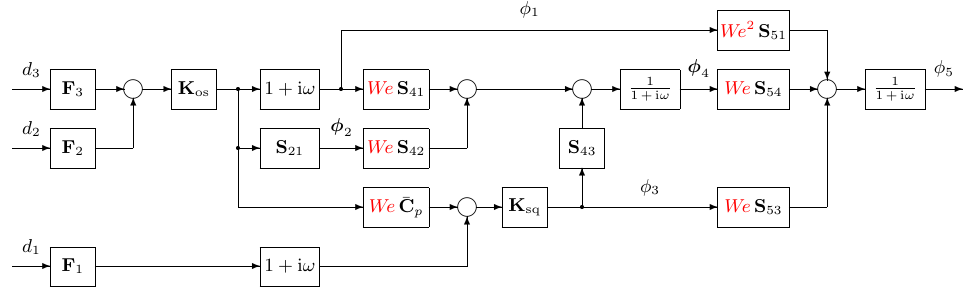}
    }
    \caption{Block diagram of the polymer stress dynamics in the streamwise-constant linearized model. The operators $\bF_j$, $\bKos$, $\bKsq$ have the same meaning as in Figure~\ref{fig.Block}, $\bar{\bC}_p = (1 + \mri \omega)^2 \, \bC_p$, and the $\bS$-operators describe the coupling between the different components in the evolution model~(\ref{eq.lnse-2d3c}). From this block diagram it follows that $d_1$ induces
    (i) zero response of $\bphi_2$;
    (ii) a $\We$-independent response of $\bphi_4$;
    and
    (iii) a linear scaling of $\phi_5$ with $\We$.
    In addition, $d_2$ and $d_3$ induce
    (i) a $\We$-independent response of $\bphi_2$;
    (ii) a linear scaling of $\bphi_4$ with $\We$;
    and
    (iii) a quadratic scaling of $\phi_5$ with $\We$.
    }
    \label{fig.blockTau}
    \end{figure}

We next examine frequency responses of polymer stress fluctuations. From the analysis presented in~\ref{sec.fr-tau}, it follows that their dynamics can be equivalently represented via the block diagram in Figure~\ref{fig.blockTau}. This representation is convenient for uncovering the $\We$-dependence of the frequency responses from the forcing components $d_1$, $d_2$, and $d_3$ to the stress components
    $
    \bphi_2 \, = \, \left[\,\tau_{22}\,\,\,\tau_{23}\,\,\,\tau_{33}\,\right]^T,
    $
    $
    \bphi_4
    \, = \,
    \left[\,\tau_{12}\,\,\,\tau_{13}\,\right]^T,
    $
and
    $
    \phi_5
    \, = \,
    \tau_{11}.
    $
In what follows, the frequency response from $d_j$ to $\bphi_i$ will be denoted by $\bGamma_{\phi_i,j}$
    \[
    \bphi_i (y,k_z,\omega;\We,\beta,\eps)
    \, = \,
    \ds{\sum_{j \, = \, 1}^{3}}
    \left[
    \bGamma_{\phi_i, j} (k_z,\omega;\We,\beta,\eps) \, d_j (\cdot,k_z,\omega)
    \right] (y),
    ~~
    i \, = \, \{ 2, 4, 5 \}.
    \]
We will also pay attention to the responses from individual forcing to individual polymer stress components. For example, $\bGamma_{12,3}$ will denote the frequency response from $d_3$ to $\tau_{12}$, and $\Sigma_{12,3}$ will denote the power spectral density of $\bGamma_{12, 3}$,
    \beq
    \Sigma_{12, 3} (k_z,\omega;\We,\beta,\eps)
    \, = \,
    \trace
    \left(
    \bGamma_{12, 3} (k_z,\omega;\We,\beta,\eps) \, \bGamma_{12,3}^*(k_z,\omega;\We,\beta,\eps)
    \right).
    \non
    \eeq
Similar notation will be used to quantify the influence of the other components of $\bd$ on the other components of $\btau$.

Since the capital letters in Figure~\ref{fig.blockTau} denote the Weissenberg-number-independent operators, the $\We$-dependence of responses from $d_j$ to the polymer stress components can be inferred by following the flow of information in this block diagram. In particular, we see that the streamwise forcing does not influence the dynamics of
    $
    \bphi_2 \, = \, \left[\,\tau_{22}\,\,\,\tau_{23}\,\,\,\tau_{33}\,\right]^T;
    $
on the other hand, this forcing creates a $\We$-independent response of
    $
    \bphi_4
    \, = \,
    \left[\,\tau_{12}\,\,\,\tau_{13}\,\right]^T,
    $
and a response of
    $
    \phi_5
    \, = \,
    \tau_{11}
    $
that scales linearly with $\We$. Furthermore, $d_2$ and $d_3$ induce
(i) a $\We$-independent response of $\bphi_2$; (ii) a response of $\bphi_4$ that depends linearly on $\We$; and (iii) a response of $\phi_5$ that scales quadratically with $\We$. Therefore, in high-Weissenberg-number flows, the wall-normal and spanwise forcing fluctuations have the strongest influence, and the impact of these forces is most powerful on the streamwise component of the polymer stress tensor, $\tau_{11}$. This follows from the observation that the frequency responses from both $d_2$ and $d_3$ to $\tau_{11}$ scale {\em quadratically\/} with the Weissenberg number; the frequency responses from all other inputs to other polymer stress components scale at most {\em linearly\/} with $\We$.

We note that almost all operators that are multiplied by the Weissenberg number in Figure~\ref{fig.blockTau} contain stretching of polymer stress fluctuations by a background shear as an integral part. The only exceptions are (i) the operators $\bS_{51}$ and $\bS_{53}$ which, respectively, arise from transport and stretching of base polymer stress by velocity fluctuations; (ii) the operator $\bS_{41}$ which captures both of these phenomena; and (iii) the operator $\bar{\bC}_p$ which, in addition to polymer stretching, also accounts for vortex tilting. In~\ref{sec.g} and~\ref{sec.Etau} we show that, in elasticity-dominated flows, vortex tilting has negligible influence on both velocity and polymer stress fluctuations.

The $\We$-scaling of the power spectral densities of the operators $\bGamma_{ik,j}$ that map $d_j$ to $\tau_{ik}$ follows directly from the above discussion, the definition of $\Sigma_{ik,j}$, and the linearity of the trace operator
    \beq
    \begin{aligned}
    &
    \left[
    \begin{array}{ccc}
    \Sigma_{22,1} (k_z,\omega;\We,\beta,\eps)
    &
    \Sigma_{23,1} (k_z,\omega;\We,\beta,\eps)
    &
    \Sigma_{33,1} (k_z,\omega;\We,\beta,\eps)
    \\[0.1cm]
    \Sigma_{12,1} (k_z,\omega;\We,\beta,\eps)
    &
    \Sigma_{13,1} (k_z,\omega;\We,\beta,\eps)
    &
    \Sigma_{11,1} (k_z,\omega;\We,\beta,\eps)
    \end{array}
    \right]
    \\
    =~
    &
    \left[ \begin{array} {ccc}
    0
    &
    0
    &
    0
    \\[0.1cm]
    {\bSigma_{12,1} (k_z,\omega;\beta,\eps)}
    &
    {\bSigma_{13,1} (k_z,\omega;\beta,\eps)}
    &
    {\bSigma_{11,1} (k_z,\omega;\beta,\eps) \, \We^2}
    \end{array}
    \right],
    \end{aligned}
    \non
    \end{equation}
    \beq
    \begin{aligned}
    &
    \left[
    \begin{array}{ccc}
    \Sigma_{22,j} (k_z,\omega;\We,\beta,\eps)
    &
    \Sigma_{23,j} (k_z,\omega;\We,\beta,\eps)
    &
    \Sigma_{33,j} (k_z,\omega;\We,\beta,\eps)
    \\[0.1cm]
    \Sigma_{12,j} (k_z,\omega;\We,\beta,\eps)
    &
    \Sigma_{13,j} (k_z,\omega;\We,\beta,\eps)
    &
    \Sigma_{11,j} (k_z,\omega;\We,\beta,\eps)
    \end{array}
    \right]
    \\
    =~
    &
    \left[ \begin{array} {ccc}
    \bSigma_{22,j} (k_z,\omega;\beta,\eps)
    &
    \bSigma_{23,j} (k_z,\omega;\beta,\eps)
    &
    \bSigma_{33,j} (k_z,\omega;\beta,\eps)
    \\[0.1cm]
    {\bSigma_{12,j} (k_z,\omega;\beta,\eps) \, \We^2}
    &
    {\bSigma_{13,j} (k_z,\omega;\beta,\eps) \, \We^2}
    &
    {\bSigma_{11,j} (k_z,\omega;\beta,\eps) \, \We^4}
    \end{array}
    \right],
    ~~
    j = 2,3.
    \end{aligned}
    \non
    \end{equation}
Here, $\bSigma_{ik,j}$ are the power spectral densities of the $\We$-independent operators $\bar{\bGamma}_{ik,j} (k_z,\omega;\beta,\eps)$, and the aggregate effect of the forcing vector $\bd$ to the six independent components of $\btau$ can be obtained using square additivity
    \beq
    \Sigma (k_z,\omega;\We,\beta,\eps)
    \, = \,
    \bSigma_a (k_z,\omega;\beta,\eps)
    \, + \,
    \bSigma_b (k_z,\omega;\beta,\eps) \, \We^2
    \, + \,
    \bSigma_c (k_z,\omega;\beta,\eps) \, \We^4,
    \non
    \eeq
with
    \[
    \ba{rcl}
    \bSigma_a
    & \! = \! &
    \bSigma_{12, 1}
    \, + \,
    \bSigma_{13, 1}
    \, + \,
    \sum_{j \, = \, 2}^{3}
    \left(
    \bSigma_{22, j} \, + \, \bSigma_{23, j} \, + \, \bSigma_{33, j}
    \right),
    \\[0.1cm]
    \bSigma_b
    & \! = \! &
    \bSigma_{11, 1}
    \, + \,
    \sum_{j \, = \, 2}^{3}
    \left(
    \bSigma_{12, j} \, + \, \bSigma_{13, j}
    \right),
    \\[0.1cm]
    \bSigma_c
    & \! = \! &
    \bSigma_{11, 2}
    \, + \,
    \bSigma_{11, 3}.
    \ea
    \]
Furthermore, the steady-state variance maintained in the independent components of $\btau$ by $\bd$ is determined by
    \beq
    E_{\tau} (k_z;\We,\beta,\eps)
    \; = \;
    a (k_z;\beta,\eps)
    \; + \;
    b (k_z;\beta,\eps) \, \We^2
    \; + \;
    c (k_z;\beta,\eps) \, \We^4,
    \tag*{(${\rm E}_{\tau}$)}
    \label{eq.Etau}
    \eeq
where, for example,
    \beq
    a (k_z;\beta,\eps)
    \, = \,
    \dfrac{1}{2 \pi}
    \int_{-\infty}^{\infty}
    \bSigma_a (k_z,\omega;\beta,\eps)
    \,
    \mrd \omega,
    \non
    \eeq
and similarly for $b (k_z;\beta,\eps)$ and $c (k_z;\beta,\eps)$.

The principal results of this section, that the remainder of the paper builds upon, are the scaling relationships~\ref{eq.Ev} and~\ref{eq.Etau} which, respectively, highlight the quadratic and quartic $\We$-dependence of the steady-state variance amplification associated with velocity and polymer stress fluctuations. We note that (i) the block diagrams in Figs.~\ref{fig.Block} and~\ref{fig.blockTau} identify polymer stretching as the key physical ingredient underlying these scaling relationships; and (ii) the scaling of the functions $f$ and $g$ in~\ref{eq.Ev} and the functions $a$, $b$, and $c$ in~\ref{eq.Etau} with $\eps$ in the high-elasticity-number limit is the topic of~\ref{sec.sp}.

\section{Main result: Variance amplification in elasticity-dominated flows}
    \label{sec.main}

In this section, we present the main result of this paper which reveals previously unknown structural similarities between velocity fluctuation dynamics in strongly elastic flows of viscoelastic fluids and strongly inertial flows of Newtonian fluids. We also provide analytical expressions for the variance amplification and discuss physical mechanisms leading to amplification from the forcing to velocity and polymer stress components. The most important mechanism involves the stretching of the polymer stress fluctuations by a background shear, and it introduces the lift-up of flow fluctuations in a similar manner as vortex tilting does in inertia-dominated flows of Newtonian fluids. Furthermore, we determine the spanwise length scales of flow structures that contribute most to the steady-state variance and show that the most energetic velocity fluctuations assume the form of high and low speed streaks. These exhibit striking similarity to the flow structures that contain the most energy in shear flows of Newtonian fluids with high Reynolds numbers.

The results presented in this section are obtained by transforming the linearized dynamics into slow and fast subsystems and then applying singular perturbation methods. For clarity of presentation, we discuss the main results here and relegate the details to the appendices.

\subsection{Variance amplification of velocity fluctuations}
    \label{sec.main-Ev}

Based on the developments in~\ref{sec.Ev} and~\ref{sec.f-eps}, it follows that in streamwise-constant Poiseuille and Couette flows of Oldroyd-B fluids with sufficiently large $\mu$, the variance maintained in $\bv$ is given by
    \beq
    E_{\mrv} (k_z;\We,\beta,\mu)
    \; = \;
    \mu \tilde{f}_0 (k_z)/\beta
    \; + \;
    \tilde{f}_1 (k_z)
    \,
    (1 - \beta)/\beta^2
    \; + \;
    \We^2 \, \tilde{g}_0 (k_z) \, (1 - \beta)^2/\beta
    \; + \;
    \cO(1/\mu).
    \label{eq.Ev-main}
    \eeq
Here, $\tilde{f}_0$, $\tilde{f}_1$, and $\tilde{g}_0$ are functions independent of $\We$, $\mu$, and $\beta$ that capture spatial frequency responses of velocity fluctuations in elasticity-dominated flows. As demonstrated in~\ref{sec.Pi_a}, the linear scaling with $\mu$ of the first term on the right-hand-side of~(\ref{eq.Ev-main}) originates from the corresponding power spectral density becoming almost uniformly distributed over the temporal frequency bandwidth which is proportional to $\mu$. Furthermore, the base-flow-independent functions $\tilde{f}_0 (k_z)$ and $\tilde{f}_1 (k_z)$ are given by (cf.\ (\ref{eq.f-high-mu}))
    \[
    \tilde{f}_0 (k_z)
    \; = \;
    f_N (k_z)
    \; = \;
    - \frac{1}{2} \, \trace \left( \bSos^{-1} \, + \, \bSsq^{-1} \right),
    ~~
    \tilde{f}_1 (k_z)
    \; = \;
    - \frac{1}{2} \, \trace \left( \bSos^{-2} \, + \, \bSsq^{-2} \right),
    \]
with $f_N (k_z)$ being the function that arises in the expression for variance amplification in Newtonian fluids~(\ref{eq.Ev-Newt}). Since this function accounts for viscous dissipation, it does not introduce any important viscoelastic physical effects. On the other hand, the function $\tilde{g}_0$ accounts for the stretching of the polymer stress fluctuations by a background shear, and it is determined by~(cf.\ (\ref{eq.g0}))
    \[
    \tilde{g}_0 (k_z)
    \, = \,
    (k_z^2 / 4)
    \,
    \trace
    \left(
    \bSsq^{-1} \tilde{\bC}_{p2} \bSos^{-2} \tilde{\bC}_{p2}^* \bSsq^{-1}
    \right).
    \]
Expression~(\ref{eq.Ev-main}) shows that the contribution of this base-flow-dependent term to the steady-state velocity variance is proportional to $\We^2$ and that it increases monotonically with a decrease in the ratio of the solvent viscosity to the total viscosity.

    \begin{figure*}
    \centering
    \subfloat[]{\includegraphics[width=0.33\textwidth]{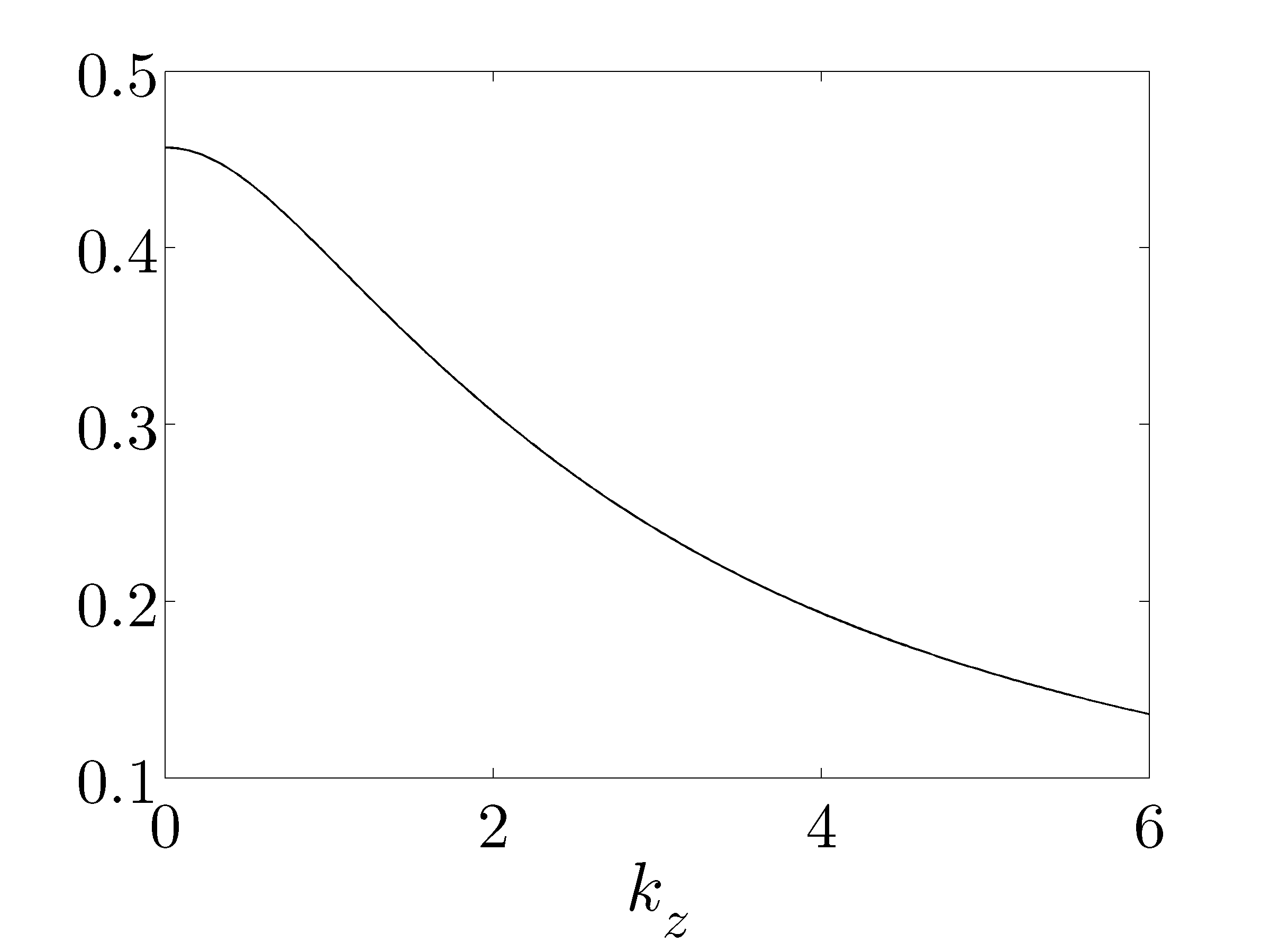}
    \label{fig.f0}}
    \subfloat[]{\includegraphics[width=0.33\textwidth]{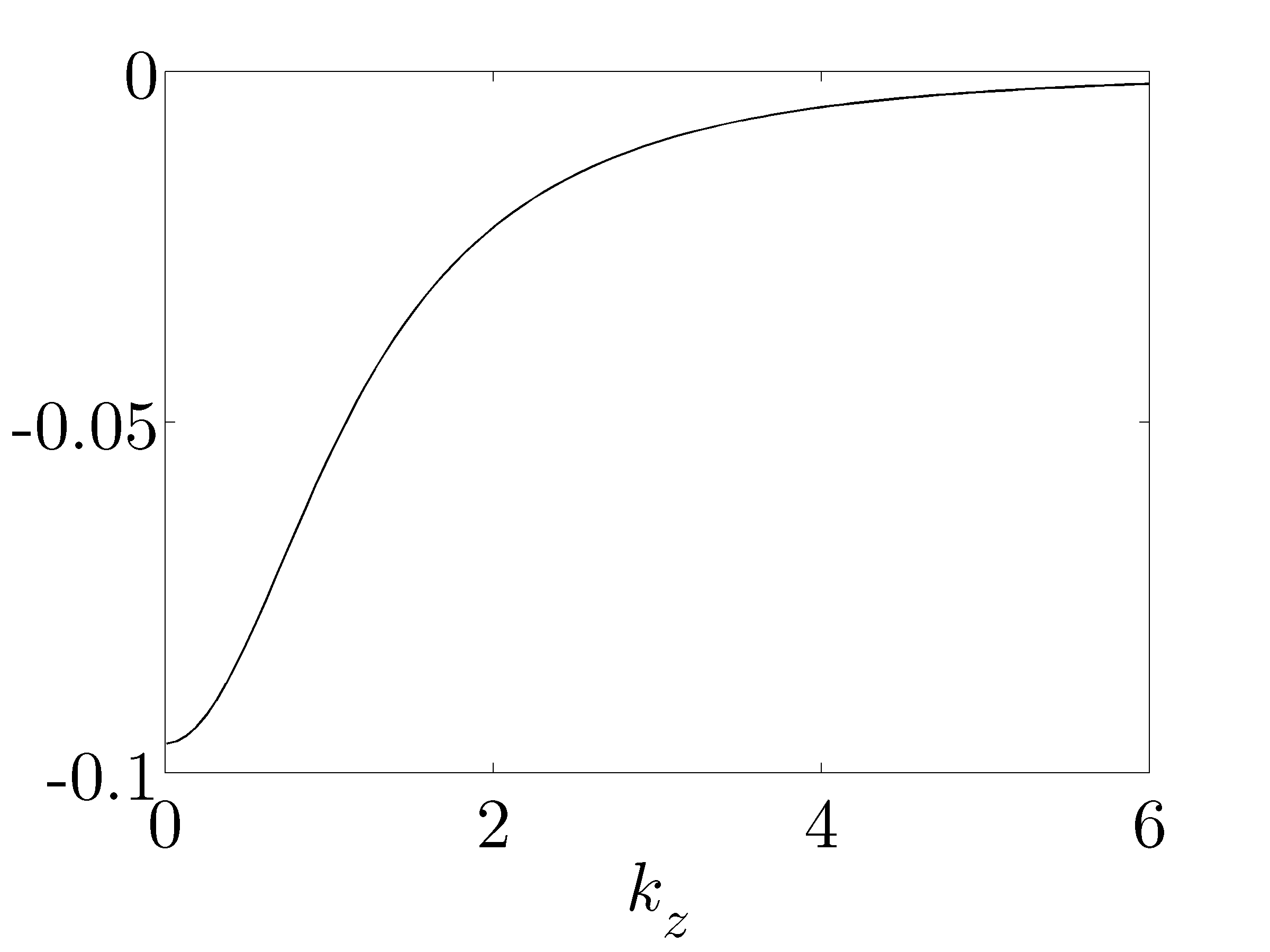}
    \label{fig.f1}}
    \subfloat[]{\includegraphics[width=0.33\textwidth]{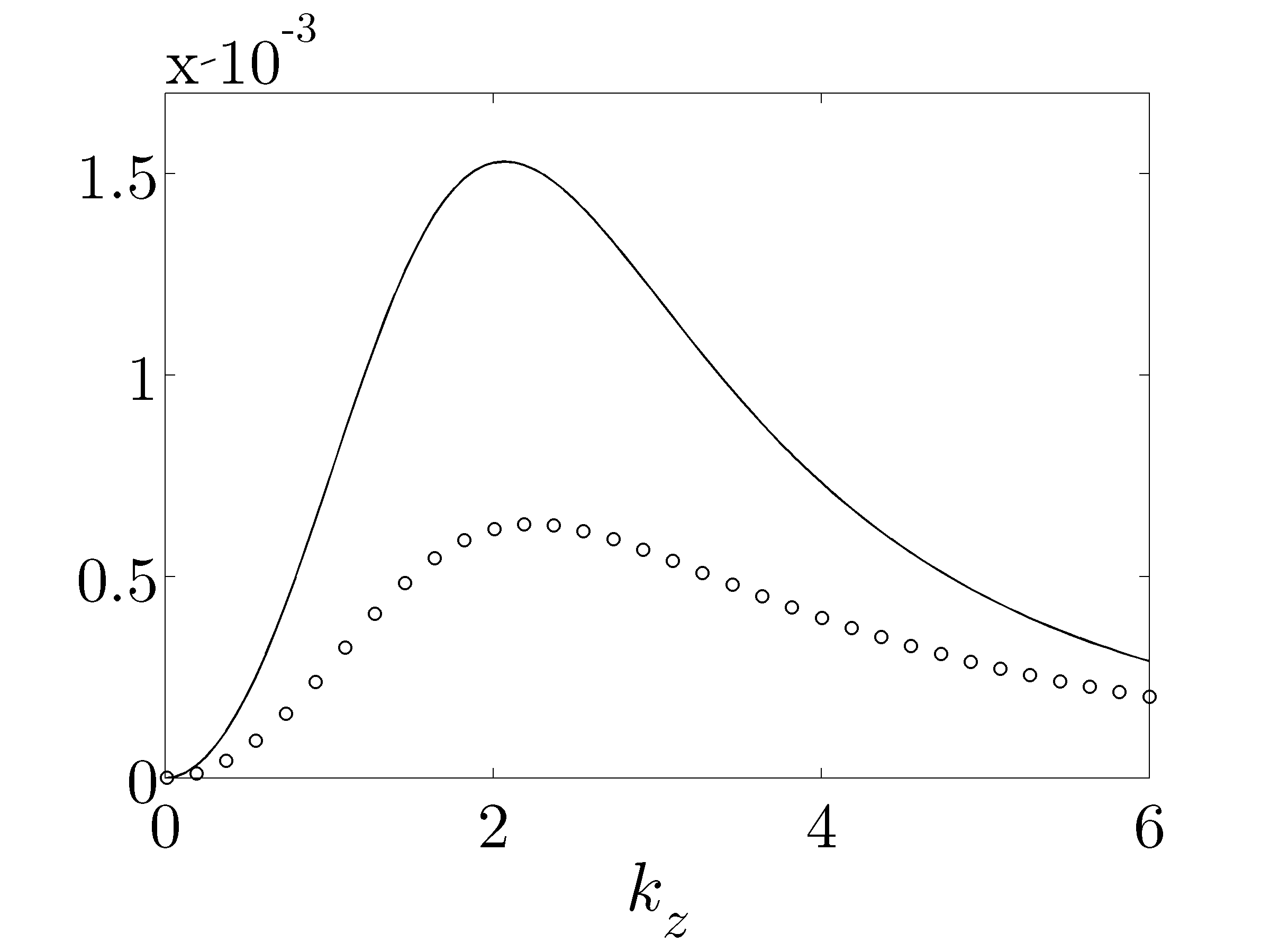}
    \label{fig.g0}}
    \caption{Plots of:
    (a) $\tilde{f}_0 (k_z)\/$;
    (b) $\tilde{f}_1 (k_z)\/$;
    (c) $\tilde{g}_0 (k_z)\/$ in both Couette (solid curve) and Poiseuille (circles) flows.}
    \label{fig.f0-g0}
    \end{figure*}

The analytical expressions for $\trace \, (\bS_{\mrk}^{-1})$ with $\mrk = \{ \mbox{os}, \mbox{sq} \}$ were derived in~\cite{Bamieh2001}; these are used to evaluate $\tilde{f}_0 (k_z) = f_N (k_z)$, which is illustrated in Figure~\ref{fig.f0}. The behavior of this function, as well as function $\tilde{f}_1 (k_z)$ in Figure~\ref{fig.f1}, is governed by viscous dissipation. In Couette flow, the expression for $\tilde{g}_0$ simplifies to
    \beq
    \tilde{g}_0 (k_z)
    \; = \;
    - (k_z^2/4) \, \trace \left( \bSos^{-2} \bSsq^{-1} \right)
    \; = \;
    - (k_z^2/4) \,
    \trace
    \left(
    \Delta^{-2} \Delta \, \Delta^{-2}
    \right),
    \label{eq.g0c}
    \eeq
and an explicit $k_z$-dependence of $\tilde{g}_0$ can be derived after some manipulation. The resulting expression for $\tilde{g}_0 (k_z)$ is used to generate the plot in Figure~\ref{fig.g0}; from this plot we observe the non-monotonic character of $\tilde{g}_0 (k_z)$, with peaks at $k_{z} \approx 2.07$ (in Couette flow) and $k_{z} \approx 2.24$ (in Poiseuille flow). In Poiseuille flow, determination of the expression for $\tilde{g}_0 (k_z)$ is considerably more involved than in Couette flow; however, the method developed in~\cite{jovbamSCL06} can be used to compute this quantity efficiently without resorting to spatial discretization. We note that, at $k_z = 0$, the function $\tilde{g}_0$ becomes equal to zero. On the other hand, at large $k_z$ both $\bSos^{-1}$ and $\bSsq^{-1}$ approximately scale as $1/k_z^2$. Therefore, the function $\tilde{g}_0$ in~(\ref{eq.g0c}) becomes negligibly small as $k_z \rightarrow \infty$. A similar argument holds in Poiseuille flow, which explains the appearance of the peaks at $k_z \neq 0$ in Figure~\ref{fig.g0}. As mentioned earlier, the values of $k_z$ where these peaks emerge determine the spanwise length scales of the most energetic response of velocity fluctuations to stochastic forcing in flows with high Weissenberg numbers.

\begin{figure}
    \centering
    {
    \begin{tabular}{c}
    \subfloat[]{ \setlength{\unitlength}{1.15cm} \begin{picture}(11.25,1.95)(0,0)
   % Block diagram for the linearized NS equations

    %first part - d2
    \put(0,1.525){\vector(1,0){0.75}}
    \put(0.375,1.725){\makebox(0,0)[b]{$ d_2 $}}
    \put(0.75,1.1){\framebox(1,0.85){$\bF_2$}}
    \put(1.75,1.525){\vector(1,0){0.6}}
    \put(2.5,1.525){\circle{0.3}}

    %second part - orr-sommerfeld
    \put(2.65,1.525){\vector(1,0){0.6}}
    \put(3.25,1.1){\framebox(2.25,0.85){$ \dfrac{- \, 1}{\beta ( \mri \omega ) \, + \, 1} \, \bSos^{-1} $}}
    \put(4.375,2.15){\makebox(0,0)[b]{\btab{c} \tc{bgblue}{\bf `diffusion'} \etab}}
    \put(5.5,1.525){\vector(1,0){0.75}}
    \put(5.875,1.725){\makebox(0,0)[b]{$ \xi $}}
    %third part - coupling
    \put(6.25,1.1){\framebox(1.25,0.85){$ \tc{red}{\We} \, \bC_{p2}$}}
    \put(6.875,2.15){\makebox(0,0)[b]{\btab{c} \tc{red}{\bf polymer} \\ \tc{red}{\bf stretching} \etab}}
    \put(7.5,1.525){\vector(1,0){0.75}}
    \put(8.25,1.1){\framebox(2.25,0.85){$ \dfrac{- \, (1 - \beta)}{\beta ( \mri \omega ) \, + \, 1} \, \bSsq^{-1}$}}
    \put(9.375,2.15){\makebox(0,0)[b]{\btab{c} \tc{bgblue}{\bf viscous} \\ \tc{bgblue}{\bf dissipation} \etab}}
    \put(10.5,1.525){\vector(1,0){0.75}}
    \put(10.875,1.725){\makebox(0,0)[b]{$ u $}}

    %part 6 - dw to v
    \put(0,0.425){\vector(1,0){0.75}}
    \put(0.375,0.625){\makebox(0,0)[b]{$ d_3 $}}
    \put(0.75,0){\framebox(1,0.85){$\bF_3$}}
    \put(1.75,0.425){\line(1,0){0.75}}
    \put(2.5,0.425){\vector(0,1){0.95}}
\end{picture}
    \label{fig.StokesOBv}}
    \\
    \\[0.35cm]
    \subfloat[]{ \setlength{\unitlength}{1.15cm} \begin{picture}(11.25,1.75)(0,0)
   % Block diagram for the linearized NS equations

    %first part - d2
    \put(0,1.375){\vector(1,0){0.75}}
    \put(0.375,1.575){\makebox(0,0)[b]{$ d_2 $}}
    \put(0.75,1){\framebox(1.,0.75){$ \bF_2$}}
    \put(1.75,1.375){\vector(1,0){0.6}}
    \put(2.5,1.375){\circle{0.3}}

    %second part - orr-sommerfeld
    \put(2.65,1.375){\vector(1,0){0.6}}
    \put(3.25,1){\framebox(2.25,0.75){$ \left(  \mri \omega \bI \, - \, \bSos \right)^{-1} $}}
    \put(4.375,1.95){\makebox(0,0)[b]{\btab{c} \tc{bgblue}{\bf `diffusion'} \etab}}
    \put(5.5,1.375){\vector(1,0){0.75}}
    \put(5.875,1.575){\makebox(0,0)[b]{$ \psi $}}
    %third part - coupling
    \put(6.25,1){\framebox(1.25,0.75){$ \tc{red}{Re} \, \bC_{p1}$}}
    \put(6.875,1.95){\makebox(0,0)[b]{\btab{c} \tc{red}{\bf vortex} \\ \tc{red}{\bf tilting} \etab}}
    \put(7.5,1.375){\vector(1,0){0.75}}
    \put(8.25,1){\framebox(2.25,0.75){$ \left(  \mri \omega \bI \, - \, \bSsq \right)^{-1} $}}
    \put(9.375,1.95){\makebox(0,0)[b]{\btab{c} \tc{bgblue}{\bf viscous} \\ \tc{bgblue}{\bf dissipation} \etab}}
    \put(10.5,1.375){\vector(1,0){0.75}}
    \put(10.875,1.575){\makebox(0,0)[b]{$ u $}}

    %part 6 - dw to v
    \put(0,0.375){\vector(1,0){0.75}}
    \put(0.375,0.575){\makebox(0,0)[b]{$ d_3 $}}
    \put(0.75,0){\framebox(1.,0.75){$ \bF_3$}}
    \put(1.75,0.375){\line(1,0){0.75}}
    \put(2.5,0.375){\vector(0,1){0.85}}
\end{picture}
    \label{fig.Newtonian}}
    \end{tabular}
    }
    \caption{Block diagrams of the frequency response operators that map the wall-normal and spanwise forces to the streamwise velocity fluctuation in streamwise-constant
    (a) creeping flows of Oldroyd-B fluids, cf.~(\ref{eq.u-sob}); and
    (b) inertial flows of Newtonian fluids, cf.~(\ref{eq.u-lnse}).
    In Newtonian fluids amplification originates from vortex tilting, i.e.\ operator
    $
    \bC_{p1},
    $
    and in viscoelastic fluids it originates from polymer stretching, i.e.\ operator
    $
    \bC_{p2}.
    $
    Note that the Weissenberg number in creeping flows of Oldroyd-B fluids takes the role of the Reynolds number in inertial flows of Newtonian fluids.
    }
    \label{fig.comparison}
    \end{figure}
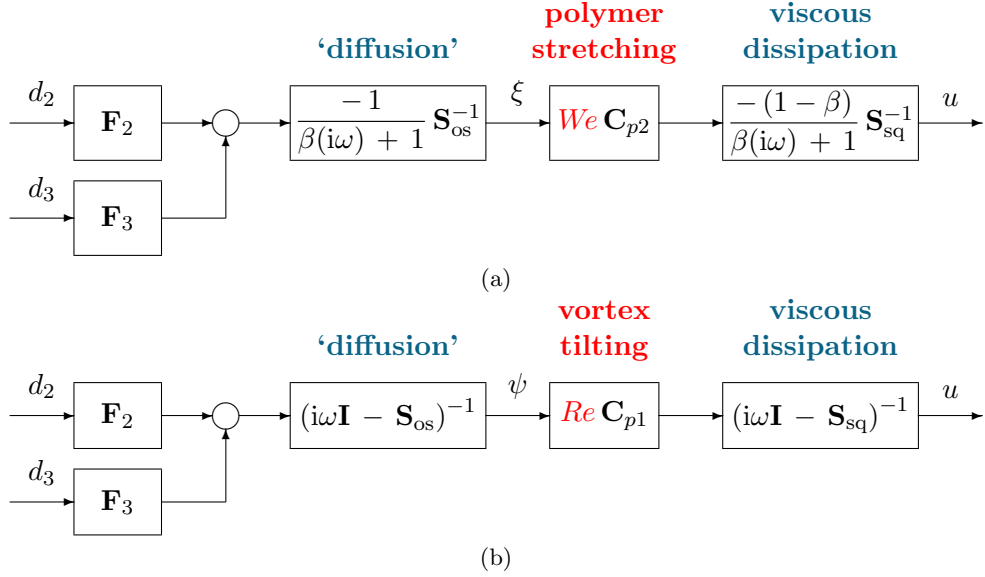

We next discuss the physical mechanisms leading to amplification from the wall-normal and spanwise forces to the streamwise velocity fluctuation. As demonstrated in~\ref{sec.g}, in flows with high elasticity numbers, the inertialess model ($\eps = 0$) captures well the responses from $d_2$ and $d_3$ to $u$. In the absence of inertia, the dynamics of the streamwise velocity are governed by
    \beq
    \ba{rcl}
    \tbo{\bSos \, \dot{\xi}}{\bSsq \, \dot{u}}
    & \!\! = \!\! &
    \tbt{- (1/\beta) \, \bSos}{0}{\We \, \frac{\beta - 1}{\beta} \, \bC_{p2}}{- (1/\beta) \, \bSsq}
    \tbo{\xi}{u}
    \, + \,
    \tbo{-(1/\beta) \bF_j}{0}
    d_j,
    ~~
    j
    \, = \,
    \{ 2,3 \},
    \ea
    \label{eq.u-sob}
    \eeq
which corresponds to the slow subsystem discussed in~\ref{sec.g}. From~\ref{sec.Hij-ss} we note that $\xi$ is obtained by filtering high temporal frequencies in the streamfunction $\psi$
    \beq
    \xi
    \, = \,
    \dfrac{1}{\mri \omega \, + \, 1} \, \psi
    ~~\Rightarrow~~
    \dot{\xi}
    \, = \,
    - \xi
    \, + \,
    \psi.
    \non
    \eeq
In comparison, by scaling time with the diffusive time $\rho L^2/\eta_s$, the responses from $d_2$ or $d_3$ to $u$ in the streamwise-constant linearized Navier-Stokes equations are captured by
    \beq
    \ba{rcl}
    \tbo{\dot{\psi}}{\dot{u}}
    & \!\! = \!\! &
    \tbt{\bSos}{0}{Re \, \bC_{p1}}{\bSsq}
    \tbo{\psi}{u}
    \, + \,
    \tbo{\bF_j}{0}
    d_j.
    \ea
    \label{eq.u-lnse}
    \eeq
Figs.~\ref{fig.StokesOBv} and~\ref{fig.Newtonian} illustrate the block diagram representations of systems~(\ref{eq.u-sob}) and~(\ref{eq.u-lnse}), respectively.

As evident from both~(\ref{eq.u-sob}) and the expression for $\tilde{g}_0 (k_z)$, the coupling term $\bC_{p2}$ plays an essential role in variance amplification (for additional illustration, see the block diagram in Figure~\ref{fig.StokesOBv}); if this term was zero, the dynamics of strongly elastic flows, at the level of velocity fluctuations, would be dominated by viscous dissipation. A careful analysis of the governing equations (see~\ref{sec.fr}) shows that
    \[
    \bC_{p2}
    \; = \;
    {\bS}_{34} \, {\bS}_{42} \, {\bS}_{21},
    \]
which demonstrates that the operator $\bC_{p2}$ emerges from
    \begin{itemize}
    \item[(i)] the wall-normal and spanwise velocity ($v,w$) gradients, $\bS_{21}$, in the equation for ($\tau_{22},\tau_{23},\tau_{33}$);

    \item[(ii)] stretching of $\tau_{22}$, $\tau_{23}$, and $\tau_{33}$ by the background shear, ${\bS}_{42}$, in the equation for ($\tau_{12},\tau_{13}$);

    \item[(iii)] the $\tau_{12}$ and $\tau_{13}$ gradients, ${\bS}_{34}$, in the equation for $u$.
    \end{itemize}

From a physical point of view, the wall-normal and spanwise forces produce weak, i.e.\ $\cO (1)$, streamwise vortices; cf.\ the $\xi$-subsystem in~(\ref{eq.u-sob}), where $\xi$ denotes the low-pass version of the streamfunction $\psi$,
    $
    \xi
    =
    \psi/(\mri \omega + 1).
    $
Spatial gradients in streamwise vortices, i.e.\ $\bS_{21} \xi$, yield $\cO (1)$ polymer stress fluctuations in the ($y,z$)-plane, ($\tau_{22},\tau_{23},\tau_{33}$). The background shear, i.e.\ $\bS_{42} \bphi_2$, stretches $\tau_{22}$ and $\tau_{23}$ ($U'(y) \tau_{22}$ and $U'(y) \tau_{23}$, respectively), thereby introducing $\cO (\We)$ fluctuations in $\tau_{12}$ and $\tau_{13}$. Finally, the wall-normal gradients of $\tau_{12}$ (i.e., $\py \tau_{12}$) and the spanwise gradients of $\tau_{13}$ (i.e., $\pz \tau_{13}$), i.e.\ $\bS_{34} \bphi_4$,  generate $\cO (\We)$ fluctuations in streamwise velocity which then get dissipated by the action of viscosity. All of these give rise to polymer stretching, leading to a transfer of energy from the base flow to fluctuations which results in large steady-state velocity variances in flows with high Weissenberg numbers.

Energy transfer from a base flow to fluctuations has been observed experimentally in elastic turbulence of swirling flow between two parallel disks~\citep{Groisman2000,Groisman2004,bursegste06,bursegste07,junste10}. As mentioned earlier, a radial pressure gradient which acts on the fluid along the curved streamlines introduces an elastic instability and promotes this energy transfer~\cite{larshamul90,Larson1992,Shaqfeh1996}. The present work demonstrates that, even in inertialess rectilinear flows, an energy transfer from a base flow to fluctuations can be initiated by high flow sensitivity. It remains an open question whether this nonmodal amplification mechanism, that arises from stretching of polymer stress fluctuations by base shear, can trigger the onset of elastic turbulence in channel flows of viscoelastic fluids. Progress in this area requires a deeper understanding of the interplay between the streak sensitivity~\cite{schhus02} and the nonlinear feedback that the streamwise-varying fluctuations induce on the streamwise rolls~\cite{wal97}. Experiments using highly viscous flows of elastic fluids in a circular pipe suggest that the pressure, and presumably other flow variables, begin to fluctuate irregularly at sufficiently large Weissenberg numbers~\cite{yes02,yes09}. However, additional experiments and calculations aimed at characterizing different stages of disturbance development are needed in order to make more definitive comparisons between theory and experiment.

\begin{figure}
    \centering
    {
    \subfloat[Couette flow with $k_z = 2.07$.]
    {\includegraphics[width=0.4\textwidth]{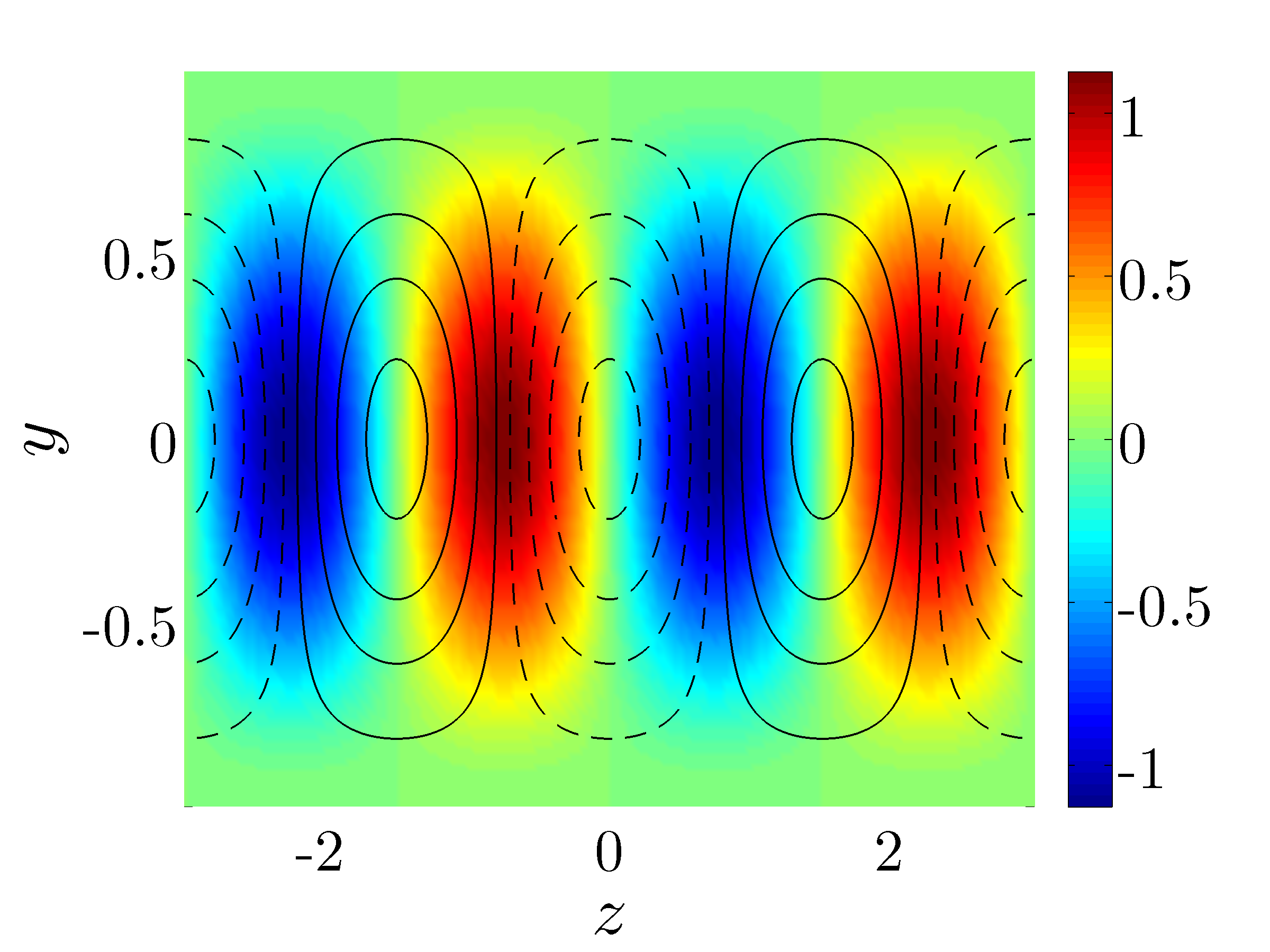}
    \label{fig.4a}}
    \subfloat[Poiseuille flow with $k_z = 2.24$.]
    {\includegraphics[width=0.4\textwidth]{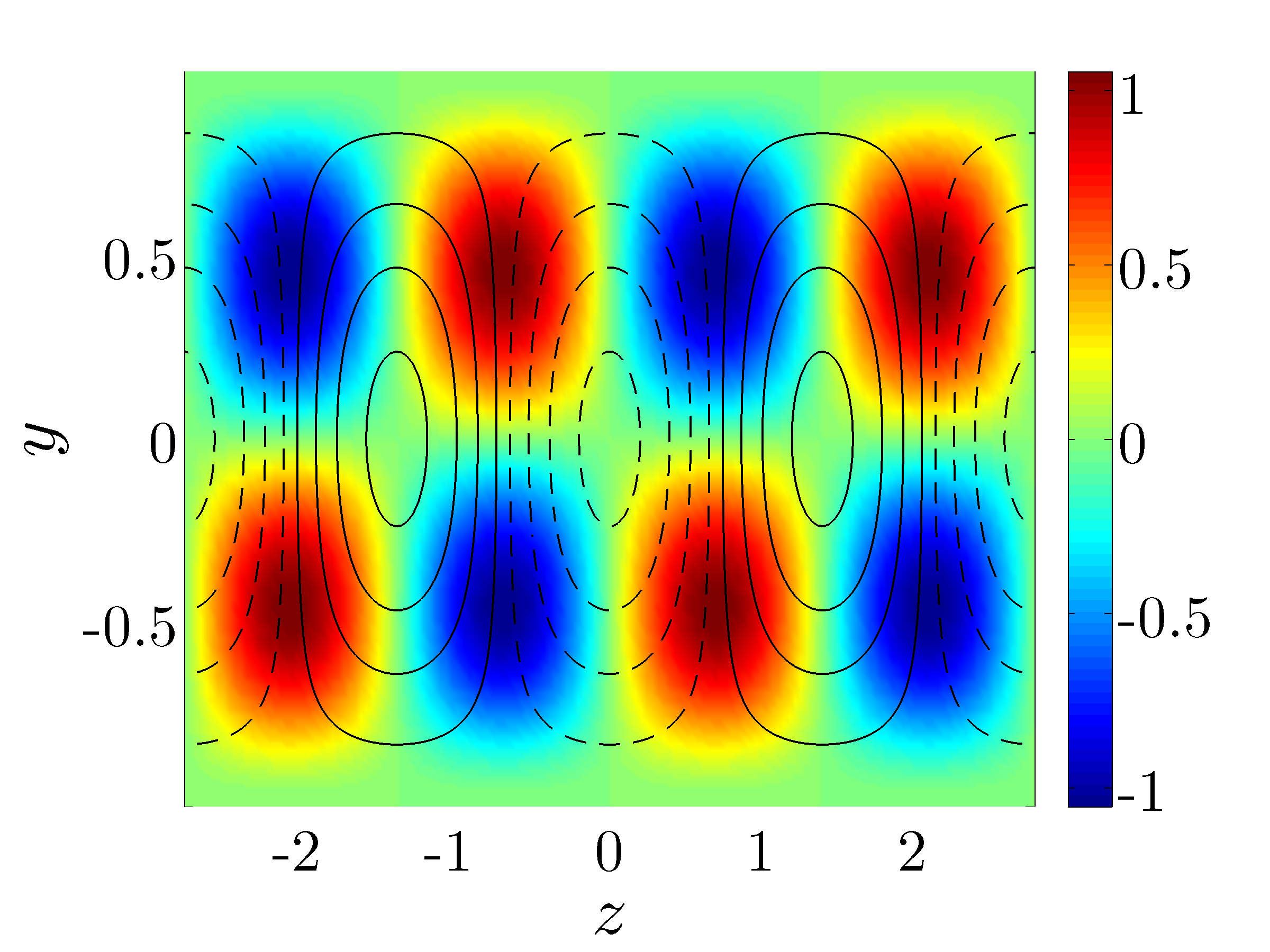}
    \label{fig.4b}}
    }
    \caption{Color plots: streamwise velocity fluctuations $u (z,y)$ containing the most variance in strongly elastic flows subject to wall-normal and spanwise stochastic forcing. Contour lines: fluctuations in the low-pass version of the streamfunction, $\xi (z,y)$. In Couette flow the most amplified set of fluctuations in $u$ accounts for $89 \, \%$ of the total variance, and in Poiseuille flow it accounts for $77 \, \%$ of the total variance.}
    \label{fig.u}
    \end{figure}

Streamwise velocity fluctuations that contain the most variance in strongly elastic flows with $k_{z} = 2.07$ (Couette) and $k_{z} = 2.24$ (Poiseuille) are shown in Figure~\ref{fig.u}. These structures are purely harmonic in $z$ and their wall-normal shapes are determined by the principal eigenfunctions of operators
    $
    (k_z^2 / 4)
    \bSsq^{-1} \tilde{\bC}_{p2} \bSos^{-2} \tilde{\bC}_{p2}^* \bSsq^{-1}
    $~\citep{Farrell1993}.
The most amplified sets of fluctuations are given by high (hot colors) and low (cold colors) speed streaks, with pairs of counter-rotating streamwise vortices in between them (contour lines). In Couette flow the streaks occupy the entire channel width, and in Poiseuille flow they are antisymmetric with respect to the channel's centerline.

These flow structures have striking resemblance to the initial conditions responsible for the largest transient growth in channel flows of Newtonian fluids~\citep{Butler1992}. Despite similarities, the fluctuations shown in Figure~\ref{fig.u} and in~\cite{Butler1992} arise from fundamentally different physical mechanisms: in high $Re$-flows of Newtonian fluids, vortex tilting is the main driving force for amplification; in high $\We$-flows of viscoelastic fluids, it is the polymer stretching mechanism described above. These two mechanisms are, respectively, captured by the action of $\bC_{p1}$ and $\bC_{p2}$ on $\psi$ and the low-pass version of $\psi$ (cf.\ the block diagrams in Figures~\ref{fig.StokesOBv} and~\ref{fig.Newtonian}). From the definitions of these operators it follows that both of them contain the background shear $U'(y)$ and the spatial variations in the flow fluctuations as their essential ingredients. In particular, in Couette flow $\bC_{p2} = U'(y) \Delta \mri k_z = \Delta \mri k_z$ and $\bC_{p1} = - U'(y) \mri k_z = - \mri k_z$. This observation in conjunction with the block diagrams in Figures~\ref{fig.StokesOBv} and~\ref{fig.Newtonian} suggests that polymer stretching in elasticity-dominated channel flows of viscoelastic fluids redistributes the mean momentum and introduces the lift-up of flow fluctuations in a similar manner as vortex tilting does in inertia-dominated flows of Newtonian fluids~\cite{Landahl1975}. In Newtonian fluids, large amplification originates from tilting of the base spanwise vorticity, $- U'(y)$, by spanwise changes in the streamfunction, $\mri k_z \psi$. In Couette flow of Oldroyd-B fluids, $U'(y)$ stretches $\tau_{22}$ and $\tau_{23}$, or equivalently it gets tilted by $\Delta \mri k_z \xi$.

Finally, we note that simple kinematics of streamwise-constant flows allow for equivalent representation of system~(\ref{eq.u-sob}) (or the block diagram in Figure~\ref{fig.StokesOBv}) in terms of a low-pass version of the wall-normal velocity,
    $
    \vartheta = \mri k_z \xi = v/(\mri \omega + 1),
    $
and the wall-normal vorticity, $\eta = \mri k_z u$. This can be achieved by replacing $\xi$ by $\vartheta$, $u$ by $\eta$, and $\bF_j$ by $\mri k_z \bF_j$ in~(\ref{eq.u-sob}), thereby yielding the wall-normal vorticity equation in inertialess streamwise-constant flows of Oldroyd-B fluids~(\ref{eq.eta-SOB}).

\subsection{Variance amplification of polymer stress fluctuations}
    \label{sec.main-Etau}

From results obtained in~\ref{sec.Etau} it follows that in streamwise-constant Poiseuille and Couette flows of Oldroyd-B fluids with sufficiently large $\mu$, the variance maintained in polymer stress fluctuations approximately becomes elasticity-number independent
    \beq
    E_{\tau} (k_z;\We,\beta,\mu)
    \; = \;
    a_0 (k_z;\beta)
    \; + \;
    \We^2
    \,
    b_0 (k_z;\beta)
    \; + \;
    \We^4
    \,
    c_0 (k_z;\beta)
    \; + \;
    \cO (1/\mu).
    \label{eq.Etau-Re0-main}
    \eeq
Here, $a_0$, $b_0$, and $c_0$ are functions independent of $\We$ and $\mu$ that capture the spatial frequency responses and $\beta$-dependence of $\btau$ in inertialess channel flows.

\begin{figure}
    \centering
    {
    \subfloat[]
    {\includegraphics[width=0.33\textwidth]{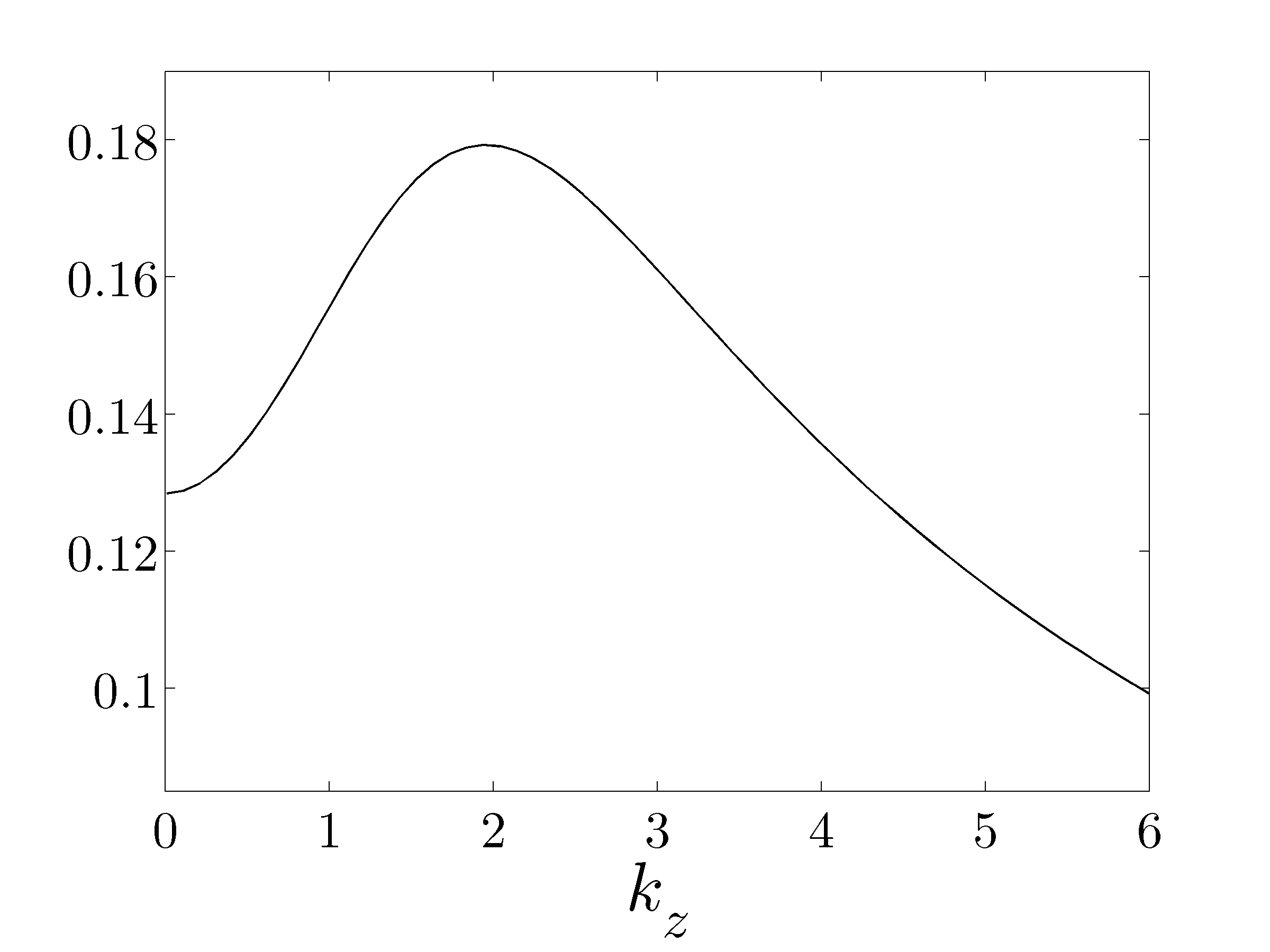}
    \label{fig.aOS0}}
    \subfloat[]
    {\includegraphics[width=0.33\textwidth]{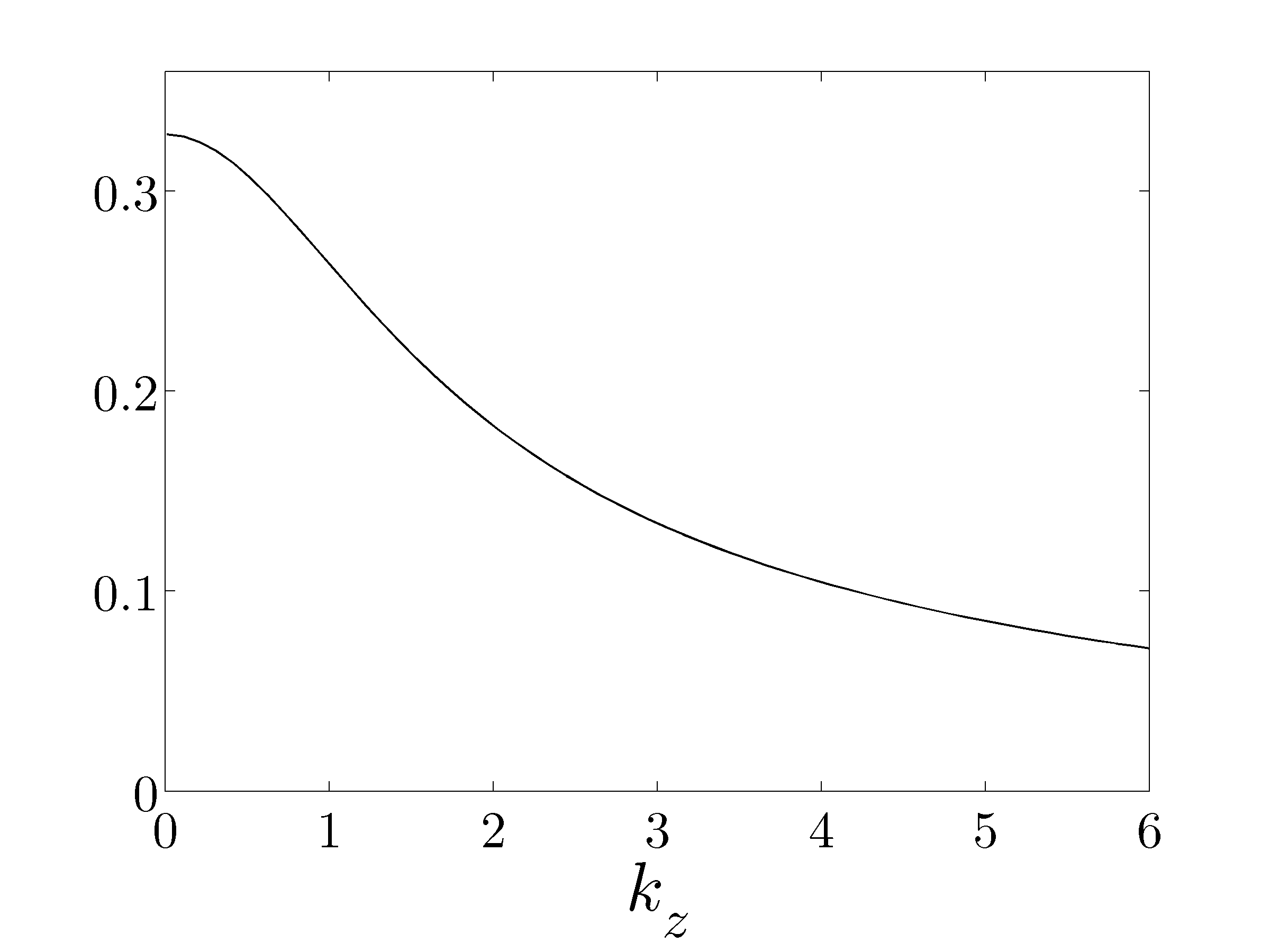}
    \label{fig.aSQ0}}
    \subfloat[]
    {\includegraphics[width=0.33\textwidth]{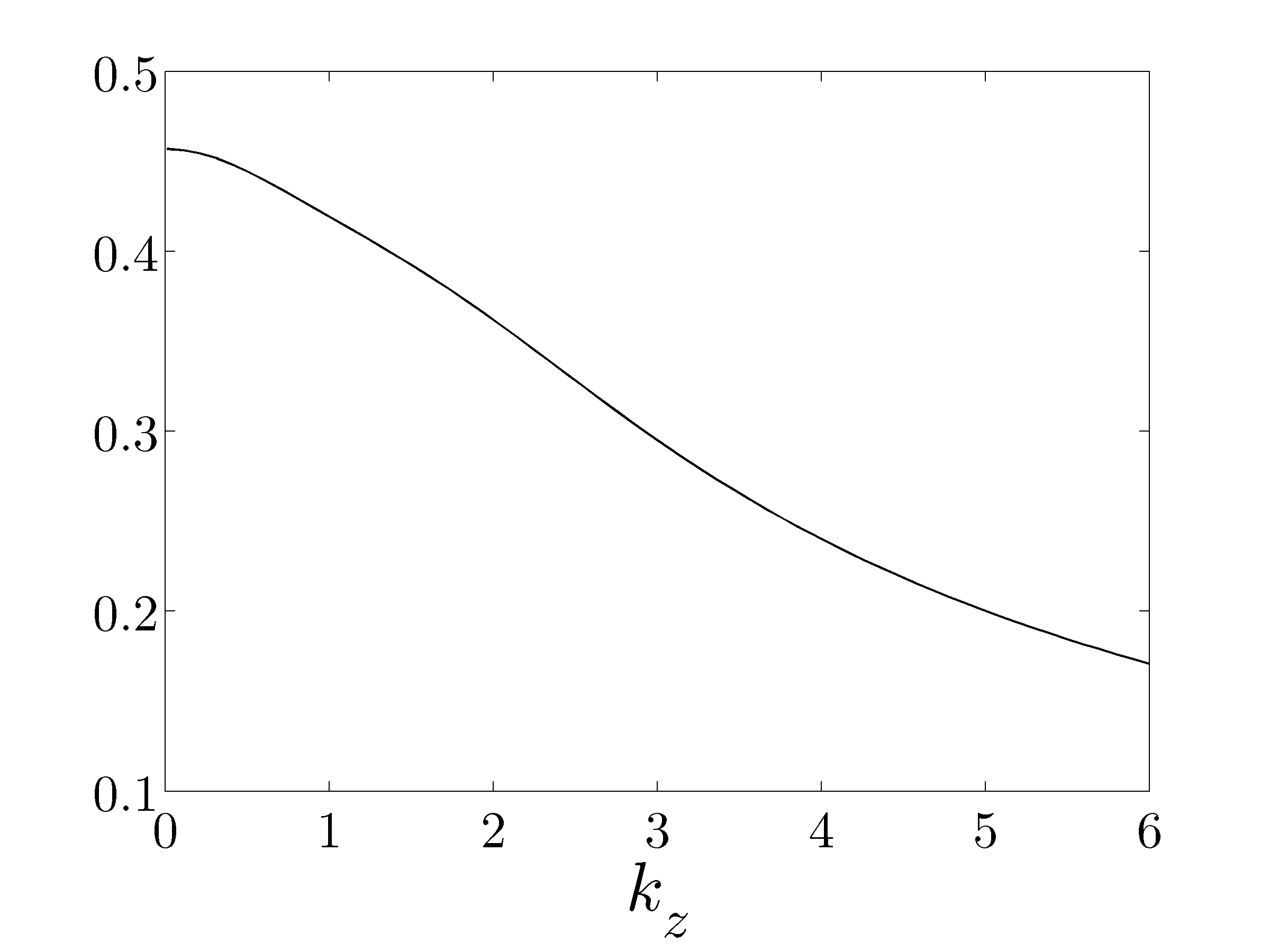}
    \label{fig.aOS0+aSQ0}}
    }
    \caption{Plots of the base-flow-independent functions
    (a) $\tilde{a}_{\mathrm{os},0} (k_z)$;
    (b) $\tilde{a}_{\mathrm{sq},0} (k_z)$;
    and
    (c) $\tilde{a}_0 (k_z) = \tilde{a}_{\mathrm{os},0} (k_z) + \tilde{a}_{\mathrm{sq},0} (k_z)$. In inertialess flows,
    the variance amplification of the operators that map
    $
    \left[\,d_2\,\,\,d_3\,\right]^T
    $
to
    $
    \bphi_2
    =
    \left[\,\tau_{22}\,\,\,\tau_{23}\,\,\,\tau_{33}\,\right]^T
    $
and
    $d_1$
    to
    $
    \bphi_4
    =
    \left[\,\tau_{12}\,\,\,\tau_{13}\,\right]^T
    $
is determined by $a_0 (k_z;\beta) = \tilde{a}_0 (k_z)/\beta$.
    }
    \label{fig.a0}
    \end{figure}

As shown in~\ref{sec.Etau}, the function $a_0$ is base-flow-independent and it is determined by
    $
    a_0 (k_z; \beta)
    =
    \tilde{a}_0 (k_z)/\beta,
    $
with (cf.\ (\ref{eq.a}))
    \beq
    \ba{rcl}
    \tilde{a}_0 (k_z)
    & \!\!\! = \!\!\! &
    \tilde{a}_{\mathrm{os},0} (k_z)
    \, + \,
    \tilde{a}_{\mathrm{sq},0} (k_z),
    \\[0.15cm]
    \tilde{a}_{\mathrm{os},0} (k_z)
    & \!\!\! = \!\!\! &
    (1/2) \,
    \trace
    \left(
    \bSos^{-2} \, \bS_{21}^* \, \bS_{21}
    \right)
    \, = \,
    2 \, k_z^2
    \,
    \trace
    \left(
    \Delta^{-2} \, \Delta \, \Delta^{-2} \, \pyy
    \right)
    \, - \,
    (1/2) \,
    \trace
    \left(
    \Delta^{-2} \Delta
    \right),
    \\[0.15cm]
    \tilde{a}_{\mathrm{sq},0} (k_z)
    & \!\!\! = \!\!\! &
    (1/2) \,
    \trace
    \left(
    \bSsq^{-2} \, \bS_{43}^* \, \bS_{43}
    \right)
    \, = \,
    - \,
    (1/2)
    \,
    \trace
    \left(
    \Delta^{-1}
    \right).
    \ea
    \non
    \eeq
Clearly, $a_0$ depends on the Orr-Sommerfeld and Squire operators, and the operators $\bS_{21}$ and $\bS_{43}$ which introduce gradients of velocity fluctuations (i.e., $\nabla \bv$) in the constitutive equations. Note that the functions $\tilde{a}_{\mathrm{os},0}$ and $\tilde{a}_{\mathrm{sq},0}$, respectively, quantify the steady-state variance amplification (as a function of the spanwise wavenumber) of the operators that map
    $
    \left[\,d_2\,\,\,d_3\,\right]^T
    $
to
    $
    \bphi_2
    =
    \left[\,\tau_{22}\,\,\,\tau_{23}\,\,\,\tau_{33}\,\right]^T
    $
and
    $d_1$
    to
    $
    \bphi_4
    =
    \left[\,\tau_{12}\,\,\,\tau_{13}\,\right]^T.
    $
Plots in Figure~\ref{fig.a0} show that $\tilde{a}_{\mathrm{os},0}$ reaches its maximum at $\cO (1)$ values of $k_z$, while $\tilde{a}_{\mathrm{sq},0}$ is characterized by viscous dissipation and it
decays monotonically with $k_z$.

\begin{figure}
    \centering
    {
    \subfloat[]
    {\includegraphics[width=0.33\textwidth]{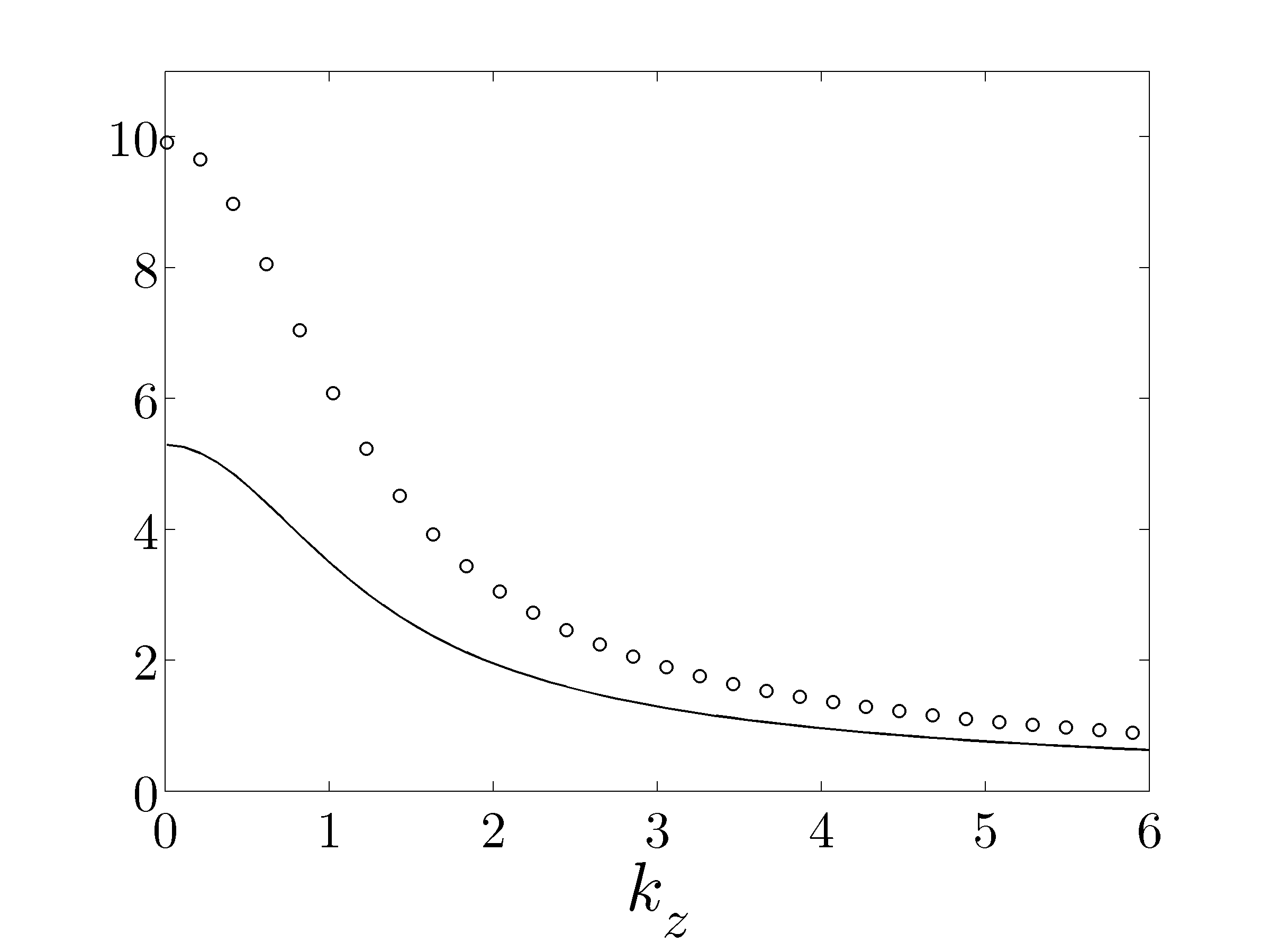}
    \label{fig.b0-phi5-d1}}
    \subfloat[]
    {\includegraphics[width=0.33\textwidth]{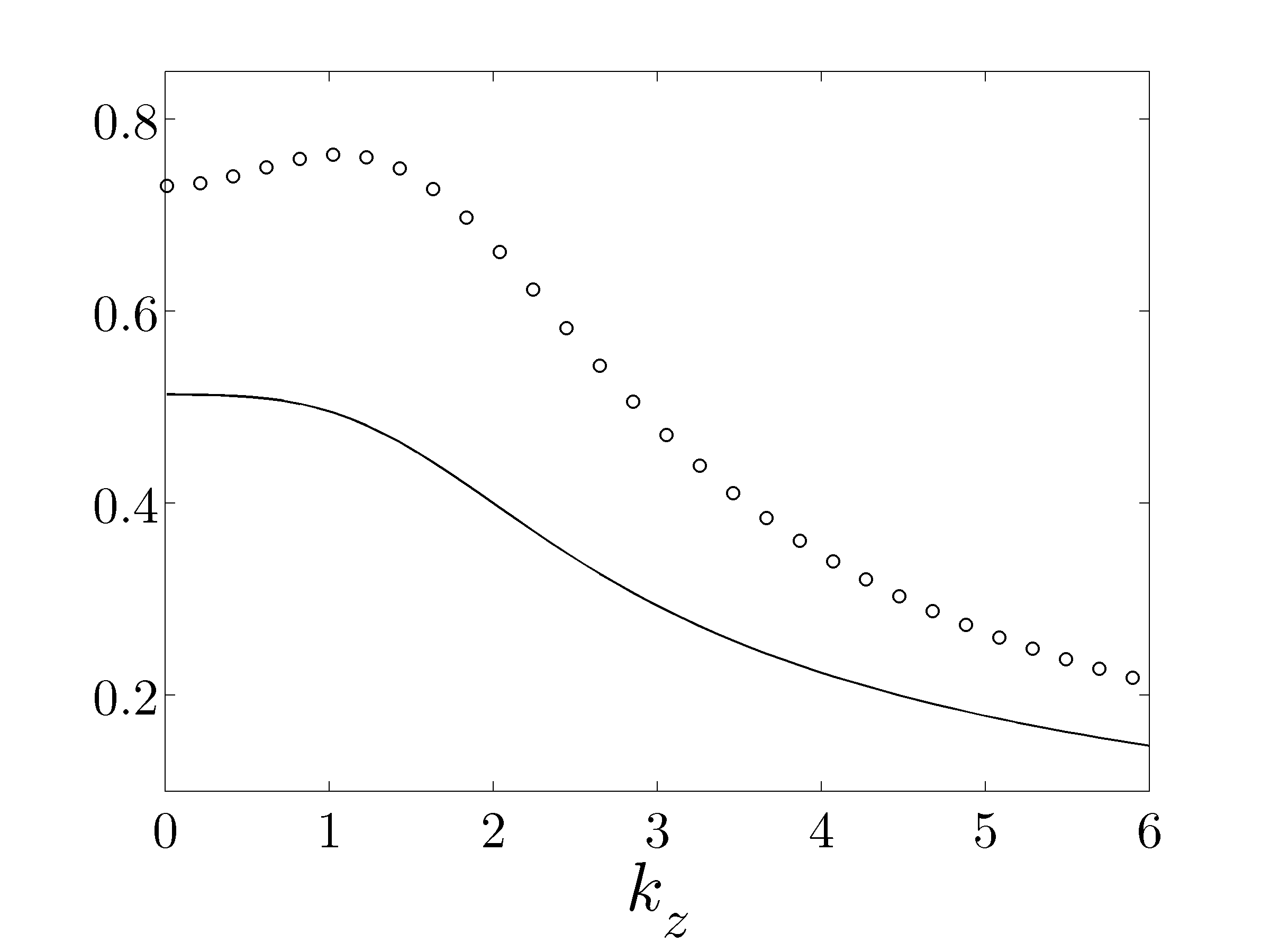}
    \label{fig.b0-phi4-d23}}
    \subfloat[]
    {\includegraphics[width=0.33\textwidth]{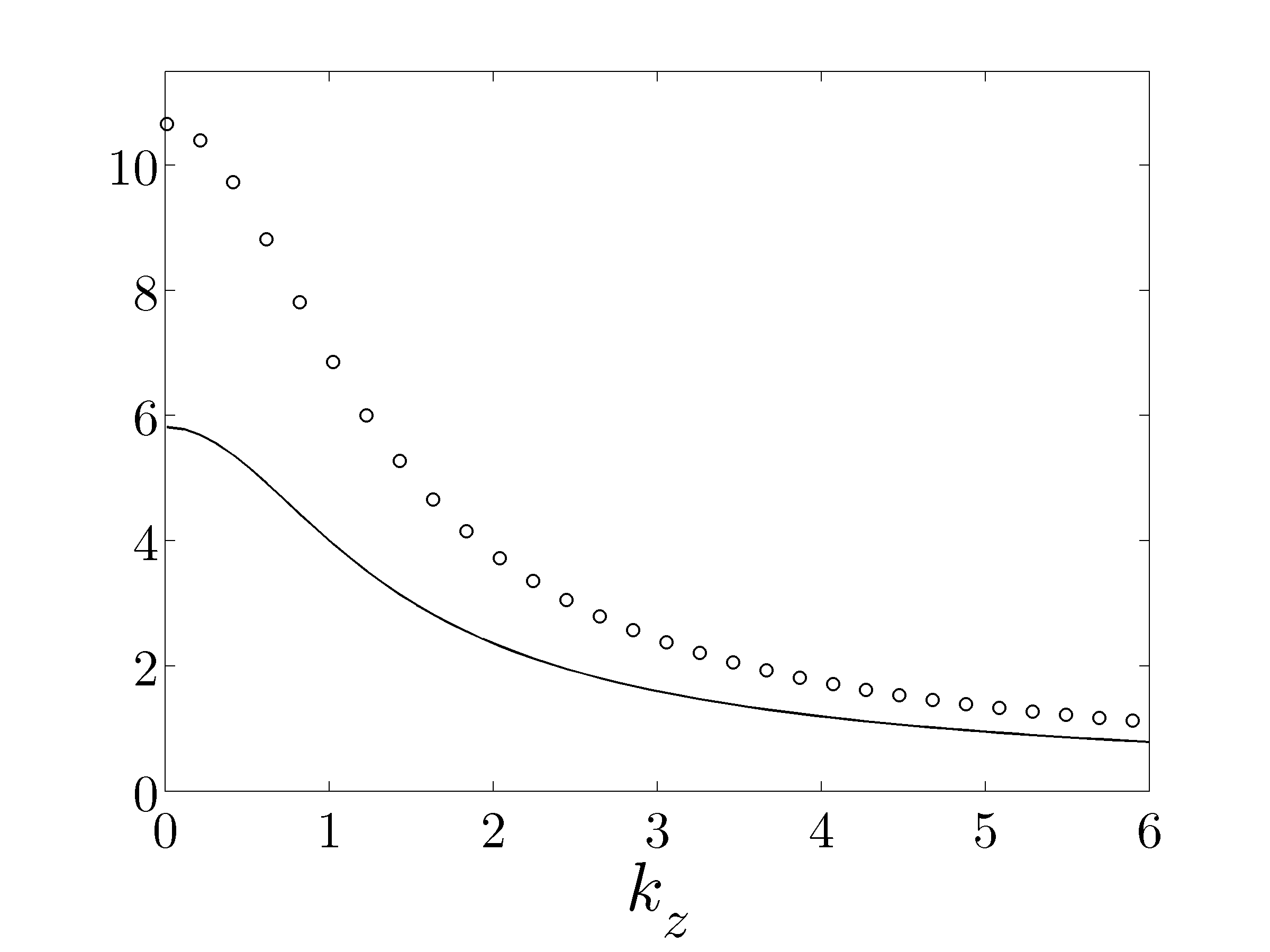}
    \label{fig.b0all}}
    }
    \caption{Plots of the functions
    (a) $b_{\phi_5} (k_z; 0.5,0)$;
    (b) $b_{\phi_4} (k_z; 0.5,0)$;
    and
    (c) $b_0 (k_z; 0.5) = b_{\phi_4} (k_z; 0.5,0) + b_{\phi_5} (k_z; 0.5,0)$ in both Couette (solid curves) and Poiseuille (circles) flows. In inertialess flows with $\We = 1$, the variance amplification of the operators that map $d_1$ to $\phi_5 = \tau_{11}$ and $[\,d_2\,\,\,d_3\,]^T$ to $\bphi_4 = \left[\,\tau_{12}\,\,\,\tau_{13}\,\right]^T$ is determined by $b_0 (k_z; \beta)$.}
    \label{fig.b0}
    \end{figure}

In contrast to $a_0$, the functions $b_0$ and $c_0$ in~(\ref{eq.Etau-Re0-main}) differ in Couette and Poiseuille flows. As shown in~\ref{sec.b}, $b_0$ determines the variance amplification from $d_1$ to $\phi_5 = \tau_{11}$ and from $[\,d_2\,\,\,d_3\,]^T$ to $\bphi_4 = \left[\,\tau_{12}\,\,\,\tau_{13}\,\right]^T$ in inertialess channel flows with $\We = 1$. To signify this, we write $b_0$ as
    \[
    b_0 (k_z; \beta)
    \, = \,
    b (k_z; \beta, \eps = 0)
    \, = \,
    b_{\phi_4} (k_z; \beta,0) \, + \, b_{\phi_5} (k_z; \beta,0),
    \]
where
    \[
    b_{\phi_4} (k_z; \beta,0)
    \, = \,
    \ds{\sum_{j \, = \, 2}^{3}}
    \left(
    b_{12, j} (k_z; \beta,0) \, + \, b_{13, j} (k_z; \beta,0)
    \right),
    ~~
    b_{\phi_5} (k_z; \beta,0)
    \, = \,
    b_{11,1} (k_z; \beta,0),
    \]
quantify the contributions of $\bphi_4$ and $\phi_5$ to the term responsible for the quadratic scaling of $E_\tau$ with the Weissenberg number (cf.\ (\ref{eq.Etau-Re0-main})). The function $b_{\phi_5}$ is given by (cf.\ (\ref{eq.b0-tau11-d1}))
    \beq
    \ba{rcl}
    b_{\phi_5} (k_z;\beta,0)
    & \!\! = \!\! &
    \dfrac{1 \, + \, 4 \beta}{2 \beta (1 \, + \, \beta)}
    \,
    \trace
    \left(
    \bSsq^{-1}
    \,
    \bS_{53}^*
    \,
    \bS_{53}
    \,
    \bSsq^{-1}
    \right)
    \\[0.35cm]
    & \!\! = \!\! &
    -
    \,
    \dfrac{2 ( 1 \, + \, 4 \beta )}{\beta (1 \, + \, \beta)}
    \,
    \trace
    \left(
    \Delta^{-1}
    \left(
    2 \, U'(y) \, U''(y) \, \py
    \, + \,
    \left( U'(y) \right)^2 \pyy
    \right)
    \Delta^{-1}
    \right),
    \ea
    \non
    \eeq
and the function $b_{\phi_4}$ can be computed using the Lyapunov equation (see~\ref{sec.Ev}) associated with~(\ref{eq.ee-phi4-d23}). Figure~\ref{fig.b0} illustrates the $k_z$-dependence of the $b$ functions in inertialess Couette and Poiseuille flows with $\beta = 0.5$ and $\We = 1$. In Poiseuille flow the function $b_{\phi_4}$ peaks at $\cO (1)$ values of $k_z$, while all the other functions in Figure~\ref{fig.b0} decay monotonically with $k_z$. Furthermore, since $b_{\phi_4}$ achieves much smaller values than $b_{\phi_5}$, the shape of $b_0$ is primarily determined by the amplification from $d_1$ to $\tau_{11}$.

In inertialess Couette flow, the variance amplification from the wall-normal and spanwise forces to the streamwise component of the polymer stress tensor is determined by (cf.\ (\ref{eq.c0-c}))
    \beq
    \ba{rcl}
    c_0 (k_z; \beta)
    & \!\! = \!\! &
    \dfrac{4 \beta^4 \, + \, 16 \beta^3 \, + \, 29 \beta^2 \, + \, 6 \beta \, + \, 1}{(\beta \, + \, 1)^3}
    \,
    \tilde{c}_0 (k_z),
    \\[0.35cm]
    \tilde{c}_0 (k_z)
    & \!\! = \!\! &
    k_z^2
    \,
    \trace
    \left(
    \py \, \Delta^{-2} \, \Delta \, \Delta^{-2} \, \py
    \right).
    \ea
    \label{eq.c0-c-main}
    \eeq
This formula separates the spanwise frequency responses from the $\beta$-dependence of the function responsible for the $\We^4$-scaling of $E_{\tau}$ in~(\ref{eq.Etau-Re0-main}). We note that $\tilde{c}_0 (k_z)$ can be efficiently evaluated using the method developed in~\cite{jovbamSCL06} that avoids the need for spatial discretization of the operators. In inertialess Poiseuille flow, the expression for $c_0 (k_z; \beta)$ is significantly more involved than in Couette flow; instead, the Lyapunov equation associated with~(\ref{eq.ee-phi5-d23}) can be used to compute this quantity.

\begin{figure}
    \centering
    {
    \subfloat[]
    {\includegraphics[width=0.4\textwidth]{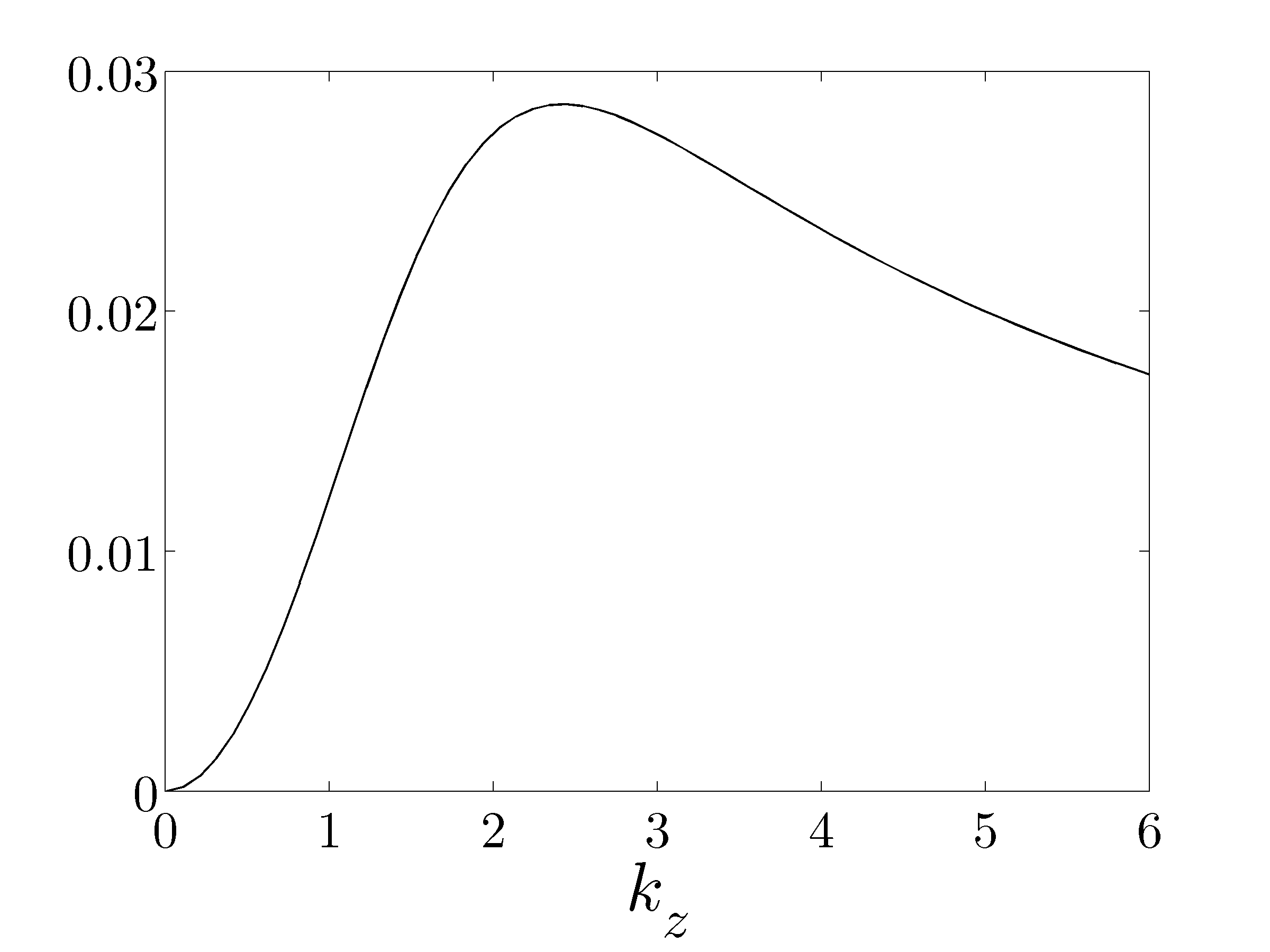}
    \label{fig.c0tilde-cou}}
    \subfloat[]
    {\includegraphics[width=0.4\textwidth]{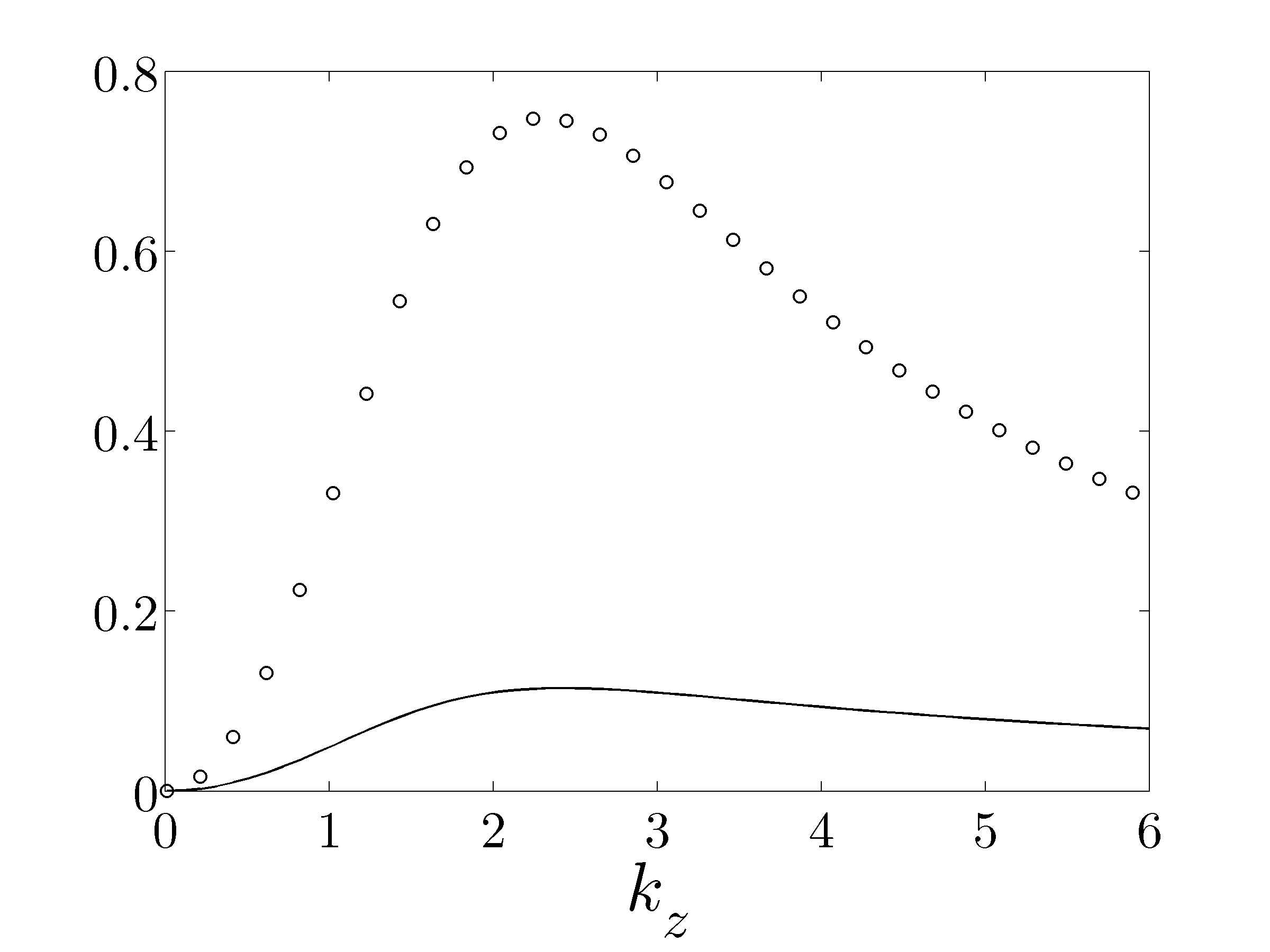}
    \label{fig.c-pos-cou-beta0p5}}
    }
    \caption{Plots of the functions
    (a) $\tilde{c}_0 (k_z)$; and
    (b) $c_0 (k_z; 0.5)$
    in both Couette (solid curves) and Poiseuille (circles) flows. In inertialess flows with $\We = 1$, the variance amplification of the operator that map $[\,d_2\,\,\,d_3\,]^T$ to $\phi_5 = \tau_{11}$ is determined by $c_0 (k_z; \beta)$.}
    \label{fig.c0}
    \end{figure}

The $k_z$-dependence of the function $\tilde{c}_0$ in inertialess Couette flow is shown in Figure~\ref{fig.c0tilde-cou}. Note that $\tilde{c}_0$ peaks at $k_{z} \approx 2.42$ which is the wavenumber determining the spanwise length scale of the most energetic response of $\tau_{11}$ to wall-normal and spanwise stochastic forcing. The non-monotonic character of $\tilde{c}_0 (k_z)$
is induced by the disappearance of this function at both $k_z = 0$ and as $k_z \rightarrow \infty$. The first assertion follows from the definition of $\tilde{c}_0 (k_z)$ in~(\ref{eq.c0-c-main}), and the second assertion follows from the observation that, at large $k_z$, $\Delta^{-2} \Delta \, \Delta^{-2}$ scales as $1/k_z^6$; consequently, for $k_z \gg 1$ we have $\tilde{c}_0 (k_z) \sim 1/k_z^4$ which justifies the existence of the peak at $k_z \neq 0$ in Figure~\ref{fig.c0tilde-cou}. Furthermore, the expression for $c_0 (k_z; \beta)$ in~(\ref{eq.c0-c-main}) shows that the term responsible for the $\We^4$-scaling of $E_{\tau}$ in inertialess Couette flow can be determined by multiplying $\tilde{c}_0 (k_z)$ with a monotonically increasing function of $\beta$.

Figure~\ref{fig.c-pos-cou-beta0p5} illustrates the variance of $\tau_{11}$ maintained by $d_2$ and $d_3$ in inertialess channel flows with $\beta = 0.5$ and $\We = 1$. The largest value of $c_0$ in Poiseuille flow, which takes place at $k_z \approx 2.32$, is about $6.5$ times larger than in Couette flow. We also see that, after reaching its peak, the function $c_0$ decays more rapidly with $k_z$ in Poiseuille flow than in Couette flow. Apart from these minor differences, most essential amplification trends are shared in both cases.

Even though analytical and physical insight into transient responses of inertialess channel flows was provided in~\cite{jovkumPOF10}, the lack of intrinsic spanwise wavelength selection in the Oldroyd-B model driven by initial conditions in stress fluctuations was observed. In fact, in transient growth analysis a high-wavenumber roll-off in $\tau_{11}$ can be obtained only upon inclusion of a small amount of stress diffusion in the constitutive equations~\cite{jovkumPOF10}. In stochastically forced problems, however, the body forces get `filtered' through the equations of motion, thereby providing both a preferred spanwise wavenumber and a roll-off at high $k_z$ even in the absence of stress diffusive terms. As block diagrams in Figures~\ref{fig.StokesOBv} and~\ref{fig.tau11d23} (or, equivalently equations~(\ref{eq.u-sob}) and~(\ref{eq.ee-phi5-d23})) illustrate, the wall-normal and spanwise forces enter into the equations for $u$ and $\tau_{11}$ through the inverse of the Orr-Sommerfeld operator, $\bSos^{-1}$, which effectively introduces `diffusion' in the dynamics of the slow subsystem.

\begin{figure}
    \centering
    {
    \subfloat[Couette flow with $k_z = 2.42$.]
    {\includegraphics[width=0.4\textwidth]{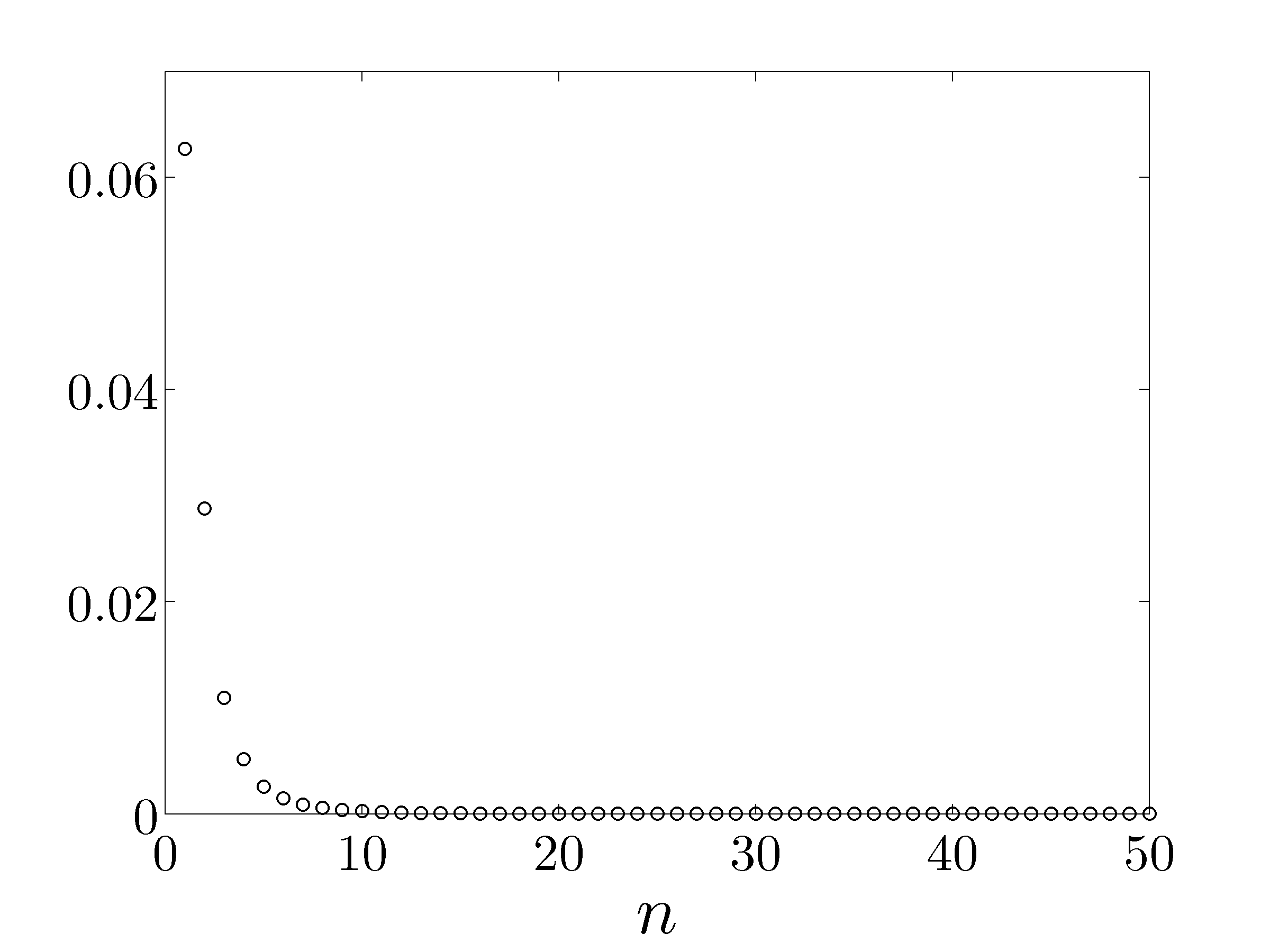}
    \label{fig.eig-P11-d23-cou}}
    \subfloat[Poiseuille flow with $k_z = 2.32$.]
    {\includegraphics[width=0.4\textwidth]{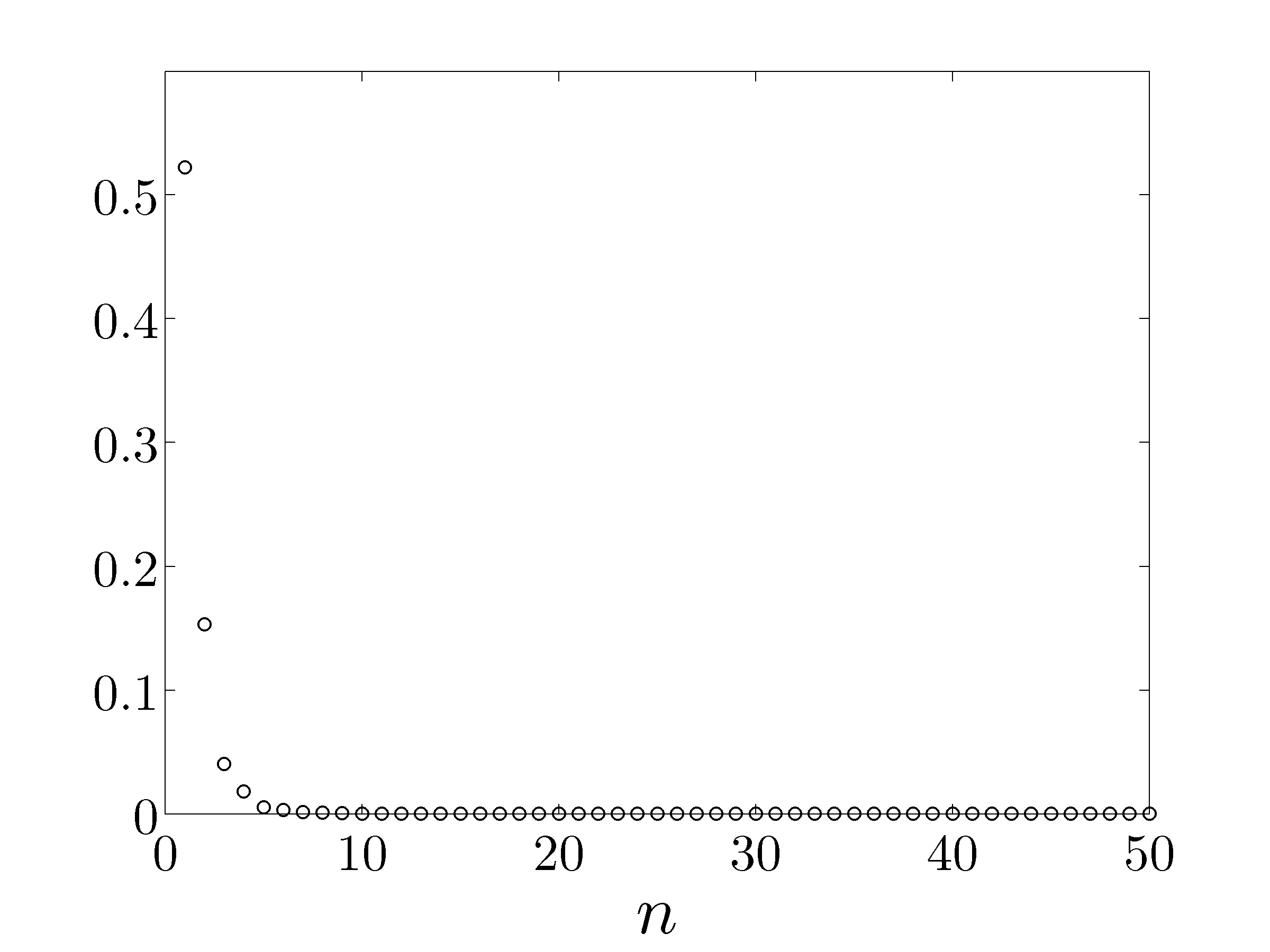}
    \label{fig.eig-P11-d23-pos}}
    }
    \caption{The eigenvalues of the autocorrelation operator of $\tau_{11}$, ordered by magnitude, in inertialess flows with $\We = 1$ and $\beta = 0.5$ subject to wall-normal and spanwise stochastic forcing. In Couette flow two principal eigenvalues contain $80 \, \%$ of the steady-state variance, and in Poiseuille flow they contain $90 \, \%$ of the steady-state variance.}
    \label{fig.eig-P11-d23}
    \end{figure}

Figure~\ref{fig.eig-P11-d23} shows the eigenvalues of the autocorrelation operator of $\tau_{11}$, arranged in descending order, in inertialess flows with $\We = 1$ and $\beta = 0.5$ subject to wall-normal and spanwise stochastic forcing. The sum of these eigenvalues determines the variance maintained in $\tau_{11}$ by $d_2$ and $d_3$~\citep{Farrell1993}. The plots in Figs.~\ref{fig.c0tilde-cou} and~\ref{fig.c-pos-cou-beta0p5} illustrate the existence of two strongly amplified fluctuation types in Couette flow with $k_z = 2.42$ and in Poiseuille flow with $k_z = 2.32$. These values of $k_z$ identify the wavenumbers for which the function $c_0 (k_z; \beta = 0.5)$ achieves its maximum. In Couette flow, the two largest eigenvalues account for $55 \, \%$ and $25 \, \%$ of the total variance, respectively; in Poiseuille flow, they account for $70 \, \%$ and $20 \, \%$ of the total variance.

\begin{figure}
    \centering
    {
    \subfloat[Couette flow with $k_z = 2.42$.]
    {\includegraphics[width=0.4\textwidth]{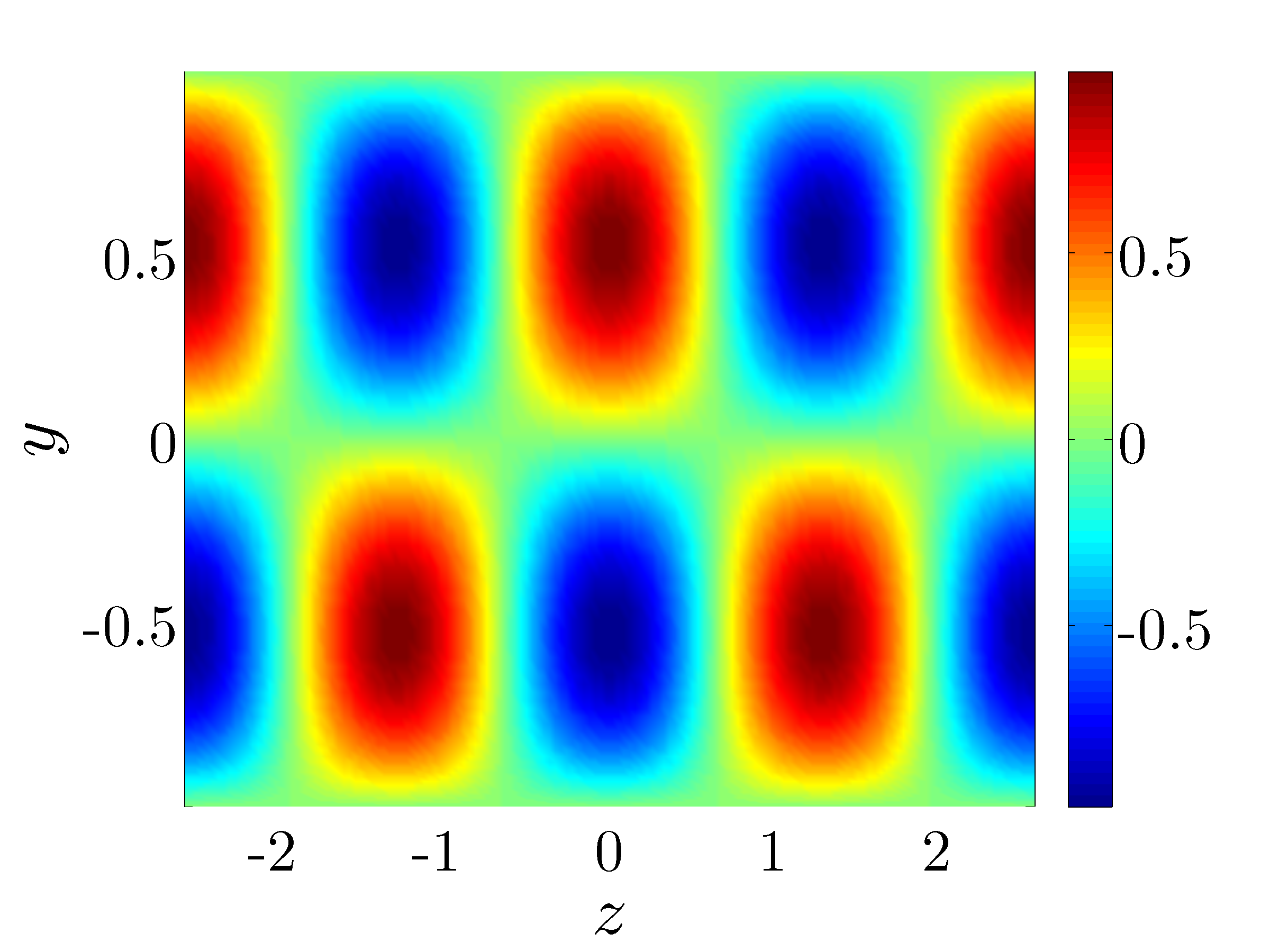}
    \label{fig.T11c-1st}}
    \subfloat[Poiseuille flow with $k_z = 2.32$.]
    {\includegraphics[width=0.4\textwidth]{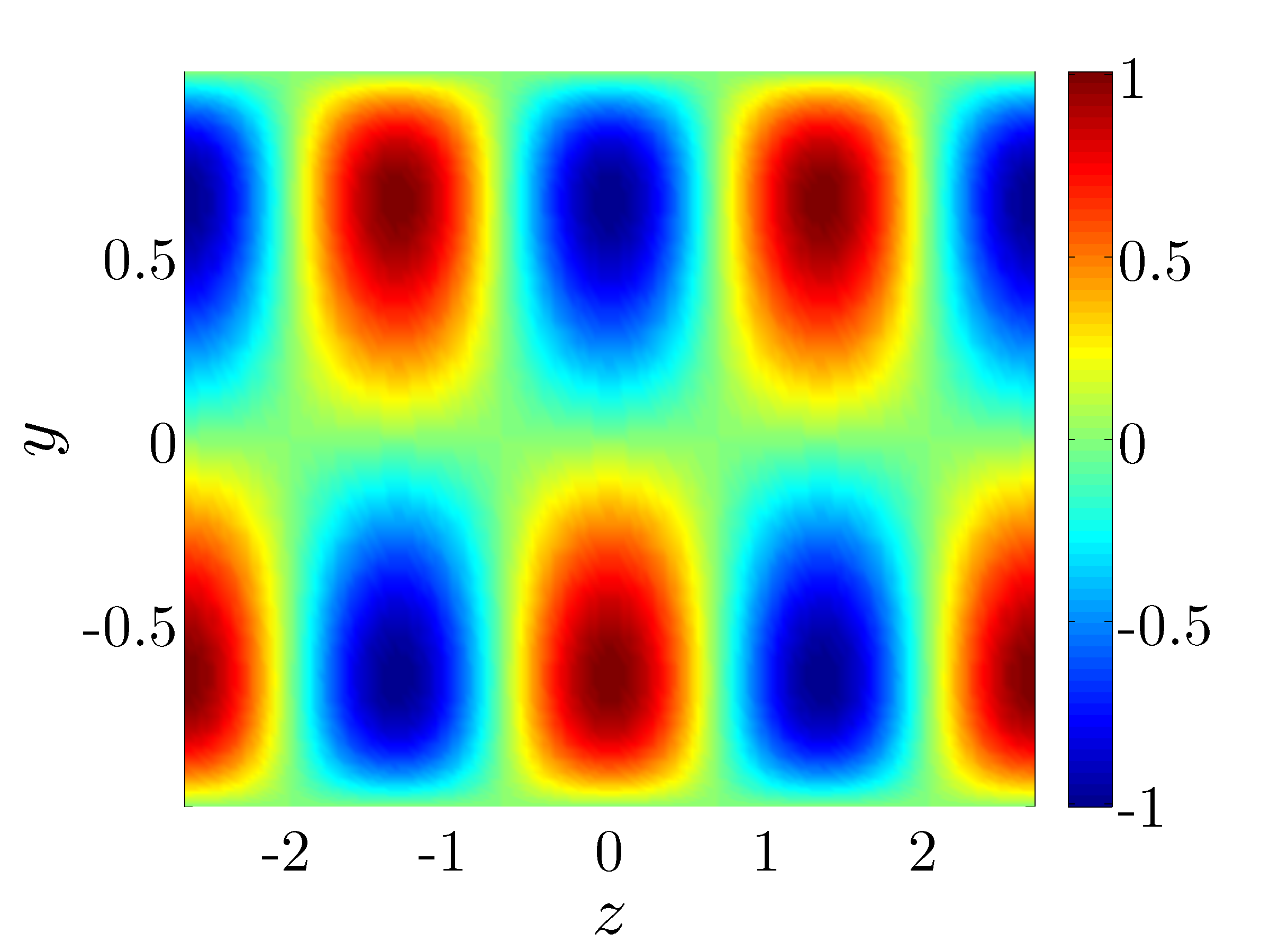}
    \label{fig.T11p-1st}}
    \\[-0.25cm]
    \subfloat[Couette flow with $k_z = 2.42$.]
    {\includegraphics[width=0.4\textwidth]{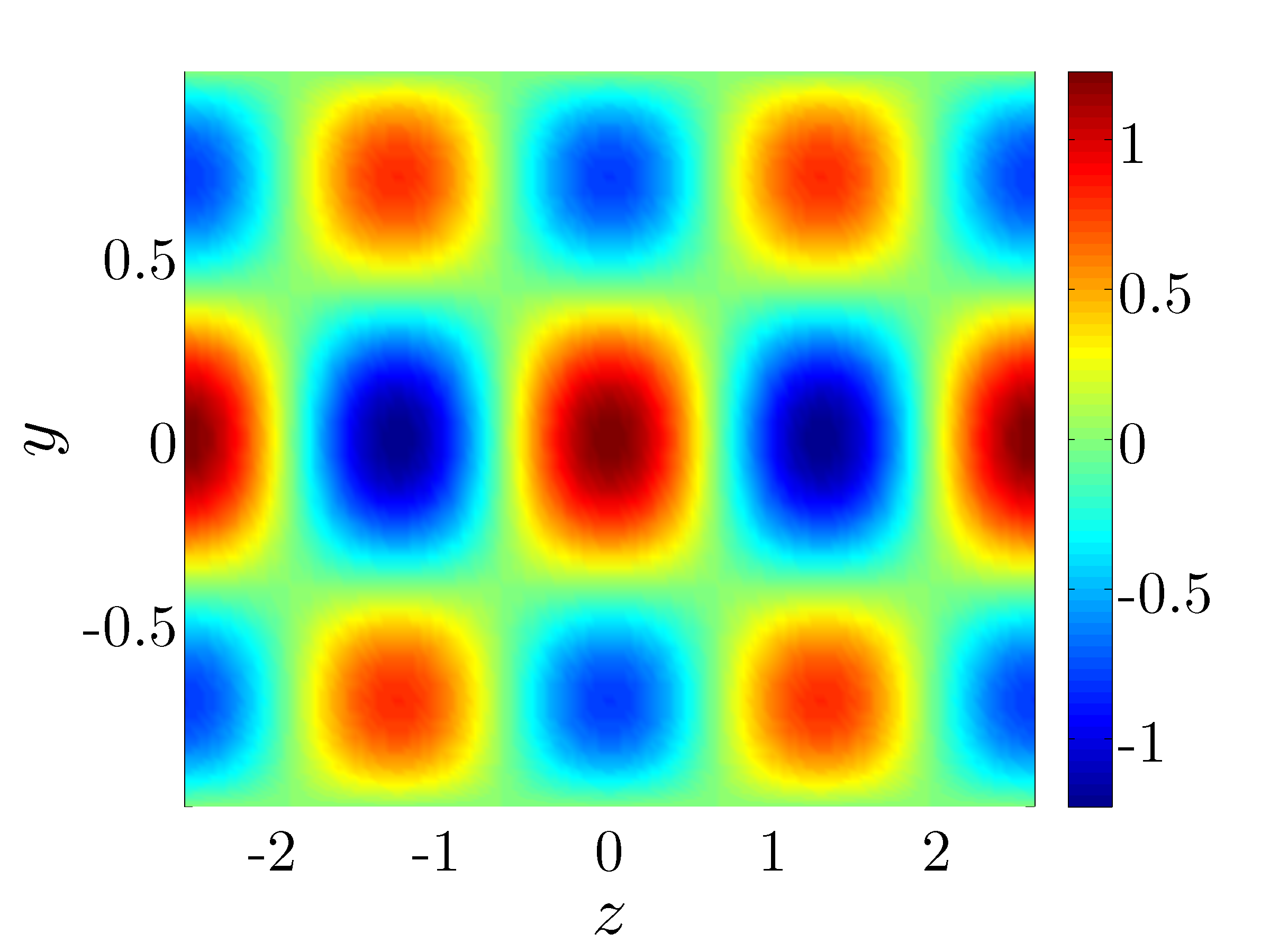}
    \label{fig.T11c-2nd}}
    \subfloat[Poiseuille flow with $k_z = 2.32$.]
    {\includegraphics[width=0.4\textwidth]{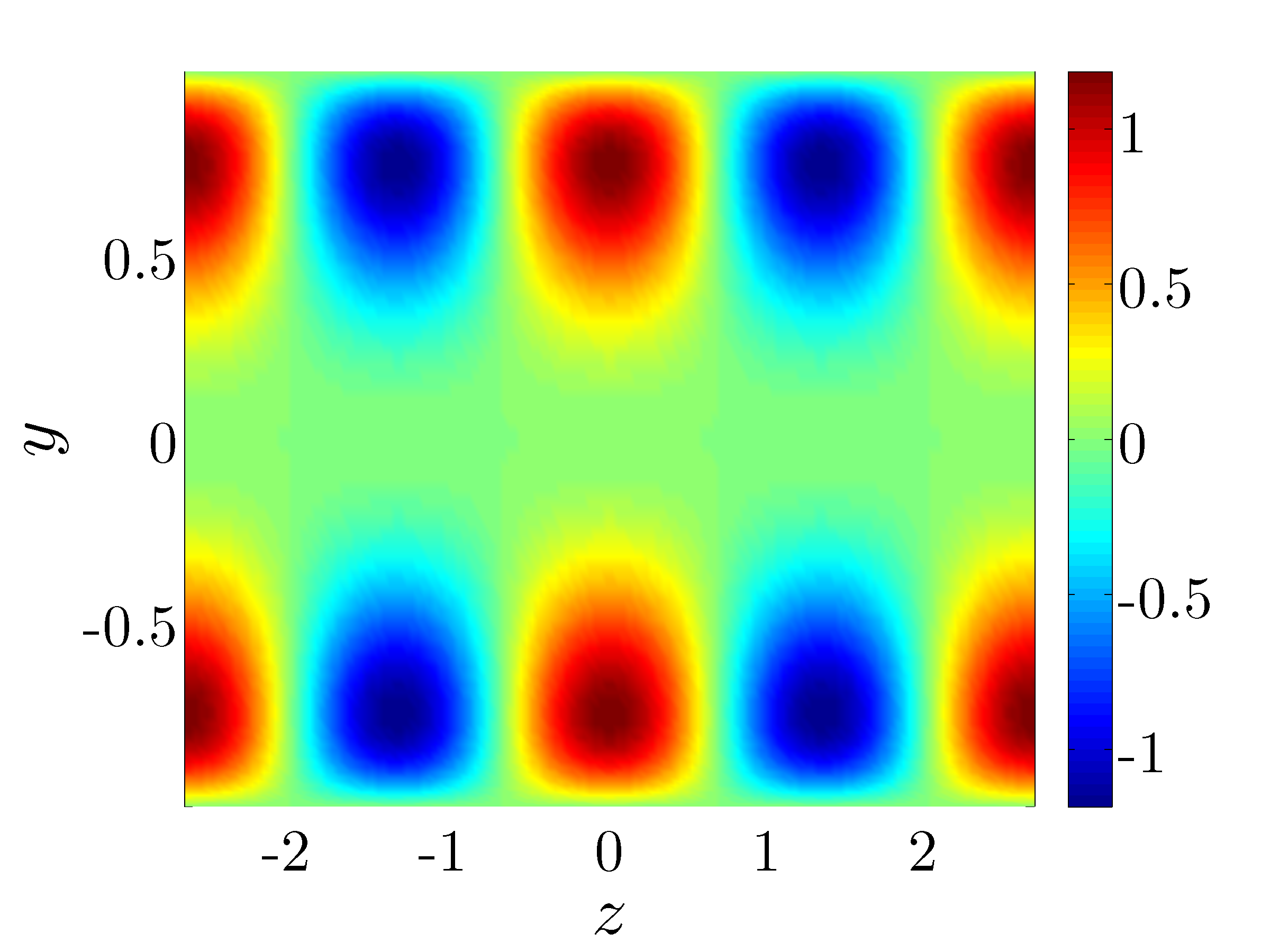}
    \label{fig.T11p-2nd}}
    }
    \caption{Polymer stress fluctuations $\tau_{11} (z,y)$ corresponding to the largest ((a)-(b)) and the second largest ((c)-(d)) eigenvalues of the autocorrelation operator of $\tau_{11}$ in inertialess flows with $\We = 1$ and $\beta = 0.5$ subject to wall-normal and spanwise stochastic forcing.}
    \label{fig.tau11}
    \end{figure}

\begin{figure}
    \centering
    {
    \includegraphics[width=0.99\textwidth]{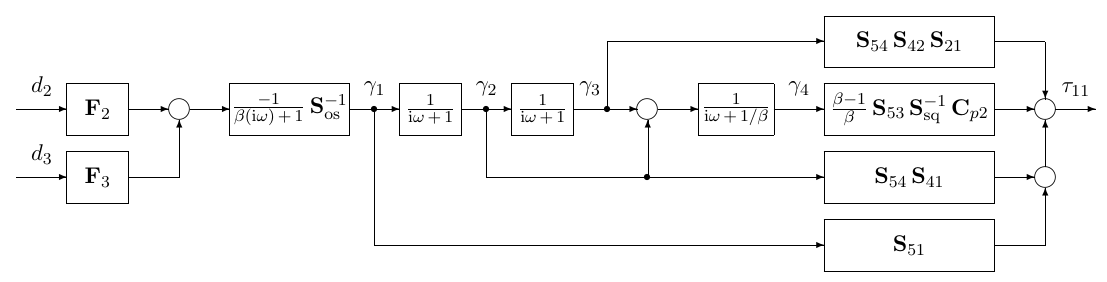}
    }
    \caption{Block diagram of the frequency response operator that maps ($d_2,d_3$) to $\tau_{11}$ in streamwise-constant creeping flows of Oldroyd-B fluids with $\We = 1$, cf.~(\ref{eq.ee-phi5-d23}).
    }
    \label{fig.tau11d23}
    \end{figure}

The flow structures with most energy, in inertialess flows with $\We = 1$ and $\beta = 0.5$ subject to wall-normal and spanwise stochastic forcing, are shown in Figure~\ref{fig.tau11}. These structures are purely harmonic in $z$ and their $y$-shapes are determined by the eigenfunctions corresponding to the two largest eigenvalues of the autocorrelation operator of $\tau_{11}$. In both Couette and Poiseuille flows, the most amplified set of fluctuations in $\tau_{11}$ is antisymmetric with respect to the channel centerline. In Couette flow $\tau_{11}$ peaks around $y \approx \pm 0.5$, while in Poiseuille flow the peaks are moved closer to the walls. The second set of most amplified fluctuations is symmetric with respect to the channel centerline and it differs vastly in shear-driven and in pressure-driven flows. In Couette flow, the eigenfunction corresponding to the second-largest eigenvalue achieves its maximum at the channel centerline, with secondary set of peaks taking place in the vicinity of the walls. In Poiseuille flow, the second set of strongly amplified fluctuations has small values in the center of the channel and the peaks occur around $y \approx \pm 0.75$. Although the stress fluctuations in experiments and nonlinear simulations are expected to be more complex than the structures presented in Figure~\ref{fig.tau11}, the flow patterns identified here are likely to play significant role in early stages of disturbance development in channel flows of viscoelastic fluids.

The block diagram of the frequency response operators that map $d_2$ and $d_3$ to $\tau_{11}$ in streamwise-constant inertialess flows of Oldroyd-B fluids with $\We = 1$ is illustrated in Figure~\ref{fig.tau11d23}, cf.~(\ref{eq.ee-phi5-d23}). In addition to exhibiting the simple aspects of the temporal responses of $\tau_{11}$ induced by the wall-normal and spanwise forces in creeping flows, this block diagram exemplifies the contribution of polymer stretching to the function $c_0$ in~(\ref{eq.Etau-Re0-main}). Namely, almost all operators that act on the $\gamma$-variables in Figure~\ref{fig.tau11d23} arise from stretching of polymer stress fluctuations by a background shear. As noted in Section~\ref{sec.PSDtau}, the only exceptions are (i) the operators $\bS_{51}$ and $\bS_{53}$ which, respectively, capture transport and stretching of a base polymer stress by velocity fluctuations; (ii) the operator $\bS_{41}$ which accounts for both of these phenomena; and (iii) the operators $\bS_{21}$ and $\bSsq$ which produce gradients of velocity fluctuations and viscous dissipation, respectively.

\section{Verification of analytical developments in stochastic simulations}
    \label{sec.sim}

In this section, we conduct stochastic simulations of the linearized flow equations in the absence of inertia. In particular, we examine responses of the fluctuations in streamwise velocity and the streamwise component of the polymer stress tensor to the wall-normal and spanwise stochastic forcing. These input-output choices are motivated by our analytical developments that identify them as the most effective way to excite the flow and the most responsive fluctuation components, respectively. The simulations presented here not only confirm our analytical developments, but they also illustrate how our results should be interpreted when compared to direct numerical simulations and experiments. We show that a proper comparison requires ensemble-averaging, rather than a comparison at the level of individual simulations or experiments.

\begin{figure}
    \centering
    {
    \subfloat[]
    {\includegraphics[width=0.33\textwidth]{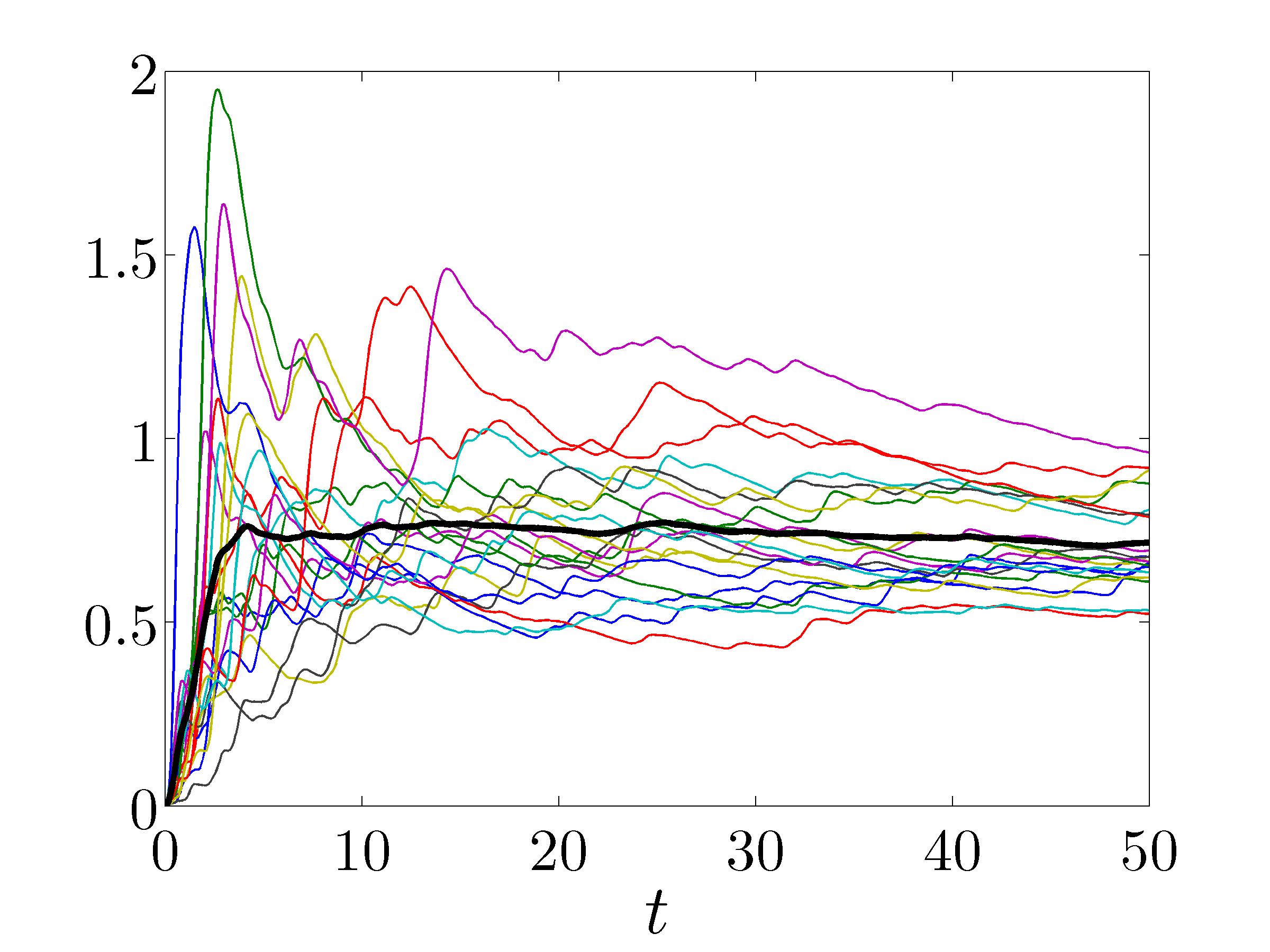}
    \label{fig.Eu(t)}}
    \subfloat[]
    {\includegraphics[width=0.33\textwidth]{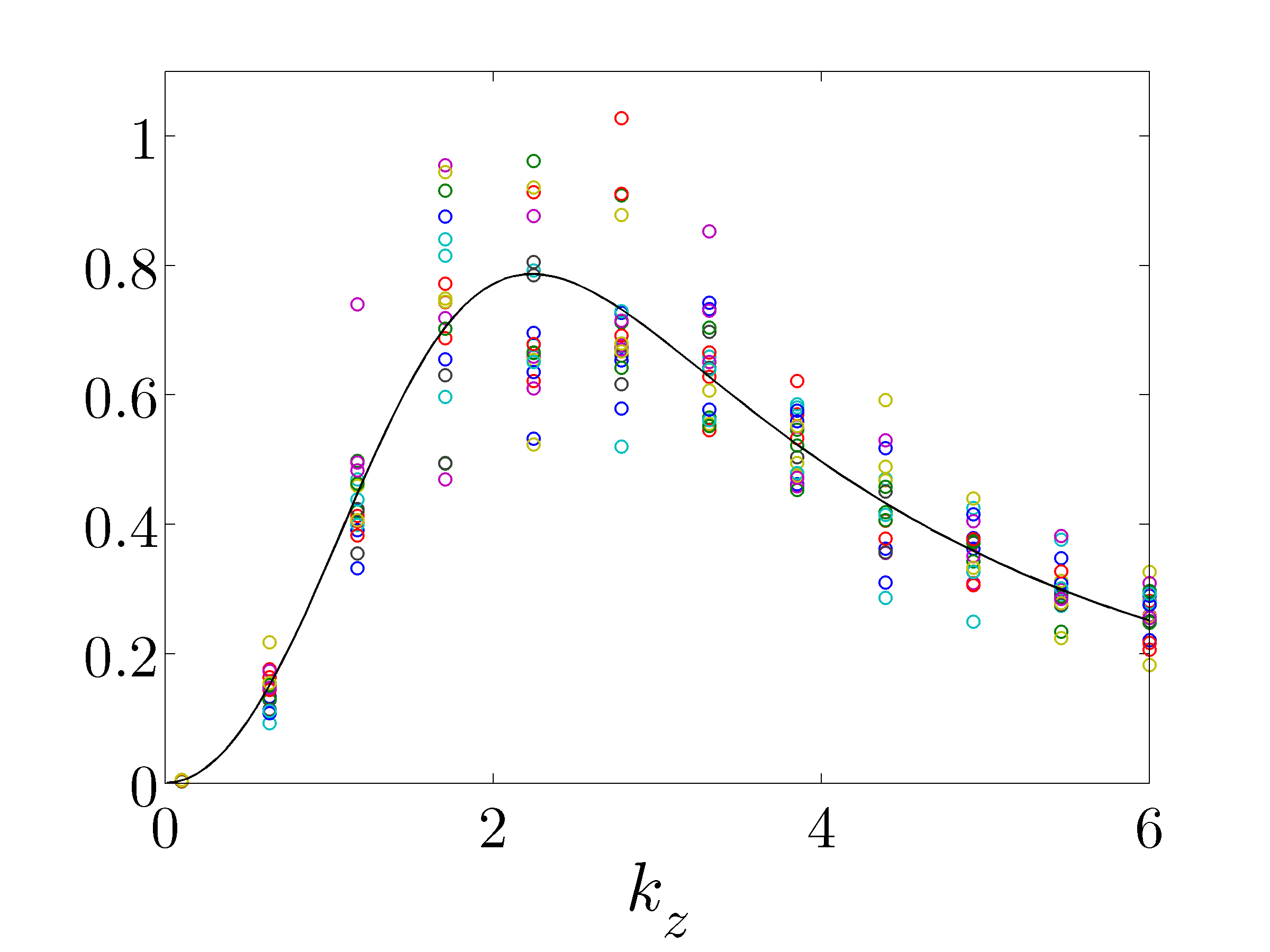}
    \label{fig.Eu(kz)-20sim}}
    \subfloat[]
    {\includegraphics[width=0.33\textwidth]{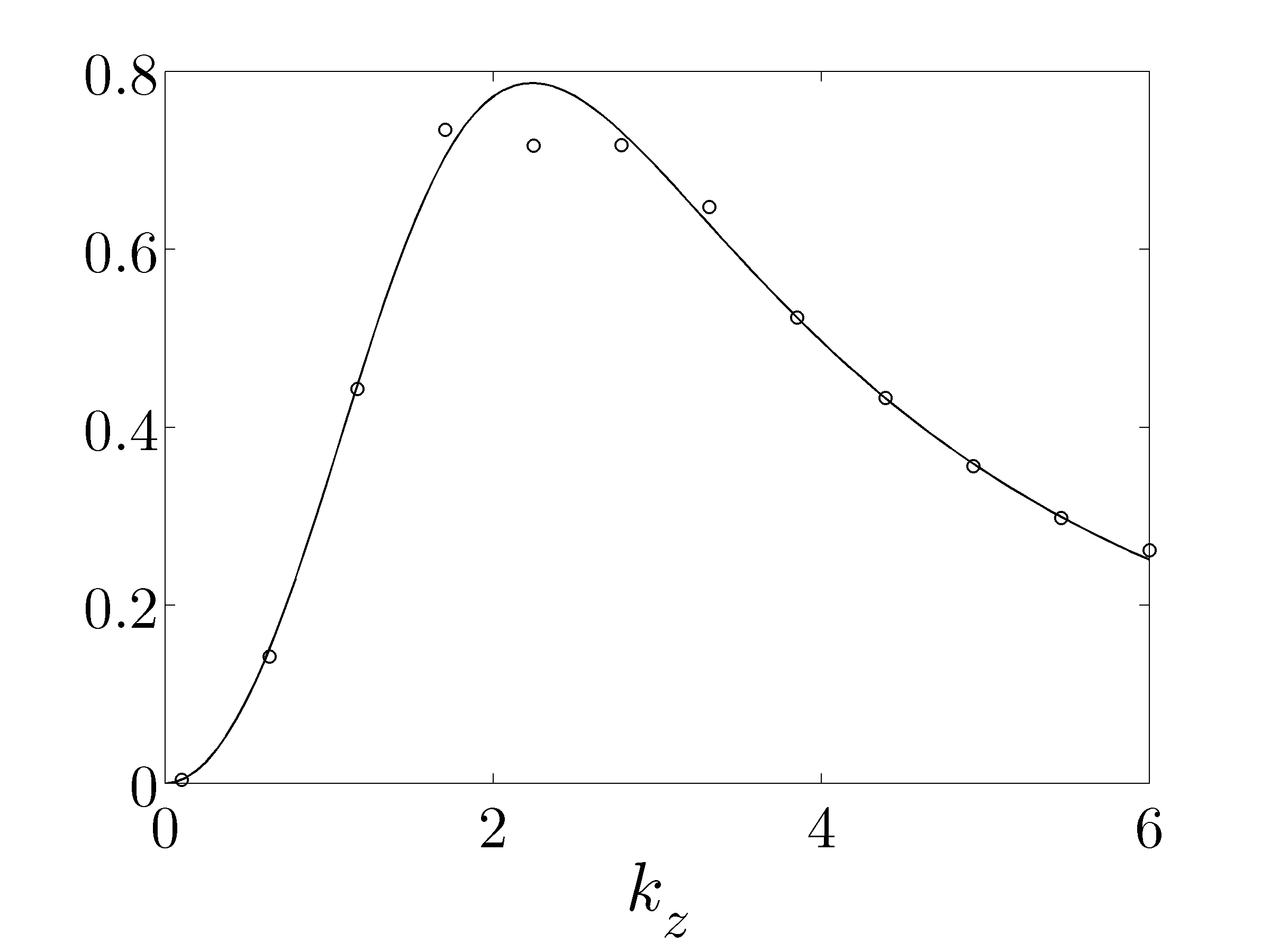}
    \label{fig.Eu(kz)-20sim-average}}
    }
    \\
    {
    \subfloat[]
    {\includegraphics[width=0.33\textwidth]{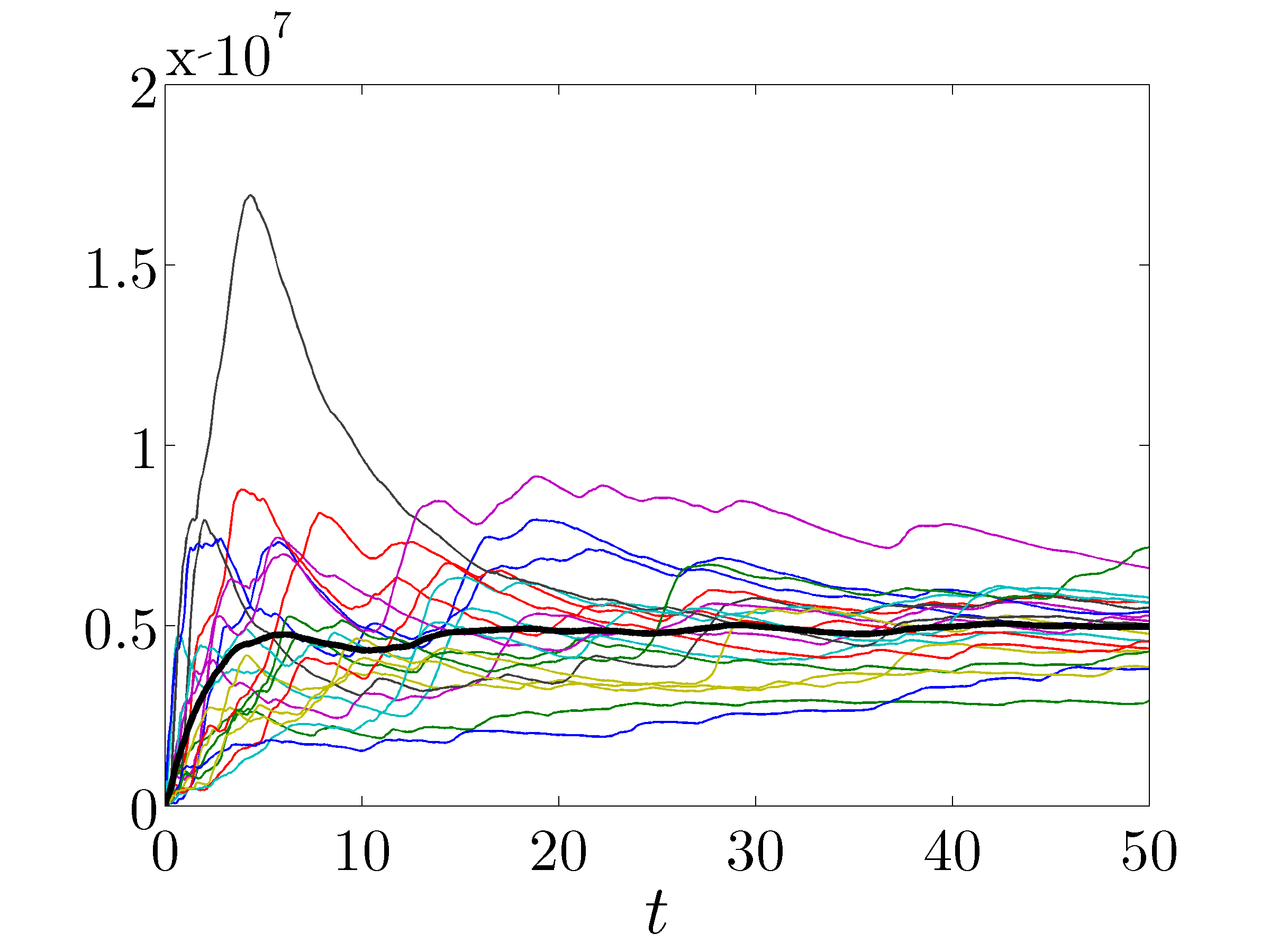}
    \label{fig.Etau11(t)}}
    \subfloat[]
    {\includegraphics[width=0.33\textwidth]{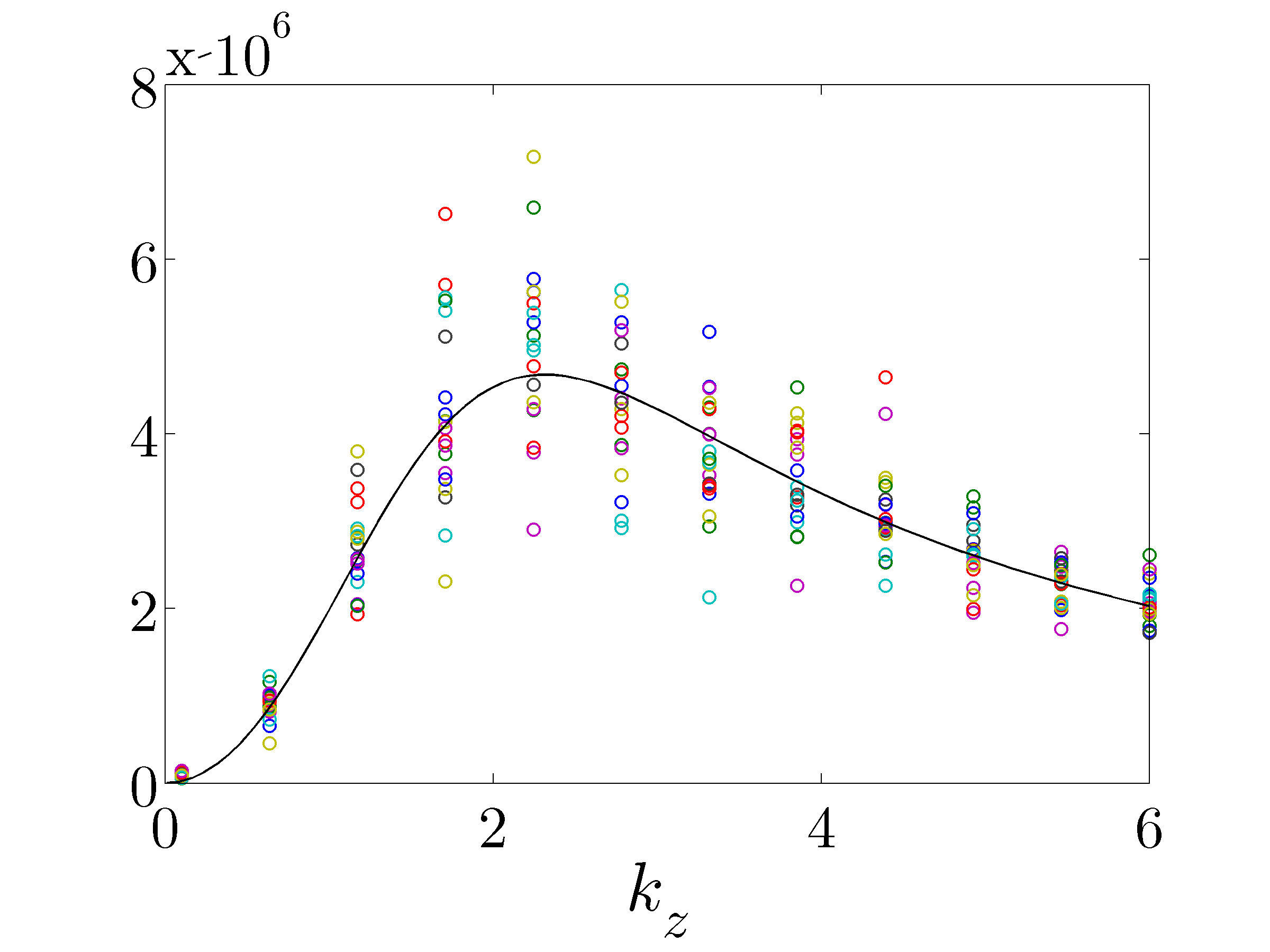}
    \label{fig.Etau11(kz)-20sim}}
    \subfloat[]
    {\includegraphics[width=0.33\textwidth]{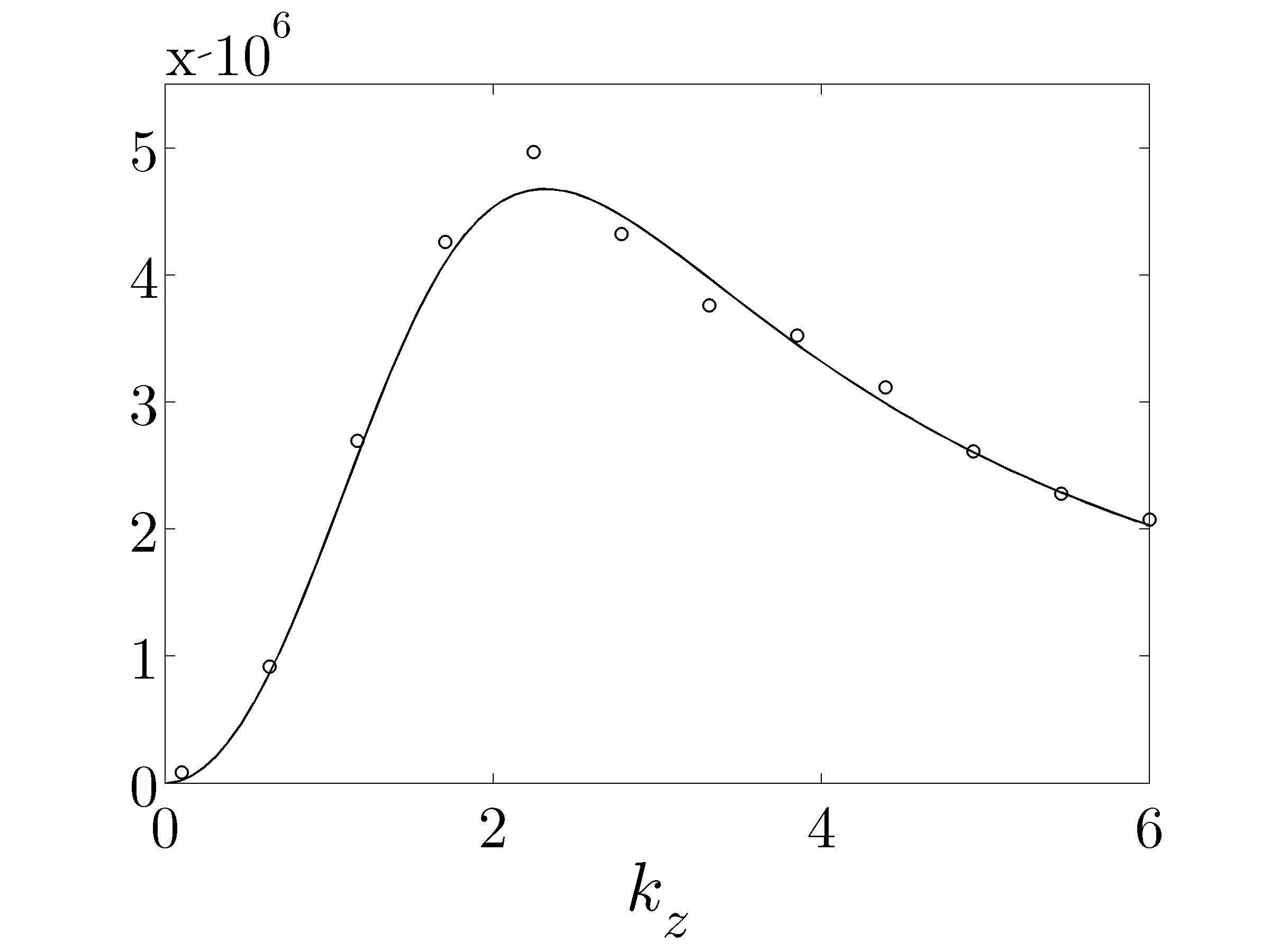}
    \label{fig.Etau11(kz)-20sim-average}}
    }
    \caption{Variance of $u$ (first row) and $\tau_{11}$ (second row) in streamwise-constant creeping Poiseuille flow with $\We = 50$ and $\beta = 0.5$ subject to stochastic forcing
    $
    \left[\,d_2\,\,\,d_3\,\right]^T.
    $
    (a), (d) The time evolution of the variance for twenty realizations of forcing with $k_z = 2.24$; the variance averaged over all simulations is shown by thick black line.
    The $k_z$-dependence of the variance at $t = 50$ resulting from (b), (e) twenty forcing realizations (circles); and (c), (f) averaging over all simulations (circles).  The solid lines in the $k_z$-dependent plots represent the steady-state variances determined from~(\ref{eq.Eu23}) with $\tilde{g}_0 (k_z)$ shown in Figure~\ref{fig.g0} (for $u$), and from
    $
    \We^4 \, c_0 (k_z; \beta)
    $
with $c_0 (k_z; 0.5)$ shown in Figure~\ref{fig.c-pos-cou-beta0p5} (for $\tau_{11}$).
    }
    \label{fig.u-tau11-sim}
    \end{figure}

We first examine variance of the streamwise velocity fluctuations in inertialess flows driven by $d_2$ and $d_3$. As described in Section~\ref{sec.main-Ev}, the aggregate effect of these forces on $u$ in statistical steady-state is captured by
    \beq
    E_u (k_z; \We,\beta)
    ~ = ~
    \We^2 \, \tilde{g}_0 (k_z) \, (1 - \beta)^2/\beta,
    \label{eq.Eu23}
    \eeq
with function $\tilde{g}_0 (k_z)$ shown in Figure~\ref{fig.g0}. The Poiseuille flow results, obtained by simulating system~(\ref{eq.u-sob}) with $\We = 50$ and $\beta = 0.5$ in the presence of a temporally stationary white Gaussian process
    $
    \left[\,d_2\,\,\,d_3\,\right]^T
    $
with zero mean and unit variance, are shown in Figures~\ref{fig.Eu(t)}-\ref{fig.Eu(kz)-20sim-average}. The wall-normal operators in~(\ref{eq.u-sob}) are approximated using the pseudo-spectral method~\cite{Reddy2000}, and twenty different simulations are performed with $50$ collocation points in $y$, and $12$ equally-spaced points between $0.1$ and $6$ in $k_z$. We have verified convergence by doubling the number of grid points in the wall-normal direction. The total simulation time is set to $50$ relaxation times. The Couette flow results exhibit similar trends and are omitted for brevity.

The time evolution of the variance of $u$, for twenty realizations of stochastic forcing with $k_z = 2.24$, is shown in Figure~\ref{fig.Eu(t)}; the variance averaged over all simulations is represented by thick black line. While individual simulations display significantly different responses, the average of twenty sample sets appears to be approaching the steady-state value predicted by theoretical analysis. This is further exemplified in Figures~\ref{fig.Eu(kz)-20sim} and~\ref{fig.Eu(kz)-20sim-average} where the $k_z$-dependence of the variance at $t = 50$ resulting from twenty forcing realizations and from averaging over these realizations are shown, respectively. The solid lines in these two figures represent the steady-state variance of $u$ determined from~(\ref{eq.Eu23}) with $\tilde{g}_0 (k_z)$ shown in Figure~\ref{fig.g0}. Even though the results of individual simulations deviate from the theoretically predicted ensemble-average energy density (cf.\ Figure~\ref{fig.Eu(kz)-20sim}), the average of all simulations displays good agreement with our analytical developments (cf.\ Figure~\ref{fig.Eu(kz)-20sim-average}).

Variance of $\tau_{11}$ in inertialess Poiseuille flow driven by the wall-normal and spanwise stochastic forcing is obtained by simulating system~(\ref{eq.ee-phi5-d23}), which corresponds to the slow subsystem discussed in~\ref{sec.c}, using a sample set of twenty forcing realizations; see Figures~\ref{fig.Etau11(t)}-\ref{fig.Etau11(kz)-20sim-average}. As in the case of the streamwise velocity fluctuations, we observe good agreement between ensemble-averaged simulations and theoretical predictions for the variance maintained in $\tau_{11}$ by
    $
    \left[\,d_2\,\,\,d_3\,\right]^T.
    $
From Section~\ref{sec.main-Etau}, we recall that the latter is determined by
    $
    \We^4 \, c_0 (k_z; \beta)
    $
with function $c_0 (k_z; 0.5)$ shown in Figure~\ref{fig.c-pos-cou-beta0p5}. Additional numerical experiments (not shown here) suggest that this agreement can be further improved by increasing the number of forcing realizations and by extending the total simulation time. We also note that the principal eigenvectors of the ensemble-averaged autocorrelation matrices of $u$ and $\tau_{11}$ at $t = 50$ closely correspond to their counterparts in Figures~\ref{fig.4b} and~\ref{fig.T11p-1st}, respectively.

The results of this section verify our theoretical predictions and demonstrate that the need for running a number of stochastic simulations with different forcing realizations can be circumvented by careful analysis of the constitutive equations. They also indicate that (i) a rather long simulation time may be required to obtain convergent statistics (at least 20 relaxation times); and that (ii) care should be exercised when comparing observations resulting from numerical simulations or experiments subject to a single forcing realization to observations resulting from ensemble-averaging. These insights are anticipated to provide guidelines for the design of numerical simulations and experiments that are well-suited for investigating transition in strongly elastic flows of polymeric fluids.

\section{Conclusions}
    \label{sec.concl}

In this paper, we have analyzed nonmodal amplification of stochastic disturbances in channel flows of Oldroyd-B fluids. For streamwise-constant fluctuations, the linearized governing equations can be cast in a compact form suitable for application of techniques from linear systems theory. Consideration of spatio-temporal frequency responses leads to the conclusion that the steady-state variances for velocity fluctuations scale quadratically with the Weissenberg number, while those for polymer stress fluctuations scale quartically with $\We$. Wall-normal and spanwise forces have the largest influence in both cases, and their effects are felt most strongly by the streamwise velocity and polymer stress fluctuations. For large elasticity numbers, the linearized governing equations can be decomposed into slow and fast subsystems, allowing application of singular perturbation methods to obtain explicit analytical expressions for the variance amplification associated with the velocity and polymer stress fields. For sufficiently large Weissenberg number, the variance amplification shows a peak at $\cO(1)$ spanwise wavenumber, and the corresponding streamwise velocity fluctuations have a structure similar to that seen in high-Reynolds-number flows of Newtonian fluids. Results from stochastic simulations confirm the validity of our analytical approach. The mechanism of the energy amplification involves polymer stretching, which gives rise to an energy transfer from the base flow to fluctuations. This transfer can be interpreted as an effective lift-up of flow fluctuations, similar to the role vortex tilting plays in inertia-dominated flows.

The results of the present work are important because they reveal the asymptotic behavior of stochastically forced channel flows in the high-elasticity-number limit. Such knowledge provides insight into the underlying physical mechanisms, and is expected to be valuable for validating and interpreting observations made in direct numerical simulations and experiments (as is the case for Newtonian fluids).  Indeed, the block diagrams presented in this paper lay bare the relationships between various inputs and outputs and the physical processes that contribute to these relationships. In addition, we have demonstrated that the inertialess limit is considerably more subtle than might be expected, for determination of the function $f$ in~(\ref{eq.Ev-intro1}) that characterizes viscous dissipation effects becomes ill-posed in this limit. In contrast, our analysis shows that the inertialess model can be used to reliably determine $E_{\tau}$, as well as the Weissenberg-number-dependent part of $E_{\mrv}$. We also note that in contrast to studies that consider transient growth phenomena arising {\em only\/} from initial conditions (i.e., with no external disturbances), a preferential spanwise length scale is selected for the stress fluctuations in stochastically forced flows. When forcing is not present, the lack of diffusive terms in the constitutive equation is manifested by the absence of a high-wavenumber roll-off in the response of polymer stress fluctuations which prevents the appearance of a preferred spanwise length scale~\citep{jovkumPOF10}. In the presence of forcing, however, the disturbances get `filtered' through the equations of motion thereby leading to a peak at $\cO (1)$ spanwise wavenumber in the variance amplification.

The present results further confirm our earlier observations~\citep{hodjovkumJFM08,hodjovkumJFM09} that stochastic disturbances can be considerably amplified by elasticity even when inertial effects are weak. This amplification can serve as an initial stage of the development of streamwise-elongated flow structures, which upon reaching a finite amplitude may undergo secondary amplification~\cite{schhus02} or instability~\cite{wal97} and thereby provide a bypass transition to elastic turbulence. It is important to point out that although we have considered a particular class of disturbances in this paper, our results raise the possibility that other types of disturbances might also be significantly amplified in elasticity-dominated flows. We note that in the present problem, energy amplification does not require the presence of curved streamlines in the base flow, which can give rise to linear instabilities in other geometries when the effects of elasticity dominate those of inertia~\cite{larshamul90,Larson1992,Shaqfeh1996}. (In the present problem, the base flow is linearly stable~\cite{Larson1992}.) However, the finite-amplitude flow structures created by the energy amplification explored here may well contain curved streamlines and be subject to further instabilities that lead to a disordered flow.

Indeed, nonlinear evolution of disturbances in viscoelastic channel flows has already been examined in several studies, but most of these have been done for two-dimensional flows~\cite{Keunings2002,Meulenbroek2004}. The present work suggests that three-dimensional effects may play a key role in the transition to elastic turbulence. Furthermore, in contrast to~\cite{Morozov2005}, where an elasticity-induced finite-amplitude instability in Couette flow was predicted, our analysis (i) identifies key physical mechanisms that enable nonmodal amplification of disturbances in the absence of inertia; and (ii) highlights the richness of the linearized constitutive equations in parallel shear flows of viscoelastic fluids. Elastic turbulence may find use in promoting mixing in microfluidic devices, where inertial effects are weak due to the small geometries~\cite{Groisman2001,Groisman2004,junste10}. In polymer processing applications, however, elastic turbulence is generally undesired~\cite{Larson1992,Larson2000}, and our work may aid the development of control strategies to maintain ordered flows.

Finally, we point out that the slow-fast decomposition of the linearized dynamics we have uncovered here does not follow {\em a priori\/} from the governing equations in their original form. Identification of this decomposition was a necessary step in the application of the singular perturbation methods that were used to develop analytical expressions for the variance amplification. The approach taken in this work may be helpful in examining the asymptotic structure of other flows at high elasticity number, especially if such flows are subject to disturbances and have nonnormal governing equations.

\section*{Acknowledgments}

M.\ R.\ J.\ would like to thank Prof.\ Petar V.\ Kokotovi\'c for stimulating discussions and insightful comments. This work was supported in part by the National Science Foundation under CAREER Award CMMI-06-44793 (to M.\ R.\ J.\/), by the Department of Energy under Award DE-FG02-07ER46415 (to S.\ K.\/), and by the University of Minnesota Digital Technology Center's 2010 Digital Technology Initiative Seed Grant (to M.\ R.\ J.\ and S.\ K.\/).

\appendix

\section{Frequency response operators}
    \label{sec.fr}

\subsection{Frequency responses of velocity fluctuations}
    \label{sec.fr-v}

The frequency response operators $\bH_{rj}$, relating the forcing and velocity components $d_j$ and $r$ with $\{ r = u, v, w$; $j = 1, 2, 3 \}$, can be obtained by applying the temporal Fourier transform to~(\ref{eq.lnse-2d3c}) subject to zero initial conditions and by eliminating polymer stresses from the equations. Equation~(\ref{eq.lnse-2d3c-tauv}) can be used to express
    $
    \bphi_2
    =
    \left[\,\tau_{22}~\,\tau_{23}~\,\tau_{33}\,\right]^T
    $
in terms of the ($y,z$)-plane streamfunction
    $
    \phi_1
    =
    \psi
    $
    \beq
    \bphi_2
    \; = \;
    \dfrac{\bS_{21}}{1 \, + \, \mri \omega} \, \phi_1,
    \label{eq.phi2}
    \eeq
where $\omega$ denotes the temporal frequency. Substitution of~(\ref{eq.phi2}) into the temporal Fourier transform of~(\ref{eq.lnse-2d3c-v}) yields
    \beq
    \ba{rcl}
    \phi_1
    & \! = \! &
    \left(
    \eps \mri \omega \bI
    \, - \,
    \beta \bS_{11}
    \, - \,
    \dfrac{1 \, - \, \beta}{1 \, + \, \mri \omega}
    \bS_{12} \bS_{21}
    \right)^{-1}
    \left(
    \bF_2 d_2 \, + \,
    \bF_3 d_3
    \right)
    \\[0.35cm]
    & \! = \! &
    (1 \, + \, \mri \omega)
    \,
    \bKos
    \left(
    \bF_2 d_2 \, + \,
    \bF_3 d_3
    \right),
    \ea
    \label{eq.phi1}
    \eeq
where the fact that
    $
    \bS_{12} \bS_{21}
    =
    \bS_{11}
    =
    \Delta^{-1} \Delta^2
    =:
    \bSos
    $
was used to define the operator $\bKos$,
    \beq
    \bKos
    \, = \,
    (
    \eps (\mri \omega)^2 \bI
    \, - \,
    ( \beta \bSos - \eps \bI)
    \mri \omega
    \, - \,
    \bSos
    )^{-1}.
    \non
    \eeq
Based on this and equation~(\ref{eq.lnse-2d3c-out}), it follows that, for streamwise-constant fluctuations, streamwise forcing does not influence the wall-normal and spanwise velocities, i.e.
    \beq
    \bH_{r1}(k_z,\omega;\We,\beta,\eps)
    \; = \;
    0,
    ~~
    r = v,w.
    \non
    \eeq
Moreover, using~(\ref{eq.lnse-2d3c-out}) and~(\ref{eq.phi1}), the operators
    $
    \bH_{r j}(k_z,\omega;\We,\beta,\eps),
    $
    $
    \{ r = v,w
    $;
    $
    j = 2,3 \},
    $
can be written as
    \beq
    \bH_{rj}(k_z,\omega;\We,\beta,\eps)
    \; = \;
    \bar{\bH}_{rj}(k_z,\omega;\beta,\eps),
    ~~
    r = v,w;
    ~
    j = 2,3,
    \non
    \eeq
where the $\We$-independent operators $\bar{\bH}_{rj}$ are given by
    \beq
    \ba{rcl}
    \bar{\bH}_{rj}(k_z,\omega;\beta,\eps)
    & \! = \! &
    (1 \, + \, \mri \omega)
    \,
    \bG_r \bKos \bF_j,
    ~~
    r = v,w;
    ~
    j = 2,3.
    \ea
    \label{eq.Huv23}
    \eeq

The following relation between the streamfunction/streamwise velocity
($\phi_1,\phi_3$) and polymer stresses
    $
    \bphi_4
    \, = \,
    \left[\,\tau_{12}~\,\tau_{13}\,\right]^T
    $
can be established by substituting~(\ref{eq.phi2}) into the temporal Fourier
transform of~(\ref{eq.lnse-2d3c-taueta})
    \begin{subequations}
    \begin{align}
    \label{eq.phi4a}
    \bphi_4
    & \; = \;
    \dfrac{1}{1 \, + \, \mri \omega}
    \left(
    \We
    \left(
    \bS_{41} \, \phi_1
    \, + \,
    \bS_{42} \, \bphi_2
    \right)
    \, + \,
    \bS_{43} \, \phi_3
    \right)
    \\[0.1cm]
    \label{eq.phi4b}
    & \; = \;
    \dfrac{\We}{1 \, + \, \mri \omega}
    \left(
    \bS_{41}
    \, + \,
    \dfrac{\bS_{42} \bS_{21}}{1 \, + \, \mri \omega}
    \right)
    \phi_1
    \; + \;
    \dfrac{1}{1 \, + \, \mri \omega}
    \,
    \bS_{43} \, \phi_3.
    \end{align}
    \end{subequations}
Substitution of this equation in the temporal Fourier transform
of~(\ref{eq.lnse-2d3c-eta}) yields
    \beq
    \ba{rcl}
    \phi_3
    & \!\! = \!\! &
    \We
    \left(
    \eps \mri \omega \bI
    -
    \beta \bS_{33}
    -
    \dfrac{(1 - \beta)}{1 + \mri \omega} \bS_{34} \bS_{43}
    \right)^{-1}
    \left(
    \eps \, \bS_{31}
    +
    \dfrac{(1 - \beta) \bS_{34}}{1 + \mri \omega}
    \left(
    \bS_{41} + \dfrac{\bS_{42} \bS_{21} }{1 + \mri \omega}
    \right)
    \right) \phi_1
    \\[0.25cm]
    & \!\! + \!\! &
    \left(
    \eps \mri \omega \bI
    -
    \beta \bS_{33}
    -
    \dfrac{(1 - \beta)}{1 + \mri \omega} \bS_{34} \bS_{43}
    \right)^{-1}
    \bF_1 d_1.
    \ea
    \label{eq.phi3}
    \eeq
Now, since $u = \bG_u \phi_3 = \phi_3$, by substituting~(\ref{eq.phi1}) into~(\ref{eq.phi3}) and using the fact that in Couette and Poiseuille flows
    \[
    {\bS}_{34} {\bS}_{43}
    \, = \,
    {\bS}_{33}
    \, = \,
    \Delta
    \, =: \,
    \bSsq,
    ~~
    {\bS}_{34} {\bS}_{41}
    \, = \,
    0,
    ~~
    {\bS}_{34} {\bS}_{42} {\bS}_{21}
    \, = \,
    \mri k_z
    \left(
    U'(y) \Delta
    \, + \,
    2 U''(y) \py
    \right),
    \]
it follows that operators $\bH_{u j}(k_z,\omega;\We,\beta,\eps)$, $\{ j = 1,2,3 \}$, are given by
    \beq
    \ba{rcl}
    \bH_{u1}(k_z,\omega;\We,\beta,\eps)
    & \! = \! &
    \bar{\bH}_{u1}(k_z,\omega;\beta,\eps),
    \\[0.1cm]
    \bH_{u j}(k_z,\omega;\We,\beta,\eps)
    & \! = \! &
    \We
    \;
    \bar{\bH}_{u j}(k_z,\omega;\beta,\eps),
    \;\;
    j = 2,3.
    \ea
    \non
    \eeq
Here, the $\We$-independent operators $\bar{\bH}_{u j}$ are determined by
    \beq
    \ba{rcl}
    \bar{\bH}_{u 1}
    (k_z,\omega;\beta,\eps)
    & \! = \! &
    (1 \, + \, \mri \omega)
    \,
    \bG_u \bKsq \bF_1
    \; = \;
    (1 \, + \, \mri \omega)
    \,
    \bKsq,
    \\[0.1cm]
    \bar{\bH}_{u j}
    (k_z,\omega;\beta,\eps)
    & \! = \! &
    \bG_u \bKsq
    \left(
    \eps (1 + \mri \omega)^2 \bC_{p1}
    \, + \,
    (1 - \beta) \bC_{p2}
    \right)
    \bKos \bF_j,
    ~~
    j = 2,3,
    \ea
    \label{eq.Huj}
    \eeq
with
    \beq
    \ba{rcl}
    \bKsq
    & \! = \! &
    (
    \eps (\mri \omega)^2 \bI
    -
    ( \beta \bSsq - \eps \bI)
    \mri \omega
    -
    \bSsq
    )^{-1},
    \\[0.1cm]
    \bC_{p1}
    & \! = \! &
    \bS_{31}
    \; = \;
    - \mri k_z U'(y),
    \\[0.1cm]
    \bC_{p2}
    & \! = \! &
    \mri k_z
    \tilde{\bC}_{p2},
    ~~
    \tilde{\bC}_{p2}
    \; = \;
    U'(y) \Delta + 2 U''(y) \py.
    \ea
    \non
    \eeq
Clearly, $\bC_{p2} = \mri k_z \tilde{\bC}_{p2} = {\bS}_{34} {\bS}_{42} {\bS}_{21}$ would vanish if background shear was absent (i.e., if $U(y)$ was constant) or if $k_z = 0$.

In summary, the input-output mappings from forcing to velocity fluctuations in channel flows of Oldroyd-B fluids are determined by
    \beq
    \ba{rcl}
    u (y,k_z,\omega;\We,\beta,\eps)
    & \!\! = \!\! &
    \left[
    \bar{\bH}_{u 1} (k_z,\omega;\beta,\eps) \, d_1 (\cdot,k_z,\omega)
    \right] (y)
    ~ +
    \\[0.1cm]
    & \!\!  \!\! &
    \We \,
    \ds{\sum_{j \, = \, 2}^{3}}
    \left[
    \bar{\bH}_{u j} (k_z,\omega;\beta,\eps) \, d_j (\cdot,k_z,\omega)
    \right] (y),
    \\[0.1cm]
    r (y,k_z,\omega;\beta,\eps)
    & \!\! = \!\! &
    \ds{\sum_{j \, = \, 2}^{3}}
    \left[
    \bar{\bH}_{r j} (k_z,\omega;\beta,\eps) \, d_j (\cdot,k_z,\omega)
    \right] (y),
    ~~
    r \, = \, v,w,
    \ea
    \label{eq.uvw}
    \eeq
where the $\We$-independent operators $\bar{\bH}_{u j}$ are given by~(\ref{eq.Huj}), and the $\We$-independent operators $\bar{\bH}_{r j}$ with
$\{ r = v,w$; $j = 2,3 \}$ are given by~(\ref{eq.Huv23}).

\subsection{Frequency responses of polymer stress fluctuations}
    \label{sec.fr-tau}

In this appendix, we determine the frequency responses, $\bGamma_{\phi_i,j}$, from forcing components $d_1$, $d_2$, and $d_3$ to polymer stress components
    $
    \bphi_2 \, = \, \left[\,\tau_{22}\,\,\,\tau_{23}\,\,\,\tau_{33}\,\right]^T,
    $
    $
    \bphi_4
    \, = \,
    \left[\,\tau_{12}\,\,\,\tau_{13}\,\right]^T,
    $
and
    $
    \phi_5
    \, = \,
    \tau_{11}.
    $
Substitution of (\ref{eq.phi1}) to (\ref{eq.phi2}) yields the $\We$-independent response of $\bphi_2$
    \beq
    \bphi_2 (y,k_z,\omega;\beta,\eps)
    \, = \,
    \ds{\sum_{j \, = \, 2}^{3}}
    \left[
    \bar{\bGamma}_{\phi_2, j} (k_z,\omega;\beta,\eps) \, d_j (\cdot,k_z,\omega)
    \right] (y),
    ~~
    \bar{\bGamma}_{\phi_2, j}
    \, = \,
    \bS_{21}
    \bKos
    \bF_j.
    \label{eq.phi2-fr}
    \eeq
Similarly, combination of~(\ref{eq.phi1}),~(\ref{eq.Huj}), and~(\ref{eq.uvw}) with~(\ref{eq.phi4b}) yields
    \beq
    \ba{c}
    \bphi_4 (y,k_z,\omega;\We,\beta,\eps)
    \, = \,
    \left[
    \bar{\bGamma}_{\phi_4, 1} (k_z,\omega;\beta,\eps) \, d_1 (\cdot,k_z,\omega)
    \right] (y)
    ~ +
    \\[0.1cm]
    \We \,
    \ds{\sum_{j \, = \, 2}^{3}}
    \left[
    \bar{\bGamma}_{\phi_4, j} (k_z,\omega;\beta,\eps) \, d_j (\cdot,k_z,\omega)
    \right] (y),
    ~~
    \bar{\bGamma}_{\phi_4, 1}
    \, = \,
    \bS_{43} \, \bKsq,
    \\[0.35cm]
    \bar{\bGamma}_{\phi_4, j}
    \, = \,
    \dfrac{1}{1 \, + \, \mri \omega}
    \left(
    \bS_{43} \, \bar{\bH}_{u j}
    +
    \left(
    (1 + \mri \omega) \bS_{41}
    +
    \bS_{42} \bS_{21}
    \right)
    \bKos \bF_j
    \right),
    ~~
    j \, = \, 2,3.
    \ea
    \label{eq.phi4-fr}
    \eeq
Finally, the temporal Fourier transform of~(\ref{eq.lnse-2d3c-tau11}) gives
    \beq
    \phi_5
    \, = \,
    \dfrac{1}{1 \, + \, \mri \omega}
    \left(
    \We^2 \, \bS_{51} \, \phi_1
    \, + \,
    \We
    \left(
    \bS_{53} \, \phi_3
    \, + \,
    \bS_{54} \, \bphi_4
    \right)
    \right),
    \non
    \label{eq.phi5}
    \eeq
which in conjunction with~(\ref{eq.phi1}),~(\ref{eq.Huj}),~(\ref{eq.uvw}), (\ref{eq.phi4-fr}) and $\bS_{54} \bS_{43} = \bS_{53}$ can be used to obtain
    \beq
    \ba{c}
    \ba{rcl}
    \phi_5 (y,k_z,\omega;\We,\beta,\eps)
    & \!\! = \!\! &
    \We
    \left[
    \bar{\bGamma}_{\phi_5, 1} (k_z,\omega;\beta,\eps) \, d_1 (\cdot,k_z,\omega)
    \right] (y)
    ~ +
    \\[0.1cm]
    & \!\!  \!\! &
    \We^2 \,
    \ds{\sum_{j \, = \, 2}^{3}}
    \left[
    \bar{\bGamma}_{\phi_5, j} (k_z,\omega;\beta,\eps) \, d_j (\cdot,k_z,\omega)
    \right] (y),
    \ea
    \\[0.35cm]
    \ba{rcl}
    \bar{\bGamma}_{\phi_5, 1}
    & \!\! = \!\! &
    \dfrac{2 \, + \, \mri \omega}{1 \, + \, \mri \omega} \,
    \bS_{53} \, \bKsq,
    \\[0.1cm]
    \bar{\bGamma}_{\phi_5, j}
    & \!\! = \!\! &
    \bS_{51} \bKos \bF_j
    \, + \,
    \dfrac{2 \, + \, \mri \omega}{(1 \, + \, \mri \omega)^2}
    \, \bS_{53} \, \bar{\bH}_{u j}
    ~ +
    \\[0.1cm]
    & \!\! \!\! &
    \dfrac{1}{(1 \, + \, \mri \omega)^2}
    \,
    \bS_{54}
    \left(
    (1 + \mri \omega) \bS_{41}
    +
    \bS_{42} \bS_{21}
    \right)
    \bKos \bF_j,
    ~~
    j \, = \, 2,3.
    \ea
    \ea
    \label{eq.phi5-fr}
    \eeq
To summarize, the frequency response operators from the forcing components to the polymer stress components $\bphi_2$, $\bphi_4$, and $\phi_5$ are, respectively, given by~(\ref{eq.phi2-fr}),~(\ref{eq.phi4-fr}), and~(\ref{eq.phi5-fr}). These expressions are utilized in Section~\ref{sec.PSDtau} and~\ref{sec.Etau} to quantify the dependence of variance amplification of polymer stress fluctuations on the Weissenberg and the elasticity numbers. While the developments of Section~\ref{sec.PSDtau} apply to both inertia- and elasticity-dominated flows, the developments of~\ref{sec.Etau} apply only to flows with $\eps = 1/\mu \ll 1$.

\section{Evolution equations for $\bar{\bH}_{r j}$}
    \label{sec.Hij-ss}

Here, we determine evolution equations for each $\bar{\bH}_{r j}$ in Section~\ref{sec.PSDv}. For a fixed temporal frequency $\omega$, each $\bar{\bH}_{r j}$ represents an operator in $y$, mapping the forcing $d_j$ to the velocity $r$ at $\We = 1$. The inverse temporal Fourier transform yields a system of PDEs in the wall-normal direction and in time which can be represented in the form of evolution equations (i.e., a coupled system of first-order in time PDEs). We show that, in elasticity-dominated flows, these equations admit a standard singularly perturbed form which is convenient for uncovering dependence of the frequency responses on elasticity number.

\subsection{Evolution equations for\/ $\bar{\bH}_{r j}$ with\/ $\{ r = u$; $j = 1 \}$ and\/ $\{ r = v,w$; $j = 2,3 \}$}
    \label{sec.f-ee}

We first determine evolution equations for the operators
    $
    \bar{\bH}_{r j}
    =
    (1 \, + \, \mri \omega)
    \,
    \bG_r \bKos \bF_j
    $
with
    $
    \{
    r = v,w
    $;
    $
    j = 2,3
    \}.
    $
From equation~(\ref{eq.phi2}), which is obtained by applying the temporal Fourier transform to~(\ref{eq.lnse-2d3c-tauv}), we see that $\bphi_2$ can be expressed as
    \beq
    \bphi_2
    \; = \;
    \bS_{21} \, \xi,
    \non
    \eeq
where $\xi$ represents a low-pass version of $\phi_1 = \psi$ (with the left-hand side denoting relations in the frequency domain, and the right-hand side denoting relations in the time domain)
    \beq
    \xi
    \, = \,
    \dfrac{1}{\mri \omega \, + \, 1} \, \psi
    ~~\Rightarrow~~
    \dot{\xi}
    \, = \,
    - \xi
    \, + \,
    \psi.
    \label{eq.xi}
    \eeq
Since $\bS_{12} \bS_{21} = \bS_{11} = \bSos$, equation~(\ref{eq.lnse-2d3c-v}) can be rewritten as
    \beq
    \eps \, \dot{\psi}
    \; = \;
    \beta \, \bSos \, \psi
    \; + \;
    (1 \,-\, \beta) \, \bSos \, \xi
    \;+\;
    \bF_2 \, d_2
    \,+\,
    \bF_3 \, d_3,
    \non
    \eeq
which in conjunction with~(\ref{eq.xi}) yields the following evolution equation for $\bH_{r j}$ with $\{ r = v,w$; $j = 2,3 \}$
    \beq
    \ba{rcl}
    \tbo{\dot{\xi}}{\eps \dot{\psi}}
    & \!\!\! = \!\!\! &
    \tbt{-\bI}{\bI}{(1 - \beta) \bSos}{\beta \bSos}
    \tbo{\xi}{\psi}
    \, + \,
    \tbo{0}{\bF_j} d_j,
    \\[0.25cm]
    r
    & \!\!\! = \!\!\! &
    \obt{0}{\bG_r}
    \tbo{\xi}{\psi}.
    \ea
    \label{eq.Hrj-ss}
    \eeq
This equation is expressed in terms of the ($y,z$)-plane streamfunction $\psi$ and the scalar field $\xi$ whose spatial gradients determine $\bphi_2$, $\bphi_2 = \bS_{21} \xi$. Since the time-derivative of $\psi$ is multiplied by a small positive parameter $\eps$ and since the operator $\beta \bSos$ is stable~\citep{schhen01}, and therefore invertible, system~(\ref{eq.Hrj-ss}) is in a standard singularly perturbed form~\citep{kokkharei99} with homogeneous Cauchy boundary conditions on both $\psi$ and $\xi$.

Furthermore, the operator from $[\,d_2\,\,\,d_3\,]^T$ to $[\,v\,\,\,w\,]^T$ can be represented by
    \beq
    \ba{rcl}
    \tbo{\dot{\xi}}{\eps \dot{\psi}}
    & \!\!\! = \!\!\! &
    \tbt{-\bI}{\bI}{(1 - \beta) \bSos}{\beta \bSos}
    \tbo{\xi}{\psi}
    \, + \,
    \tbt{0}{0}{\bF_2}{\bF_3}
    \tbo{d_2}{d_3},
    \\[0.35cm]
    \tbo{v}{w}
    & \!\!\! = \!\!\! &
    \tbt{0}{\bG_v}{0}{\bG_w}
    \tbo{\xi}{\psi}.
    \ea
    \non
    \eeq

An evolution equation for the operator $\bar{\bH}_{u 1}$ that maps $d_1$ to $u$ can be obtained using a similar procedure. Namely, since $d_1$ does not influence the dynamics of $\phi_1$ and $\bphi_2$, setting $\{ \phi_1 = 0$, $\bphi_2 = 0 \}$ in~(\ref{eq.lnse-2d3c-eta}) and~(\ref{eq.lnse-2d3c-taueta}) yields the following
system of equations
    \begin{subequations}
    \label{eq.phi4-u}
    \begin{align}
    \label{eq.phi4-u-a}
    \dot{\bphi}_4
    & \;=\;
    - \, \bphi_4
    \;+\;
    \bS_{43} \, u,
    \\
    \label{eq.phi4-u-b}
    \eps \, \dot{u}
    & \;=\;
    \beta \, \bS_{33} \, u
    \;+\;
    (1 \,-\, \beta) \, \bS_{34} \, \bphi_4
    \;+\;
    \bF_1 \, d_1.
    \end{align}
    \end{subequations}
Now, $\bphi_4$ can be expressed as
    \beq
    \bphi_4
    \; = \;
    \bS_{43} \, \zeta,
    \label{eq.phi4-eta}
    \eeq
where $\zeta$ denotes a low-pass version of $u$,
    \beq
    \zeta
    \, = \,
    \dfrac{1}{\mri \omega \, + \, 1} \, u
    ~~\Rightarrow~~
    \dot{\zeta}
    \, = \,
    - \zeta
    \, + \,
    u.
    \label{eq.eta}
    \eeq
Since $\bS_{34} \bS_{43} = \bS_{33} = \bSsq$, substitution of~(\ref{eq.phi4-eta}) into~(\ref{eq.phi4-u-b}) in conjunction with~(\ref{eq.eta}) yields the following evolution equation for $\bar{\bH}_{u1}$
    \beq
    \ba{rcl}
    \tbo{\dot{\zeta}}{\eps \dot{u}}
    & \!\!\! = \!\!\! &
    \tbt{-\bI}{\bI}{(1 - \beta) \bSsq}{\beta \bSsq}
    \tbo{\zeta}{u}
    \, + \,
    \tbo{0}{\bF_1} d_1,
    \\[0.25cm]
    u
    & \!\!\! = \!\!\! &
    \obt{0}{\bG_u}
    \tbo{\zeta}{u},
    \ea
    \label{eq.Hu1-ss}
    \eeq
with homogeneous Dirichlet boundary conditions on both $u$ and $\zeta$. Owing to invertibility of the operator $\beta \bSsq$~\citep{schhen01} and multiplication of $\dot{u}$ by a small parameter $\eps$, system~(\ref{eq.Hu1-ss}) is in a standard singularly perturbed form~\citep{kokkharei99}. We note that the expression for $\bphi_4$ in~(\ref{eq.phi4-eta}) holds only in the absence of wall-normal and spanwise forces (i.e., for $d_2 = d_3 = 0$). The evolution equations capturing the influence of these forces on the streamwise velocity are determined in~\ref{sec.g-ee}.

In summary, the operators
    $
    \{
    \bar{\bH}_{u 1}
    =
    (1 \, + \, \mri \omega)
    \,
    \bG_u \bKsq \bF_1;
    $
    $
    \bar{\bH}_{r j}
    =
    (1 \, + \, \mri \omega)
    \,
    \bG_r \bKos \bF_j,
    $
    $
    r = v,w
    $;
    $
    j = 2,3
    \}
    $
can be represented by the following evolution equation
    \beq
    \ba{rcl}
    \tbo{\dot{x}_{r j}}{\eps \dot{z}_{r j}}

    & \!\!\! = \!\!\! &
    \tbt{-\bI}{\bI}{(1 - \beta) \bS_{\mrk}}{\beta \bS_{\mrk}}
    \tbo{x_{r j}}{z_{r j}}
    \, + \,
    \tbo{0}{\bF_j} d_j,
    \\[0.25cm]
    r
    & \!\!\! = \!\!\! &
    \obt{0}{\bG_r}
    \tbo{x_{r j}}{z_{r j}},
    \ea
    \label{eq.f-ss}
    \eeq
with $\{ \mrk = \mathrm{sq}$ for $r = u$; $\mrk = \mathrm{os}$ for $r = v,w \}$, homogeneous Dirichlet  boundary conditions on $x_{u 1} = \zeta$ and $z_{u 1} = u$, and homogeneous Cauchy boundary conditions on $x_{r j} = \xi$ and $z_{r j} = \psi$ for $\{ r = v,w$; $j = 2,3 \}$.

\subsection{Evolution equations for\/ $\bar{\bH}_{u j}$ with\/ $j = 2,3$}
    \label{sec.g-ee}

We next determine evolution equations for the operators that map $d_2$ and $d_3$ to $u$,
    \[
    \bar{\bH}_{uj}
    \, = \,
    \bKsq
    \left(
    \eps (1 + \mri \omega)^2 \bC_{p1}
    \, + \,
    (1 - \beta) \bC_{p2}
    \right)
    \bKos \bF_j,
    \,\,
    j \, = \,2,3.
    \]
Acting on equation~(\ref{eq.phi4a}) with $\bS_{34}$ and using $\bS_{34} \bS_{41} = 0$, $\bS_{34} \bS_{43} = \bS_{33} = \bSsq$, $\bphi_2 = \bS_{21} \xi$, and $\bC_{p2} = {\bS}_{34} {\bS}_{42} {\bS}_{21}$ we obtain
    \beq
    \varphi
    \, = \,
    \dfrac{1}{\mri \omega \, + \, 1}
    \left(
    \We
    \,
    \bC_{p2} \, \xi
    \, + \,
    \bSsq \, u
    \right)
    ~~ \Rightarrow ~~
    \dot{\varphi}
    \, = \,
    - \varphi
    \, + \,
    \We
    \,
    \bC_{p2} \, \xi
    \, + \,
    \bSsq \, u,
    \label{eq.varphi-app}
    \eeq
where
    \beq
    \varphi
    \, = \,
    \bS_{34} \, \bphi_4.
    \non
    \eeq
Consequently, equation~(\ref{eq.lnse-2d3c-eta}), governing the evolution of $u$ in flows with $d_1 = 0$, can be written as
    \beq
    \eps \, \dot{u}
    \, = \,
    \beta \, \bSsq \, u
    \, + \,
    \eps \, \We \, \bC_{p1} \, \psi
    \, + \,
    (1 \,-\, \beta) \, \varphi.
    \label{eq.u-app}
    \eeq
Thus, equations~(\ref{eq.Hrj-ss}),~(\ref{eq.varphi-app}), and~(\ref{eq.u-app}) with homogeneous Cauchy boundary conditions on $\psi$ and $\xi$, homogeneous Dirichlet boundary conditions on $u$, and no boundary conditions on $\varphi$ determine evolution model for $\bH_{uj}$ with $j = 2,3$. By selecting
$\bx = [\,\xi~~\varphi\,]^T$, $\bz = [\,\psi~~u\,]^T$, we obtain a singularly perturbed realization of $\bH_{u j}$, $j = 2,3$,
    \beq
    \ba{rcl}
    \tbo{\dot{\bx}}{\eps \dot{\bz}}
    & \!\!\! = \!\!\! &
    \tbt{\bA_{11}}{\bA_{12}}{\bA_{21}}{\bA_{22} (\eps)}
    \tbo{\bx}{\bz}
    \, + \,
    \tbo{0}{\bB_2} d_j,
    \\[0.25cm]
    u
    & \!\!\! = \!\!\! &
    \obt{0}{\bC_2}
    \tbo{\bx}{\bz},
    \ea
    \label{eq.g-ss}
    \eeq
where all operators are partitioned conformably with the elements of $\bx$ and $\bz$,
    \beq
    \ba{c}
    \bA_{11}
    \, = \,
    \tbt{-\bI}{0}{\We \, \bC_{p2}}{-\bI},
    ~
    \bA_{12}
    \, = \,
    \tbt{\bI}{0}{0}{\bSsq},
    ~
    \bB_2
    \, = \,
    \tbo{\bF_j}{0},
    \\[0.35cm]
    \bA_{21}
    \, = \,
    \tbt{(1 - \beta) \, \bSos}{0}{0}{(1 - \beta) \, \bI},
    ~
    \bA_{22} (\eps)
    \, = \,
    \tbt{\beta \, \bSos}{0}{\eps \, \We \, \bC_{p1}}{\beta \, \bSsq},
    ~
    \bC_2 \, = \, \obt{0}{\bI}.
    \ea
    \label{eq.ABCuj}
    \eeq
The evolution equations for $\bar{\bH}_{u 2}$ and $\bar{\bH}_{u 3}$ are determined by~(\ref{eq.g-ss}) and~(\ref{eq.ABCuj}) with $\We = 1$. This system of equations is in the standard singularly perturbed form~\citep{kokkharei99} as the time-derivative of $\bz$ is multiplied by a small positive parameter $\eps$ and the operator $\bA_{22}$ is invertible (this follows from the lower-block-triangular structure of $\bA_{22}$ and invertibility of both $\beta \bSos$ and $\beta \bSsq$).

\section{Dependence of variance amplification on the elasticity number: Proof of the main result}
    \label{sec.sp}

We next examine how the $\We$-independent functions in the expressions for variance amplification of velocity and polymer stress fluctuations depend on $\eps = 1/\mu$. The mathematical developments that follow have been used in Section~\ref{sec.main} to gain insight into the conditions under which strong elasticity amplifies stochastic disturbances. Considering the case of high $\mu$,
    $
    \eps
    =
    1/\mu
    \ll
    1,
    $
we employ singular perturbation methods~\citep{kokkharei99} to show that function $g$ in~\ref{eq.Ev} and functions $a$, $b$, and $c$ in~\ref{eq.Etau} approximately become elasticity-number-independent. In elasticity-dominated flows, we demonstrate that these functions are correctly predicted by the analysis of creeping flows. In contrast, the function $f$ that quantifies variance amplification from $d_1$ to $u$ and from ($d_2,d_3$) to ($v,w$) in~\ref{eq.Ev} is inversely proportional to $\eps$. Furthermore, while the inertialess model correctly predicts behavior of the operators $\bH_{r j}$ with $\{r = u$; $j = 1 \}$ and $\{ r = v,w$; $j = 2,3 \}$ at low temporal frequencies, it provides a poor approximation at high temporal frequencies (see~\ref{sec.Pi_a}). We also show that, from a physical point of view, no important viscoelastic effects take place in the contribution of the function $f$ to the variance amplification.

The developments that follow make heavy use of singular perturbation techniques for stochastically forced linear systems~\citep{kokkharei99}. As summarized in Section~\ref{sec.main}, the use of such techniques provides (i) important physical insight about the dynamics of strongly elastic fluids; and (ii) the asymptotic forms of the functions ($f,g$) in~\ref{eq.Ev} and ($a,b,c$) in~\ref{eq.Etau} at high elasticity number.

\subsection{Variance amplification of velocity fluctuations}
    \label{sec.Ev}

As described in~\ref{sec.Hij-ss}, each $\bar{\bH}_{r j}$ (cf.\ Section~\ref{sec.PSDv}) represents an operator in $y$, mapping the forcing $d_j$ to the velocity $r$ at $\We = 1$. Since the inverse temporal Fourier transform yields a system of PDEs in $y$ and $t$, the $\We$-independent contributions to the variance amplification of velocity fluctuations can be determined by recasting each $\bar{\bH}_{r j}$ in the evolution form
    \beq
    \ba{rcl}
    \dot{\bx}_{rj}(y,k_z,t)
    & \! = \! &
    \bA_{rj} (k_z) \, \bx_{rj} (y,k_z,t)
    \, + \,
    \bB_j (k_z) \, d_j (y,k_z,t),
    \\[0.1cm]
    r (y,k_z,t)
    & \! = \! &
    \bC_r (k_z) \, \bx_{rj} (y,k_z,t),
    \ea
    \non
    \eeq
where $\bx_{rj}$ is a vector of state variables, and ($d_j,r$) is the input-output pair for the frequency response operator $\bar{\bH}_{r j}$, $\{ r = u, v, w$; $j = 1, 2, 3 \}$. Note that $\bx_{r j}$ and operators $\bA_{r j}$, $\bB_j$, and $\bC_r$ will, in general, be different for each $\bar{\bH}_{r j}$. It is a standard fact~\citep{Farrell1993} that the variance of $r$ sustained by $d_j$ is determined by
    \[
    E_{rj}
    \, = \,
    \trace \left( \bP_{r j} \bC_r^* \bC_r \right),
    \]
where $\bP_{rj}$ denotes the steady-state auto-correlation operator of $\bx_{rj}$, which is found by solving the Lyapunov equation,
    \[
    \bA_{rj} \bP_{rj}
    \, + \,
    \bP_{rj} \bA_{rj}^*
    \, = \,
    -
    \bB_j \bB_j^*.
    \]

From~\ref{sec.Hij-ss} it follows that the evolution equations of each $\bar{\bH}_{r j}$ assume the form
    \beq
    \ba{rcl}
    \tbo{\dot{\bx}}{\eps \dot{\bz}}
    & \!\!\! = \!\!\! &
    \tbt{\bA_{11}}{\bA_{12}}{\bA_{21}}{\bA_{22} (\eps)}
    \tbo{\bx}{\bz}
    \, + \,
    \tbo{0}{\bB_2} d_j,
    \\[0.25cm]
    r
    & \!\!\! = \!\!\! &
    \obt{0}{\bC_2}
    \tbo{\bx}{\bz},

    \ea
    \label{eq.Hij-ss}
    \eeq
with appropriate boundary conditions on $\bx$ and $\bz$. To simplify notation we have omitted the $r$ and $j$ indices in~(\ref{eq.Hij-ss}); it is to be noted, however, that $\bx$, $\bz$ and the $\bA$-operators are indexed by both $r$ and $j$, the $\bB$-operators are indexed by $j$, and the $\bC$-operators are indexed by $r$. Equations~(\ref{eq.f-ss}) and~(\ref{eq.g-ss}) (and consequently~(\ref{eq.Hij-ss})) are in the standard {\em singularly perturbed\/} form~\citep{kokkharei99} as the time-derivative of the second part of the state is multiplied by a small positive parameter $\eps$ and the lower-right-hand-corner blocks of the dynamical generators in both~(\ref{eq.f-ss}) and~(\ref{eq.g-ss}) are invertible. Furthermore, this representation gives evolution equations for different components of the frequency response operator $\bar{\bH}$ with a lower number of states compared to the original evolution model~(\ref{eq.lnse-2d3c}). In particular, there are two state variables in the evolution equations for operators $\bar{\bH}_{r j}$ with $\{r = u$; $j = 1 \}$ and $\{ r = v,w$; $j = 2,3 \}$ (cf.\ (\ref{eq.f-ss})), and four state variables in the evolution equations for operators $\bar{\bH}_{u 2}$ and $\bar{\bH}_{u 3}$ (cf.\ (\ref{eq.g-ss})). In comparison, there are eight states in the evolution model~(\ref{eq.lnse-2d3c}).

We next exploit the structure of equations~(\ref{eq.f-ss}) and~(\ref{eq.g-ss}) to uncover a slow-fast decomposition of each $\bar{\bH}_{r j}$, identify the physics of the slow and fast subsystems, and provide explicit analytical expressions for the variance amplification of velocity fluctuations in flows of strongly elastic polymeric fluids. These analytical developments have been utilized in Section~\ref{sec.main} to clearly identify important physical mechanisms leading to amplification from different forcing to different velocity components.

\subsubsection{Scaling of function\/ $f$ in\/~\mbox{\rm \ref{eq.Ev}} with $\eps$}
    \label{sec.f}

We first examine how function $f (k_z;\beta,\eps)$ in the expression for variance amplification of velocity fluctuations~\ref{eq.Ev} depends on $\eps$. From Section~\ref{sec.PSDv} we recall that $f$ is determined by
    \[
    f
    \, = \,
    f_{u 1} \, + \, \sum_{j \, = \, 2}^{3} \left( f_{v j} \, + \, f_{w j} \right),
    \]
where functions $f_{r j}$ quantify the variance amplification of the frequency response operators from $d_j$ to $r$, with $\{r = u$; $j = 1 \}$ and $\{ r = v,w$; $j = 2,3 \}$. The analysis presented in~\ref{sec.f-eps} reveals that $f$ is determined by
    \beq
    \ba{rcl}
    f (k_z;\beta,\eps)
    & \! = \! &
    \dfrac{1}{2}
    \,
    \trace
    \left(
    \bSos^{-1}
    ( \beta \bSos - \eps \bI )^{-1}
    \, + \;
    \bSsq^{-1}
    ( \beta \bSsq - \eps \bI )^{-1}
    \right)
    \; -
    \\[0.35cm]
    & \!  \! &
    \dfrac{1}{2 \eps}
    \,
    \trace
    \left(
    ( \beta \bSos - \eps \bI )^{-1}
    \, + \;
    ( \beta \bSsq - \eps \bI )^{-1}
    \right).
    \ea
    \label{eq.f}
    \eeq
This expression for $f (k_z;\beta,\eps)$ is valid for {\em all\/} $k_z \in (- \infty, \, \infty)$, $\beta \in (0,1)$, and $\eps > 0$. Furthermore, in strongly elastic flows, i.e.\ for $0 < \eps \ll 1$, $f (k_z;\beta,\eps)$ can be expressed as~(for details, see~\ref{sec.f-eps})
    \beq
    \ba{c}
    \ba{rcl}
    f (k_z;\beta,\eps)
    & \!\! = \!\! &
    (1/\eps)
    \ds{\sum_{n \, = \, 0}^{\infty}}
    \eps^n f_n (k_z;\beta)
    \\[0.15cm]
    & \!\! = \!\! &
    (1/\eps)
    f_0 (k_z;\beta)
    \, + \,
    f_1 (k_z;\beta)
    \, + \,
    \eps f_2 (k_z;\beta)
    \, + \,
    \ldots,
    \ea
    \\[0.75cm]
    f_0 (k_z;\beta)
    \, = \,
    \tilde{f}_0 (k_z)/\beta,
    ~
    \tilde{f}_0 (k_z)
    \, = \,
    - (1/2)
    \,
    \trace
    \left(
    \bSos^{-1}
    + \,
    \bSsq^{-1}
    \right),
    \\[0.1cm]
    f_n (k_z;\beta)
    \, = \,
    (1-\beta) \tilde{f}_n (k_z)/\beta^{n + 1},
    ~
    \tilde{f}_n (k_z)
    \, = \,
    - (1/2)
    \,
    \trace
    \left(
    \bSos^{-(n+1)}
    +
    \,
    \bSsq^{-(n+1)}
    \right),
    ~
    n \geq 1.
    \ea
    \label{eq.f-high-mu}
    \eeq

There are two key results of this section that quantify the dependence of the function $f$ in~\ref{eq.Ev} on $\eps = 1/\mu$. While scaling relation~(\ref{eq.f}) holds for flows with arbitrary but finite elasticity number, scaling relation~(\ref{eq.f-high-mu}) holds only for flows with high elasticity numbers, $1 \ll \mu < \infty$. The latter relation shows that, in elasticity-dominated flows, the traces of the inverses of the Orr-Sommerfeld and Squire operators in streamwise-constant flows of Newtonian fluids with $Re = 1$ specify the spatial frequency content of the function $f$. In~\ref{sec.Pi_a}, we demonstrate that the $1/\eps$-scaling of this function originates from the corresponding power spectral density becoming almost uniformly distributed over the temporal frequency bandwidth which is inversely proportional to $\eps$. This broad temporal spectrum of the frequency response operators from $d_j$ to $r$, with $\{r = u$; $j = 1 \}$ and $\{ r = v,w$; $j = 2,3 \}$, is accompanied by viscous dissipation in $k_z$ and it does not change the value of the peaks in the power spectral densities.

\subsubsection{Scaling of function\/ $g$ in\/~\mbox{\rm \ref{eq.Ev}} with $\eps$}
    \label{sec.g}

By examining $\bar{\bH}_{u2}$ and $\bar{\bH}_{u3}$ we can determine the $\eps$-dependence of terms responsible for the $\We^2$-scaling of the steady-state velocity variance in~\ref{eq.Ev}. As shown in Section~\ref{sec.PSDv},
    \[
    g (k_z;\beta,\eps)
    \, = \,
    g_{u 2} (k_z;\beta,\eps)
    \, + \,
    g_{u 3} (k_z;\beta,\eps),
    \]
where $g_{u j}$ denotes the steady-state variance of system~(\ref{eq.Hij-ss}) with $\{ r = u$; $j = 2,3 \}$ and~(for details, see~\ref{sec.g-ee})
    \beq
    \ba{c}
    \bA_{11}
    \, = \,
    \tbt{-\bI}{0}{\bC_{p2}}{-\bI},
    ~
    \bA_{12}
    \, = \,
    \tbt{\bI}{0}{0}{\bSsq},
    ~
    \bB_2
    \, = \,
    \tbo{\bF_j}{0},
    \\[0.35cm]
    \bA_{21}
    \, = \,
    \tbt{(1 - \beta) \, \bSos}{0}{0}{(1 - \beta) \, \bI},
    ~
    \bA_{22} (\eps)
    \, = \,
    \tbt{\beta \, \bSos}{0}{\eps \, \bC_{p1}}{\beta \, \bSsq},
    ~
    \bC_2 \, = \, \obt{0}{\bI}.
    \ea
    \non
    \eeq
Setting $\eps = 0$ in the $\bz$-equation of system~(\ref{eq.Hij-ss}) yields
    \beq
    \bar{\bz}
    \, = \,
    - \bA_{22}^{-1} (0)
    \left(
    \bA_{21} \, \bar{\bx}
    \, + \,
    \bB_2 \, d_j
    \right),
    \non
    \eeq
which in conjunction with the definitions of $\bx = [\,\xi~\,\varphi\,]^T$ and $\bz = [\,\psi~\,u\,]^T$ leads to the following expressions for the streamfunction and the streamwise velocity,
    \beq
    \tbo{\bar{\psi}}{\bar{u}}
    \, = \,
    \dfrac{1}{\beta}
    \left(
    \tbt{-(1 - \beta) \, \bI}{0}{0}{-(1 - \beta) \, \bSsq^{-1}}
    \tbo{\bar{\xi}}{\bar{\varphi}}
    \, + \,
    \tbo{- \bSos^{-1} \bF_j}{0} d_j
    \right).
    \non
    \eeq
Here, we use the overbar to denote the solution of system~(\ref{eq.Hij-ss}) with $\eps = 0$. As shown in~\ref{sec.g-ee}, the components of $\bx$ account for a low-pass version of the streamfunction and the spanwise/wall-normal gradients in the components of $\bphi_4$, i.e.,
    \[
    \xi
    \; = \;
    \psi/(\mri \omega + 1),
    ~~
    \varphi
    \; = \;
    \bS_{34} \bphi_4.
    \]

Note that $\bar{\psi}$ is not a valid approximation of $\psi$; this is because of the white noise component $d_j$ in the expression for $\bar{\psi}$, which yields infinite variance of the difference between $\psi$ and $\bar{\psi}$ irrespective of how small $\eps$ is. Nevertheless, $\bar{\psi}$ can still be employed as an approximation of an input $\psi$ to the $\bx$-subsystem in~(\ref{eq.Hij-ss}) as the slow system filters out the white noise component in $\bar{\psi}$. On the other hand, the absence of $d_j$ in the expression of $\bar{u}$ makes
    $
    ( 1 - 1/\beta ) \, \bSsq^{-1} \bar{\varphi}
    $
a valid approximation of the streamwise velocity fluctuation. Furthermore, the approximate dynamics of the slow subsystem are obtained by substituting the above expression for $\bar{\bz}$ into the $\bx$-equation of system~(\ref{eq.Hij-ss}),
    \beq
    \ba{rcl}
    \dot{\bx}_{uj,s}
    & \! = \! &
    \bA_{uj,s}
    \,
    \bx_{uj,s}
    \; + \;
    \bB_{j,s}
    \,
    d_j,
    \\[0.1cm]
    u
    & \! = \! &
    \bC_{u,s}
    \,
    \bx_{uj,s},
    \ea
    \non
    \eeq
with
    \beq
    \ba{c}
    \bA_{uj,s}
    \, = \,
    \bA_{11}
    \, - \,
    \bA_{12}
    \,
    \bA_{22}^{-1} (0)
    \,
    \bA_{21}
    \, = \,
    \tbt{- (1/\beta) \, \bI}{0}{\bC_{p2}}{- (1/\beta) \, \bI},
    \\[0.35cm]
    \bB_{j,s}
    \, = \,
    -
    \bA_{12} \, \bA_{22}^{-1} (0) \, \bB_2
    \, = \,
    \tbo{-(1/\beta) \bSos^{-1} \bF_j}{0},
    ~~
    \bC_{u,s}
    \, = \,
    \obt{0}{\frac{\beta - 1}{\beta} \, \bSsq^{-1}}.
    \ea
    \non
    \eeq
Thus, we have
    \[
    g_{uj}
    \, = \,
    \trace
    \left(
    \bP_{u j, s} \bC_{u,s}^* \bC_{u,s}
    \right)
    \, + \,
    \cO (\eps),
    \]
where $\bP_{uj,s}$ denotes the auto-correlation operator of $\bx_{uj,s}$,
    \[
    \bA_{uj,s} \bP_{uj,s}
    \, + \,
    \bP_{uj,s}
    \bA_{uj,s}^*
    \, = \,
    -
    \bB_{j,s} \bB_{j,s}^*.
    \]
A bit of algebra along with the self-adjointness of $\bSos$ and $\bSsq$ can be used to obtain
    \beq
    g_{uj} (k_z;\beta,\eps)
    \, = \,
    \dfrac{(1-\beta)^2}{4 \beta}
    \,
    \trace
    \left(
    \bSsq^{-1} \bC_{p2} \bSos^{-1} \bF_j \bF_j^* \bSos^{-1} \bC_{p2}^* \bSsq^{-1}
    \right)
    \, + \,
    \cO (\eps).
    \label{eq.guj}
    \eeq
In fact, a closer examination of the evolution equations of $\bar{\bH}_{u2}$ and $\bar{\bH}_{u3}$ (cf.~(\ref{eq.g-ss})) in conjunction with the singular perturbation methods of~\cite{kokkharei99} can be used to show that
    \[
    g (k_z; \beta, \eps)
    \, = \,
    \sum_{n \, = \, 0}^{\infty}
    \eps^n
    g_n (k_z; \beta)
    \, = \,
    g_0 (k_z; \beta)
    \, + \,
    \eps \, g_1 (k_z; \beta)
    \, + \,
    \cO (\eps^2),
    ~~
    \eps \, \ll \, 1,
    \]
where
    \[
    g_0 (k_z; \beta)
    \, = \,
    \sum_{j \, = \, 2}^{3}
    \trace
    \left(
    \bP_{uj,s} \bC_{u,s}^* \bC_{u,s}
    \right)
    \, = \,
    g_{u 2} (k_z; \beta, 0)
    \, + \,
    g_{u 3} (k_z; \beta, 0).
    \]
Now, since $\bC_{p2} = \mri k_z \tilde{\bC}_{p2}$ (cf.~(\ref{eq.Cp2})) and $\bF_2 \bF_2^* + \bF_3 \bF_3^* = \bI$ (see~\cite{jovbamJFM05}), we can use~(\ref{eq.guj}) to obtain
    \beq
    g_0 (k_z; \beta)
    \, = \,
    \tilde{g}_0 (k_z) (1 - \beta)^2/\beta,
    ~~
    \tilde{g}_0 (k_z)
    \, = \,
    (k_z^2 / 4)
    \,
    \trace
    \left(
    \bSsq^{-1} \tilde{\bC}_{p2} \bSos^{-2} \tilde{\bC}_{p2}^* \bSsq^{-1}
    \right).
    \label{eq.g0}
    \eeq
An in-depth study of function $\tilde{g}_0 (k_z)$ and its importance in the early stages of transition to elastic turbulence has been provided in Section~\ref{sec.main}.

Finally, we note that in the absence of inertia the operators $\bar{\bH}_{u 2}$ and $\bar{\bH}_{u 3}$ simplify to
    \beq
    \bar{\bH}_{u j} (k_z,\omega;\beta,0)
    \, = \,
    \dfrac{1 \, - \, \beta}{(1 \, + \, \beta \, \mri \omega )^2}
    \,
    \bSsq^{-1}
    \bC_{p2}
    \bSos^{-1}
    \bF_j,
    ~~
    j = 2,3.
    \non
    \eeq
Using the separation of the temporal and the spatial responses in $\bar{\bH}_{u j} (k_z,\omega;\beta,0)$ it is now straightforward to show that $g_{u j} (k_z;\beta,0)$ in~(\ref{eq.guj}) is determined by
    \beq
    g_{u j} (k_z;\beta,0)
    \, = \,
    \dfrac{1}{2 \pi}
    \int_{-\infty}^{\infty}
    \trace
    \left(
    \bar{\bH}_{u j} (k_z,\omega;\beta,0)
    \,
    \bar{\bH}_{u j}^* (k_z,\omega;\beta,0)
    \right)
    \,
    \mrd \omega.
    \non
    \eeq
As a matter of fact, creeping flow of an Oldroyd-B fluid captures well the responses from the wall-normal and spanwise forces to the streamwise velocity at both low and high temporal frequencies in elasticity-dominated regimes. This is in contrast to the analysis conducted in~\ref{sec.Pi_a} where it was shown that a creeping-flow model poorly approximates the responses from $d_1$ to $u$ and from ($d_2,d_3$) to ($v,w$) at high temporal frequencies.

To summarize, in streamwise-constant channel flows of Oldroyd-B fluids with $\eps = 1/\mu \ll 1$, the function $g$ contributing to the $\We^2$-scaling of the steady-state velocity variance in~\ref{eq.Ev} is approximately $\eps$-independent and it is determined by~(\ref{eq.g0}).

Finally, we note that the following scaling of the variance amplification associated with the velocity field
    \beq
    \hat{E}_{\mrv} (k_z;Re,\beta,\mu)
    \, \approx \,
    Re
    \,
    \hat{f} (k_z; \beta)
    \, + \,
    \mu
    Re^3
    \,
    \hat{g} (k_z; \beta),
    ~~
    1 \, \ll \, \mu \, < \, \infty,
    \label{eq.Ev-hjk}
    \eeq
was hypothesized by~\cite{hodjovkumJFM09} on the basis of numerical data. Even though~(\ref{eq.Ev-hjk}) appears to be at odds with~(\ref{eq.Ev-main}), we next furnish a proof of its validity. The evolution model~(\ref{eq.lnse-2d3c}),
    \beq
    \ba{rcl}
    \partial_{t_{\mre}} \bx (y,k_z,t)
    & \!\! = \!\! &
    \bA_{\mre} (k_z) \, \bx (y,k_z,t)
    \, + \,
    \bB_{\mre} (k_z) \, \bd_{\mre} (y,k_z,t),
    \\[0.1cm]
    \bv (y,k_z,t)
    & \!\! = \!\! &
    \bC (k_z) \, \bx (y,k_z,t),
    \ea
    \label{eq.ee-new}
    \eeq
and the evolution model in~\cite{hodjovkumJFM09},
    \beq
    \ba{rcl}
    \partial_{t_{\mri}} \bx (y,k_z,t)
    & \!\! = \!\! &
    \bA_{\mri} (k_z) \, \bx (y,k_z,t)
    \, + \,
    \bB_{\mri} (k_z) \, \bd_{\mri} (y,k_z,t),
    \\[0.1cm]
    \bv (y,k_z,t)
    & \!\! = \!\! &
    \bC (k_z) \, \bx (y,k_z,t),
    \ea
    \label{eq.ee-hjk}
    \eeq
are obtained using different time and forcing scalings. In~(\ref{eq.ee-new}), $t_{\mre}$ denotes time normalized by $\lambda$, and $\bd_{\mre}$ denotes forcing per unit mass normalized by $(\eta_s + \eta_p) U_o/\rho L^2$; in~(\ref{eq.ee-hjk}), time is normalized by the convective time scale $L/U_o$, and forcing per unit mass is normalized by $U_o^2/L$. It is easy to show that the $\bA$ and $\bB$ operators in~(\ref{eq.ee-hjk}) and~(\ref{eq.ee-new}) are related by
    \beq
    \bA_{\mri}
    \, = \,
    (1/\We) \, \bA_{\mre},
    ~~
    \bB_{\mri}
    \, = \,
    (Re/\We) \, \bB_{\mre}.
    ~~
    \non
    \eeq
Therefore, the solutions to the corresponding Lyapunov equations
    \[
    \bA_{\mrk} \bP_{\mrk}
    \, + \,
    \bP_{\mrk} \bA_{\mrk}^*
    \, = \,
    -
    \bB_{\mrk} \bB_{\mrk}^*,
    ~~
    \mrk
    \, = \,
    \{ \mre, \, \mri \},
    \]
are related to each other by
    \beq
    \bP_{\mri}
    \, = \,
    (Re^2/\We)
    \,
    \bP_{\mre}
    \, = \,
    (Re/\mu)
    \,
    \bP_{\mre},
    \non
    \eeq
which in conjunction with (\ref{eq.Ev-hjk}) and (\ref{eq.Ev-main}) can be used to obtain the following expression for variance amplification in elasticity-dominated flows
    \beq
    \ba{rcl}
    \hat{E}_{\mrv} (k_z;Re,\beta,\mu)
    & \!\!\! = \!\!\! &
    \trace
    \left(
    \bP_{\mri} \bC^* \bC
    \right)
    \, = \,
    (Re/\mu)
    \,
    \trace
    \left(
    \bP_{\mre} \bC^* \bC
    \right)
    \, = \,
    (Re/\mu)
    \,
    E_{\mrv} (k_z; \We,\beta,\mu)
    \\[0.15cm]
    & \!\!\! = \!\!\! &
    Re \, \tilde{f}_0 (k_z)/\beta
    \; + \;
    \mu Re^3 \, \tilde{g}_0 (k_z) \, (1 - \beta)^2/\beta
    \; + \;
    \cO(1/\mu).
    \ea
    \non
    \eeq
This establishes validity of the scaling conjectured in~\cite{hodjovkumJFM09} and shows that the functions $\hat{f}$ and $\hat{g}$ in~(\ref{eq.Ev-hjk}) are, respectively, determined by $\hat{f} (k_z; \beta) = \tilde{f}_0 (k_z)/\beta$ and $\hat{g} (k_z; \beta) = \tilde{g}_0 (k_z) (1 - \beta)^2/\beta$.

\subsection{Variance amplification of polymer stress fluctuations}
    \label{sec.Etau}

In~\ref{sec.Ev}, we have studied how the elasticity number influences frequency responses of velocity fluctuations in strongly elastic channel flows of Oldroyd-B fluids. Here, we examine the elasticity number scaling of the functions $a$, $b$, and $c$ in the expression for the steady-state variance of polymer stresses~\ref{eq.Etau}. In flows with $\eps = 1/\mu \ll 1$, we show that these functions approximately become $\eps$-independent, thereby implying that $E_{\tau}$ in~\ref{eq.Etau} scales as
    \beq
    E_{\tau} (k_z;\We,\beta,\mu)
    \; = \;
    a_0 (k_z;\beta)
    \; + \;
    \We^2
    \,
    b_0 (k_z;\beta)
    \; + \;
    \We^4
    \,
    c_0 (k_z;\beta)
    \; + \;
    \cO(1/\mu).
    \label{eq.Etau-Re0}
    \eeq

One of the key results of this section is our finding that, in flows with high elasticity numbers, the analysis of the inertialess Oldroyd-B model correctly approximates dynamics of polymer stress fluctuations. This follows directly from the observation that the evolution model~(\ref{eq.lnse-2d3c}) is in a standard singularly perturbed form. Namely, setting $\eps$ to zero in~(\ref{eq.lnse-2d3c-v}) and~(\ref{eq.lnse-2d3c-eta}) yields the expressions for $\phi_1 = \psi$ and $\phi_3 = u$ in terms of the polymer stress fluctuation tensor $\btau$ and the stochastic forcing $\bd$. As explained in~\ref{sec.Pi_a}, even though these expressions do not represent valid approximations of $\psi$ and $u$ (see the discussion following equation~(\ref{eq.zbar})) they can still be used to approximate these two fields as an input into the equations for polymer stresses~(\ref{eq.lnse-2d3c-tauv}),~(\ref{eq.lnse-2d3c-taueta}), and~(\ref{eq.lnse-2d3c-tau11}). This is because the error in approximating $\psi$ and $u$ by white noise forcing is filtered out by the dynamics of the slow subsystem. Although this is a viable approach to the analysis of the functions $a$, $b$, and $c$ in~\ref{eq.Etau}, a more convenient representation for determination of these functions is laid out next.

\subsubsection{Scaling of function $a$ in\/~\mbox{\rm \ref{eq.Etau}} with $\eps$}
    \label{sec.a}

We begin this section by examining the $\eps$-dependence of the operators that map $d_1$ to
    $
    \bphi_4
    =
    \left[\,\tau_{12}\,\,\,\tau_{13}\,\right]^T
    $
and $d_2$ or $d_3$ to
    $
    \bphi_2
    =
    \left[\,\tau_{22}\,\,\,\tau_{23}\,\,\,\tau_{33}\,\right]^T.
    $
From Section~\ref{sec.PSDtau} we recall that the steady-state variance of these operators, which are respectively denoted by $\bar{\bGamma}_{\phi_4,1}$ and $\bar{\bGamma}_{\phi_2,j}$ with $j = \{ 2,3 \}$, is quantified by the function $a (k_z; \beta,\eps)$ in~\ref{eq.Etau}. Based on the developments in~\ref{sec.f-ee} and~\ref{sec.Pi_a} we conclude that $\bar{\bGamma}_{\phi_4,1}$ and $\bar{\bGamma}_{\phi_2,j}$ admit the following evolution equations
    \beq
    \ba{rcl}
    \tbo{\dot{x}_{i j}}{\eps \dot{z}_{i j}}
    & \!\!\! = \!\!\! &
    \tbt{-\bI}{\bI}{(1 - \beta) \bS_{\mrk}}{\beta \bS_{\mrk}}
    \tbo{x_{i j}}{z_{i j}}
    \, + \,
    \tbo{0}{\bF_j} d_j,
    \\[0.25cm]
    \bphi_i
    & \!\!\! = \!\!\! &
    \obt{\bC_i}{0}
    \tbo{x_{i j}}{z_{i j}},
    \ea
    \label{eq.Sa-ss}
    \eeq
with $\bC_2 = \bS_{21}$, $\bC_4 = \bS_{43}$, $\{ \mrk = \mathrm{os}$ for $i = 2$; $\mrk = \mathrm{sq}$ for $i = 4 \}$, homogeneous Dirichlet  boundary conditions on $x_{41} = \zeta = u/(\mri \omega + 1)$ and $z_{4 1} = u$, and homogeneous Cauchy boundary conditions on $x_{2 j} = \xi = \psi/(\mri \omega + 1)$ and $z_{2 j} = \psi$ for $j = \{ 2,3 \}$. The analysis presented in~\ref{sec.a-eps} develops the following formula for the function $a$
    \beq
    \ba{rcl}
    \!\!
    a (k_z; \beta,\eps)
    & \!\!\! = \!\!\! &
    a_{\mathrm{sq}} (k_z; \beta,\eps)
    \, + \,
    a_{\mathrm{os}} (k_z; \beta,\eps)
    \\[0.15cm]
    \!\!
    & \!\!\! = \!\!\! &
    \dfrac{1}{2}
    \,
    \trace
    \left(
    \bSsq^{-1}
    ( \beta \bSsq - \eps \bI )^{-1}
    \bS_{43}^* \, \bS_{43}
    \right)
    \, + \,
    \dfrac{1}{2}
    \,
    \trace
    \left(
    \bSos^{-1}
    ( \beta \bSos - \eps \bI )^{-1}
    \bS_{21}^* \, \bS_{21}
    \right),
    \ea
    \non
    \eeq
which holds for all $k_z$, non-negative values of $\eps$, and $\beta \in (0,1)$. Furthermore, in flows with $\eps = 1/\mu \ll 1$, the function $a$ in~\ref{eq.Etau} is approximately $\eps$-independent, i.e.
    \beq
    \ba{rcl}
    a (k_z; \beta, \eps)
    & \!\! = \!\! &
    a_0 (k_z; \beta)
    \, + \,
    \cO (\eps)
    \, = \,
    \tilde{a}_0 (k_z)/\beta
    \, + \,
    \cO (\eps),
    \\[0.15cm]
    \tilde{a}_0 (k_z)
    & \!\! = \!\! &
    \tilde{a}_{\mathrm{sq},0} (k_z)
    \, + \,
    \tilde{a}_{\mathrm{os},0} (k_z)
    \, = \,
    (1/2) \,
    \trace
    \left(
    \bSsq^{-2} \, \bS_{43}^* \, \bS_{43}
    \, + \,
    \bSos^{-2} \, \bS_{21}^* \, \bS_{21}
    \right),
    \ea
    \label{eq.a}
    \eeq
and the aggregate variance amplification of the operators $\bar{\bGamma}_{\phi_4,1}$, $\bar{\bGamma}_{\phi_2,2}$, and $\bar{\bGamma}_{\phi_2,3}$ in inertialess flows of an Oldroyd-B fluid is determined by $a (k_z; \beta, 0) = \tilde{a}_0 (k_z)/\beta$.

We note that similar arguments as in~\ref{sec.Pi_a} can be employed to show that the dynamics of the slow subsystems in (\ref{eq.Sa-ss}) are, respectively, given by
    \beq
    \ba{rcl}
    \dot{\xi}
    & \!\! = \!\! &
    - (1/\beta) \, \xi
    \, - \,
    (1/\beta) \, \bSos^{-1} \, \bF_{j} \, d_j,
    \\[0.1cm]
    \bphi_2
    & \!\! = \!\! &
    \bS_{21}
    \,
    \xi,
    \ea
    \label{eq.xi-slow}
    \eeq
with $j = \{ 2,3 \}$, and
    \beq
    \ba{rcl}
    \dot{\zeta}
    & \!\! = \!\! &
    - (1/\beta) \, \zeta
    \, - \,
    (1/\beta) \, \bSsq^{-1} \, d_1,
    \\[0.1cm]
    \bphi_4
    & \!\! = \!\! &
    \bS_{43}
    \,
    \zeta.
    \ea
    \label{eq.eta-slow}
    \eeq
It is easy to show that the frequency responses of slow subsystems~(\ref{eq.xi-slow}) and~(\ref{eq.eta-slow}) are fully captured by those determined from the inertialess model, i.e.,
    \beq
    \ba{rcl}
    \bar{\bGamma}_{\phi_2,j} (\omega,k_z;\beta,0)
    & \!\! = \!\! &
    -
    \dfrac{1/\beta}{\mri \omega \, + \, 1/\beta}
    \,
    \bS_{21} \, \bSos^{-1} \, \bF_{j},
    ~~
    j \, = \, 2,3,
    \\[0.25cm]
    \bar{\bGamma}_{\phi_4,1} (\omega,k_z;\beta,0)
    & \!\! = \!\! &
    -
    \dfrac{1/\beta}{\mri \omega \, + \, 1/\beta}
    \,
    \bS_{43} \, \bSsq^{-1},
    \ea
    \non
    \eeq
and that the aggregate variance amplification of these operators is obtained by setting $\eps$ to zero in~(\ref{eq.a}). This separates the temporal and the spatial parts of the responses and suggests simple temporal dynamics of $\bar{\bGamma}_{\phi_2,j}$ and $\bar{\bGamma}_{\phi_4,1}$ in inertialess flows. The simple features of the temporal responses would not be obvious if the singular perturbation techniques were instead applied directly to the original evolution model~(\ref{eq.lnse-2d3c}).

\subsubsection{Scaling of function $b$ in~\mbox{\rm \ref{eq.Etau}} with $\eps$}
    \label{sec.b}

Singular perturbation techniques can be employed to show that $\eps$ has a negligible influence on the function $b$ in elasticity-dominated flows. Since this analysis follows a similar path to what was already presented, here we only derive the expression for the function $b_0$,
    \beq
    b (k_z; \beta, \eps)
    \, = \,
    b_0 (k_z; \beta)
    \, + \,
    \cO (\eps),
    \non
    \eeq
which determines the steady-state variance amplification from $d_1$ to $\phi_5 = \tau_{11}$ and from $[\,d_2\,\,\,d_3\,]^T$ to $\bphi_4 = \left[\,\tau_{12}\,\,\,\tau_{13}\,\right]^T$ in inertialess flows.

Since the streamwise forcing does not influence the dynamics of $\phi_1$ and $\bphi_2$, the response of $\phi_5$ arising from $d_1$ is determined by (see~\ref{sec.fr-tau})
    \beq
    \phi_5
    \, = \,
    \dfrac{\We}{\mri \omega \, + \, 1}
    \left(
    \bS_{53} \, u
    \, + \,
    \bS_{54} \, \bphi_4
    \right)
    \, = \,
    \We
    \,
    \dfrac{\mri \omega \, + \, 2}{\mri \omega \, + \, 1}
    \,
    \bS_{53}
    \,
    \zeta.
    \non
    \eeq
In arriving at this expression, we have used (i) the definition of $\zeta$, $\zeta = u/(\mri \omega + 1)$; (ii) the fact that $\bphi_4 = \bS_{43} \zeta$ when $\phi_1 = 0$ and $\bphi_2 = 0$; and (iii) $\bS_{54} \bS_{43} = \bS_{53}$. Now, in inertialess flows the dynamics of $\zeta$ are governed by~(\ref{eq.eta-slow}), and we thus have
    \beq
    \bar{\bGamma}_{\phi_5,1} (\omega,k_z;\beta,0)
    \, = \,
    \bar{\bGamma}_{11,1} (\omega,k_z;\beta,0)
    \, = \,
    -
    \,
    \dfrac{\mri \omega \, + \, 2}
    {\beta (\mri \omega \, + \, 1) (\mri \omega \, + \, 1/\beta)}
    \,
    \bS_{53}
    \,
    \bSsq^{-1}.
    \non
    \eeq
A bit of algebra yields the expression for the variance amplification of this operator
    \beq
    b_{11,1} (k_z;\beta,0)
    \, = \,
    \dfrac{1 \, + \, 4 \beta}{2 \beta (1 \, + \, \beta)}
    \,
    \trace
    \left(
    \bSsq^{-1}
    \,
    \bS_{53}^*
    \,
    \bS_{53}
    \,
    \bSsq^{-1}
    \right).
    \label{eq.b0-tau11-d1}
    \eeq

We next examine variance amplification of operators from $d_2$ or $d_3$ to $\bphi_4$ in inertialess flows. Using~(\ref{eq.phi4b}) and the definition of $\xi$, $\xi = \psi/(\mri \omega + 1)$, we can express $\bphi_4$ as
    \beq
    \ba{rcl}
    \bphi_4
    & \!\! = \!\! &
    \dfrac{\We}{\mri \omega \, + \, 1}
    \left(
    \bS_{41}
    \, + \,
    \dfrac{\bS_{42} \bS_{21}}{\mri \omega \, + \, 1}
    \right)
    \psi
    \; + \;
    \dfrac{1}{\mri \omega \, + \, 1}
    \,
    \bS_{43} \, u
    \\[0.25cm]
    & \!\! = \!\! &
    \We
    \left(
    \bS_{41} \, \xi
    \, + \,
    \bS_{42} \, \bS_{21}
    \,
    \dfrac{1}{\mri \omega \, + \, 1}
    \,
    \xi
    \right)
    \; + \;
    \dfrac{1}{\mri \omega \, + \, 1}
    \,
    \bS_{43} \, u
    \ea
    \non
    \eeq
In the absence of inertia, $u$ arising from $d_2$ or $d_3$ is determined by (for details, see~\ref{sec.g})
    \beq
    u
    \, = \,
    \dfrac{\beta - 1}{\beta} \, \bSsq^{-1} \, \varphi
    \, = \,
    \We \, \dfrac{\beta - 1}{\beta} \, \bSsq^{-1} \, \bC_{p2} \,
    \dfrac{1}{\mri \omega \, + \, 1/\beta} \, \xi,
    \non
    \eeq
which in conjunction with the above expression for $\bphi_4$ and~(\ref{eq.xi-slow}) yields
    \beq
    \ba{rcl}
    \bphi_4
    & \!\! = \!\! &
    \We
    \left(
    \bS_{41} \, \xi
    \, + \,
    \bS_{42} \, \bS_{21}
    \,
    \dfrac{1}{\mri \omega \, + \, 1}
    \,
    \xi
    \, + \,
    \dfrac{\beta - 1}{\beta} \, \bS_{43} \, \bSsq^{-1} \, \bC_{p2}
    \,
    \dfrac{1}{(\mri \omega \, + \, 1/\beta) (\mri \omega \, + \, 1)}
    \,
    \xi
    \right),
    \\[0.25cm]
    \xi
    & \!\! = \!\! &
    - \dfrac{1/\beta}{\mri \omega \, + \, 1/\beta}
    \, \bSos^{-1} \, \bF_j \, d_j,
    ~~
    j \, = \, 2,3.
    \ea
    \label{eq.phi4-xi}
    \eeq
By selecting $\xi_1 = \xi$, $\xi_2 = \xi_1/(\mri \omega + 1)$, $\xi_3 = \xi_2/(\mri \omega + 1/\beta)$, the operator $\bar{\bGamma}_{\phi_4,j}$ that relates $d_2$ or $d_3$ to $\bphi_4$ in~(\ref{eq.phi4-xi}) can be represented by the following evolution equation
    \beq
    \ba{rcl}
    \thbo{\dot{\xi}_1}{\dot{\xi}_2}{\dot{\xi}_3}
    & \!\! = \!\! &
    \left[
    \ba{crc}
    {-1/\beta} & {0} & {0}
    \\
    {1} & {-1} & {0}
    \\
    {0} & {1} & {-1/\beta}
    \ea
    \right]
    \thbo{\xi_1}{\xi_2}{\xi_3}
    \, + \,
    \thbo{-(1/\beta) \, \bSos^{-1} \, \bF_j}{0}{0}
    d_j,
    \\[0.5cm]
    \bphi_4
    & \!\! = \!\! &
    \obth{\bS_{41}~}{\bS_{42} \, \bS_{21}}{~\frac{\beta - 1}{\beta} \, \bS_{43} \, \bSsq^{-1} \, \bC_{p2}}
    \thbo{\xi_1}{\xi_2}{\xi_3},
    \ea
    \label{eq.ee-phi4-d23}
    \eeq
with homogeneous Cauchy boundary conditions on $\xi_1$, $\xi_2$, and $\xi_3$. In inertialess Poiseuille and Couette flows, the Lyapunov equation associated with~(\ref{eq.ee-phi4-d23}) can be employed to compute the variance amplification of the frequency response operators that map $d_2$ or $d_3$ to $\bphi_4$. By defining $\bxi = \left[\,\xi_1\,\,\,\xi_2\,\,\,\xi_3\,\right]^{T}$, equation~(\ref{eq.ee-phi4-d23}) can be rewritten as
    \beq
    \ba{rcl}
    \dot{\bxi} (y,k_z,t)
    & \! = \! &
    \bA_{\xi} (k_z) \, \bxi (y,k_z,t)
    \, + \,
    \bB_j (k_z) \, d_j (y,k_z,t),
    \\[0.1cm]
    \bphi_4 (y,k_z,t)
    & \! = \! &
    \bC_4 (k_z) \, \bxi (y,k_z,t),
    \ea
    \non
    \eeq
and the function which determines the variance amplification from $d_2$ and $d_3$ to $\bphi_4 = \left[\,\tau_{12}\,\,\,\tau_{13}\,\right]^T$ in inertialess flows is obtained from $\trace \left( \bP_{\xi j} \bC_4^* \bC_4 \right)$, where $\bP_{\xi j}$ solves the Lyapunov equation
    \[
    \bA_{\xi} \bP_{\xi j}
    \, + \,
    \bP_{\xi j} \bA_{\xi}^*
    \, = \,
    -
    \bB_j \bB_j^*.
    \]

\subsubsection{Scaling of function $c$ in\/~\mbox{\rm \ref{eq.Etau}} with $\eps$}
    \label{sec.c}

The dependence of the steady-state variance amplification from $[\,d_2\,\,\,d_3\,]^T$ to $\phi_5 = \tau_{11}$ on $\eps$ in elasticity-dominated flows of Oldroyd-B fluids can be ascertained using singular perturbation techniques,
    \beq
    c (k_z; \beta, \eps)
    \, = \,
    c_0 (k_z; \beta)
    \, + \,
    \cO (\eps).
    \non
    \eeq
As shown in Section~\ref{sec.PSDtau}, this function is responsible for a quartic scaling of $E_{\tau}$ with $\We$ (cf.\ \ref{eq.Etau}).

In this section we only present the procedure that can be used to compute the function $c_0$ which quantifies the variance sustained in $\tau_{11}$ by $d_2$ and $d_3$ in inertialess channel flows. In the absence of inertia, the operator that maps the wall-normal and spanwise forces to the streamwise component of the polymer stress tensor can be expressed as
    \beq
    \ba{rcl}
    \phi_5
    & \!\! = \!\! &
    \We^2
    \left(
    \bS_{51} \, \xi
    \, + \,
    \bS_{54} \, \bS_{41}
    \,
    \dfrac{1}{\mri \omega \, + \, 1}
    \,
    \xi
    \, + \,
    \bS_{54} \, \bS_{42} \, \bS_{21}
    \dfrac{1}{(\mri \omega \, + \, 1)^2}
    \,
    \xi
    \right)
    ~ +
    \\[0.35cm]
    & \!\!  \!\! &
    \We^2
    \,
    \dfrac{\beta - 1}{\beta} \, \bS_{53} \, \bSsq^{-1} \, \bC_{p2}
    \,
    \dfrac{1}{\mri \omega \, + \, 1/\beta}
    \left(
    \dfrac{1}{\mri \omega \, + \, 1}
    \,
    \xi
    \, + \,
    \dfrac{1}{(\mri \omega \, + \, 1)^2}
    \,
    \xi
    \right),
    \\[0.35cm]
    \xi
    & \!\! = \!\! &
    - \dfrac{1/\beta}{\mri \omega \, + \, 1/\beta}
    \, \bSos^{-1} \, \bF_j \, d_j,
    ~~
    j \, = \, 2,3.
    \ea
    \label{eq.phi5-xi}
    \eeq
Now, by selecting $\gamma_1 = \xi$, $\gamma_2 = \gamma_1/(\mri \omega + 1)$, $\gamma_3 = \gamma_2/(\mri \omega + 1)$, $\gamma_4 = (\gamma_2 + \gamma_3)/(\mri \omega + 1/\beta)$, the operator $\bar{\bGamma}_{\phi_5,j}$ that relates $d_2$ or $d_3$ to $\phi_5$ in~(\ref{eq.phi5-xi}) can be represented by the following evolution equation
    \beq
    \ba{rcl}
    \fbo{\dot{\gamma}_1}{\dot{\gamma}_2}{\dot{\gamma}_3}{\dot{\gamma}_4}
    & \!\! = \!\! &
    \left[
    \ba{crrc}
    {-1/\beta} & {0} & {0} & {0}
    \\
    {1} & {-1} & {0} & {0}
    \\
    {0} & {1} & {-1} & {0}
    \\
    {0} & {1} & {1} & {-1/\beta}
    \ea
    \right]
    \fbo{\gamma_1}{\gamma_2}{\gamma_3}{\gamma_4}
    \, + \,
    \fbo{-(1/\beta) \, \bSos^{-1} \, \bF_j}{0}{0}{0}
    d_j,
    \\[0.5cm]
    \phi_5
    & \!\! = \!\! &
    \obf{\bS_{51}~}{\bS_{54} \, \bS_{41}}{~\bS_{54} \, \bS_{42} \, \bS_{21}}
    {~\frac{\beta - 1}{\beta} \, \bS_{53} \, \bSsq^{-1} \, \bC_{p2}}
    \fbo{\gamma_1}{\gamma_2}{\gamma_3}{\gamma_4},
    \ea
    \label{eq.ee-phi5-d23}
    \eeq
with homogeneous Cauchy boundary conditions on $\gamma_1$, $\gamma_2$, $\gamma_3$, and $\gamma_4$. This equation is in a form suitable for computing the variance sustained in $\phi_5 = \tau_{11}$ by wall-normal and spanwise forces in inertialess flows (i.e., the function $c_0 (k_z; \beta)$). In Poiseuille flow, the explicit expression for $c_0 (k_z; \beta)$ is rather involved, but in Couette flow some tedious algebraic manipulations can be used to derive the following formula,
    \beq
    \ba{rcl}
    c_0 (k_z; \beta)
    & \!\! = \!\! &
    \dfrac{4 \beta^4 \, + \, 16 \beta^3 \, + \, 29 \beta^2 \, + \, 6 \beta \, + \, 1}{(\beta \, + \, 1)^3}
    \,
    \tilde{c}_0 (k_z),
    \\[0.35cm]
    \tilde{c}_0 (k_z)
    & \!\! = \!\! &
    k_z^2
    \,
    \trace
    \left(
    \py \, \Delta^{-2} \, \Delta \, \Delta^{-2} \, \py
    \right).
    \ea
    \label{eq.c0-c}
    \eeq

The analysis of the functions $a_0$, $b_0$, and $c_0$ in~(\ref{eq.Etau-Re0}) that determine spatial frequency responses of polymer stress fluctuations in inertialess Couette and Poiseuille flows of Oldroyd-B fluids has been conducted in Section~\ref{sec.main-Etau}.

\section{Scaling of functions $f$ in~$\mbox{\rm \ref{eq.Ev}}$ and $a$ in~$\mbox{\rm \ref{eq.Etau}}$ with $\eps$}
    \label{sec.f-a-eps}

We outline here the procedure that is most convenient for uncovering explicit $\eps$-scaling of the $\We$-independent functions $f$ and $a$ in the expressions for variance amplification of velocity~\ref{eq.Ev} and polymer stress~\ref{eq.Etau} fluctuations. This approach utilizes the fact that the variance amplification can be determined from the solution of the corresponding Lyapunov equation.

\subsection{Scaling of function $f$ in\/~\mbox{\rm \ref{eq.Ev}} with $\eps$}
    \label{sec.f-eps}

From Section~\ref{sec.PSDv} we recall that the function $f$ in~\ref{eq.Ev} is determined by
    \[
    f
    \, = \,
    f_{u 1} \, + \, \sum_{j \, = \, 2}^{3} \left( f_{v j} \, + \, f_{w j} \right)
    \, = \,
    f_{\mathrm{sq}} \, + \, f_{\mathrm{os}},
    \]
where functions $f_{\mathrm{sq}}$ and $f_{\mathrm{os}}$, respectively, quantify the variance amplification of the frequency response operators from $d_1$ to $u$ and from $[\,d_2\,\,\,d_3\,]^T$ to $[\,v\,\,\,w\,]^T$. From~\ref{sec.f-ee} it follows that these operators admit evolution representations with
    \beq
    \bA_{\mrk}
    \, = \,
    \tbt{-\bI}{\bI}
    {\frac{1-\beta}{\eps} \, \bS_{\mrk}}{\frac{\beta}{\eps} \, \bS_{\mrk}},
    ~
    \bB_{\mrk}
    \, = \,
    \tbo{0}{\frac{1}{\eps} \, \bF_{\mrk}},
    ~
    \bC_{\mrk}
    \, = \,
    \obt{0}{\bG_{\mrk}},
    ~
    \mrk \, = \, \{ \mbox{os}, \, \mbox{sq} \},
    \non
    \eeq
where
    \beq
    \bF_{\mathrm{sq}}
    \, = \,
    \bF_1
    \, = \,
    \bI,
    ~~
    \bG_{\mathrm{sq}}
    \, = \,
    \bG_u
    \, = \,
    \bI,
    ~~
    \bF_{\mathrm{os}}
    \, = \,
    \obt{\bF_2}{\bF_3},
    ~~
    \bG_{\mathrm{os}}
    \, = \,
    \tbo{\bG_v}{\bG_w}.
    \non
    \eeq
Each $f_{\mrk}$ is then determined by
    \[
    f_{\mrk}
    \, = \,
    \trace \left( \bP_{\mrk} \bC_{\mrk}^* \bC_{\mrk} \right),
    ~~
    \mrk \, = \, \{ \mbox{os}, \, \mbox{sq} \},
    \]
where $\bP_{\mrk}$ denotes the solution to the Lyapunov equation
    \[
    \bA_{\mrk} \bP_{\mrk}
    \, + \,
    \bP_{\mrk} \bA_{\mrk}^*
    \, = \,
    -
    \bB_{\mrk} \bB_{\mrk}^*.
    \]

Now, since both $\bSos$ and $\bSsq$ are self-adjoint, and since $\{ \bF_1 \bF_1^* = \bI$; $\bF_2 \bF_2^* + \bF_3 \bF_3^* = \bI \}$, $\{ \bG_u^* \bG_u = \bI$; $\bG_v^* \bG_v + \bG_w^* \bG_w = \bI \}$ we have
    \beq
    \bA_{\mrk}^*
    \, = \,
    \left[
    \ba{rc}
    -\bI & \frac{1-\beta}{\eps} \, \bS_{\mrk}
    \\[0.1cm]
    \bI & \frac{\beta}{\eps} \, \bS_{\mrk}
    \ea
    \right],
    ~
    \bB_{\mrk} \bB_{\mrk}^*
    \, = \,
    \tbt{0}{0}{0}{\frac{1}{\eps^2} \, \bI},
    ~
    \bC_{\mrk}^* \bC_{\mrk}
    \, = \,
    \tbt{0}{0}{0}{\bI},
    ~
    \mrk \, = \, \{ \mbox{os}, \, \mbox{sq} \}.
    \non
    \eeq
We represent the self-adjoint operators $\bP_{\mrk} (\eps)$ as
    \beq
    \bP_{\mrk} (\eps)
    \, = \,
    \tbt{\bX_{\mrk} (\eps)}{\bY_{\mrk}^* (\eps)}
    {\bY_{\mrk} (\eps)}{\bZ_{\mrk} (\eps)},
    ~~
    \mrk \, = \, \{ \mbox{os}, \, \mbox{sq} \},
    \label{eq.Pk}
    \eeq
where the components of $\bP_{\mrk}$ are determined from the following system of equations
    \begin{subequations}
    \label{eq.XYZ}
    \begin{align}
    \label{eq.XYZa}
    - 2 \bX_{\mrk} (\eps) \, + \, \bY_{\mrk} (\eps) \, + \, \bY_{\mrk}^* (\eps)
    & \; = \;
    0,
    \\[0.cm]
    \label{eq.XYZb}
    (1 - \beta) \, \bS_{\mrk} \bX_{\mrk} (\eps)
    \, + \,
    \beta \, \bS_{\mrk} \bY_{\mrk} (\eps)
    \, - \,
    \eps \bY_{\mrk} (\eps)
    \, + \,
    \eps \, \bZ_{\mrk} (\eps)
    & \; = \;
    0,
    \\[0.cm]
    \label{eq.XYZc}
    \bS_{\mrk}
    \left(
    \beta \, \bZ_{\mrk} (\eps)
    +
    (1 - \beta) \bY_{\mrk}^* (\eps)
    \right)
    \, + \,
    \left(
    \beta \, \bZ_{\mrk} (\eps)
    +
    (1 - \beta) \bY_{\mrk} (\eps)
    \right)
    \bS_{\mrk}
    & \; = \;
    -
    (1/\eps)
    \,
    \bI.
    \end{align}
    \end{subequations}
Since the operators $\bS_{\mrk}$ with $\mrk = \{ \mbox{os}, \, \mbox{sq} \}$ in~(\ref{eq.XYZ}) are self-adjoint, they can be diagonalized using their respective eigenfunctions as the orthonormal basis of the underlying function space (see Appendix~B of~\cite{jovbamJFM05}). Thus, the solutions $\bX_{\mrk}$, $\bY_{\mrk}$, and $\bZ_{\mrk}$ of~(\ref{eq.XYZ}) also admit diagonal representation. This observation in conjunction with the fact that all coefficients in~(\ref{eq.XYZ}) are real can be used to obtain $\bY_{\mrk}^* (\eps) = \bY_{\mrk} (\eps)$, which consequently simplifies system~(\ref{eq.XYZ}) to
    \begin{subequations}
    \label{eq.XYZ1}
    \begin{align}
    \label{eq.XYZa1}
    \bY_{\mrk} (\eps)
    & \; = \;
    \bX_{\mrk} (\eps),
    \\[0.cm]
    \label{eq.XYZb1}
    \bS_{\mrk} \bX_{\mrk} (\eps)
    \, - \,
    \eps \bX_{\mrk} (\eps)
    \, + \,
    \eps \, \bZ_{\mrk} (\eps)
    & \; = \;
    0,
    \\[0.cm]
    \label{eq.XYZc1}
    \bS_{\mrk}
    \left(
    \beta \, \bZ_{\mrk} (\eps)
    +
    (1 - \beta) \bX_{\mrk} (\eps)
    \right)
    \, + \,
    \left(
    \beta \, \bZ_{\mrk} (\eps)
    +
    (1 - \beta) \bX_{\mrk} (\eps)
    \right)
    \bS_{\mrk}
    & \; = \;
    -
    (1/\eps)
    \,
    \bI.
    \end{align}
    \end{subequations}

Now, Lemma~1 from~\cite{bamdah03} can be utilized to solve~(\ref{eq.XYZc1}),
    \beq
    \beta \, \bZ_{\mrk} (\eps)
    \, + \,
    (1 - \beta) \bX_{\mrk} (\eps)
    \, = \,
    - \dfrac{1}{2 \eps} \, \bS_{\mrk}^{-1},
    \non
    \eeq
which in combination with~(\ref{eq.XYZb1}) yields the following expressions for the operators $\bX_{\mrk} (\eps)$ and $\bZ_{\mrk} (\eps)$,
    \begin{subequations}
    \label{eq.XZ}
    \begin{align}
    \label{eq.X}
    \bX_{\mrk} (\eps)
    & \; = \;
    \dfrac{1}{2} \,
    \bS_{\mrk}^{-1}
    ( \beta \bS_{\mrk} - \eps \bI )^{-1},
    \\[0.15cm]
    \label{eq.Z}
    \bZ_{\mrk} (\eps)
    & \; = \;
    \dfrac{1}{2} \,
    \bS_{\mrk}^{-1}
    ( \beta \bS_{\mrk} - \eps \bI )^{-1}
    \, - \,
    \dfrac{1}{2 \eps}
    \,
    ( \beta \bS_{\mrk} - \eps \bI )^{-1}.
    \end{align}
    \end{subequations}

From the above decomposition of $\bP_{\mrk} (\eps)$, the definitions of $f_{\mrk}$ and $\bC_{\mrk}^* \bC_{\mrk}$, and~(\ref{eq.Z}) it follows that
the $\We$-independent functions $f_{\mrk}$ are determined by
    \beq
    f_{\mrk}
    \, = \,
    \trace
    \left(
    \bZ_{\mrk} (\eps)
    \right)
    \, = \,
    \dfrac{1}{2}
    \,
    \trace
    \left(
    \bS_{\mrk}^{-1}
    ( \beta \bS_{\mrk} - \eps \bI )^{-1}
    \, - \,
    (1/\eps)
    \,
    ( \beta \bS_{\mrk} - \eps \bI )^{-1}
    \right),
    ~
    \mrk
    =
    \{ \mbox{os}, \, \mbox{sq} \}.
    \non
    \eeq
Since both $\bSos$ and $\bSsq$ are stable self-adjoint operators, the operators $( \beta \bS_{\mrk} - \eps \bI )$ are invertible. Consequently, this expression for $f_{\mrk}$ holds for all positive values of $\eps$ and for all $\beta \in (0,1)$. Furthermore, for $\eps \ll 1$, the Neumann series can be utilized to rewrite the inverse of the operator $( \beta \bS_{\mrk} - \eps \bI )$ as
    \beq
    \ba{rcl}
    \left( \beta \bS_{\mrk} - \eps \bI \right)^{-1}
    & \! = \! &
    (1/\beta) \, \bS_{\mrk}^{-1}
    \left(
    \bI \, - \, \eps \, (1/\beta) \, \bS_{\mrk}^{-1}
    \right)^{-1}
    \, = \,
    (1/\beta) \, \bS_{\mrk}^{-1}
    \ds{\sum_{n \, = \, 0}^{\infty}}
    \left( (\eps/\beta)  \, \bS_{\mrk}^{-1} \right)^n
    \\[0.25cm]
    & \! = \! &
    (1/\beta) \, \bS_{\mrk}^{-1}
    \, + \,
    \eps \, (1/\beta^2) \, \bS_{\mrk}^{-2}
    \, + \,
    \eps^2 \, (1/\beta^3) \, \bS_{\mrk}^{-3}
    \, + \,
    \cO (\eps^3).
    \ea
    \non
    \eeq
Hence, for $0 < \eps \ll 1$, the function $f$ in~\ref{eq.Ev} can be expressed as
    \beq
    \ba{c}
    f (k_z;\beta,\eps)
    \, = \,
    \dfrac{1}{\eps}
    \,
    \ds{\sum_{n \, = \, 0}^{\infty}}
    \eps^n f_n (k_z;\beta),
    ~~
    f_0 (k_z;\beta)
    \, = \,
    - \dfrac{1}{2 \beta}
    \,
    \trace
    \left(
    \bSos^{-1}
    \, + \,
    \bSsq^{-1}
    \right),
    \\[0.5cm]
    f_n (k_z;\beta)
    \, = \,
    - \dfrac{(1-\beta)}{2 \beta^{n + 1}}
    \,
    \trace
    \left(
    \bSos^{-(n+1)}
    \, + \,
    \bSsq^{-(n+1)}
    \right),
    ~~
    n \, = \, 1, 2, \ldots.
    \ea
    \non
    \eeq

\subsection{Scaling of function $a$ in\/~\mbox{\rm \ref{eq.Etau}} with $\eps$}
    \label{sec.a-eps}

From Section~\ref{sec.PSDtau} we recall that the function $a$ in~\ref{eq.Etau} is determined by
    \[
    a
    \, = \,
    a_{\mathrm{sq}} \, + \, a_{\mathrm{os}},
    \]
where $a_{\mathrm{sq}} = a_{12, 1} + a_{13, 1}$ and $a_{\mathrm{os}} = \sum_{j \, = \, 2}^{3} \left( a_{22, j} + a_{23, j} + a_{33, j} \right)$, respectively, quantify the variance amplification of the frequency response operators from $d_1$ to $\bphi_4 = \left[\,\tau_{12}\,\,\,\tau_{13}\,\right]^T$ and from $[\,d_2\,\,\,d_3\,]^T$ to $\bphi_2 = \left[\,\tau_{22}\,\,\,\tau_{23}\,\,\,\tau_{33}\,\right]^T$. Since these operators admit evolution equations given by~(\ref{eq.Sa-ss}) we conclude that the autocorrelation operator of the state in~(\ref{eq.Sa-ss}) is determined by $\bP_{\mrk} (\eps)$ in~(\ref{eq.Pk}). Therefore,
    \beq
    \ba{rcl}
    a (k_z; \beta,\eps)
    & \!\! = \!\! &
    a_{\mathrm{sq}} (k_z; \beta,\eps)
    \, + \,
    a_{\mathrm{os}} (k_z; \beta,\eps)
    \\[0.15cm]
    & \! = \! &
    \trace
    \left(
    \bX_{\mathrm{sq}} (\eps) \, \bS_{43}^* \, \bS_{43}
    \right)
    \, + \,
    \trace
    \left(
    \bX_{\mathrm{os}} (\eps) \, \bS_{21}^* \, \bS_{21}
    \right)
    \\[0.15cm]
    & \! = \! &
    \dfrac{1}{2} \,
    \trace
    \left(
    \bSsq^{-1}
    ( \beta \bSsq - \eps \bI )^{-1}
    \bS_{43}^* \, \bS_{43}
    \, + \,
    \bSos^{-1}
    ( \beta \bSos - \eps \bI )^{-1}
    \bS_{21}^* \, \bS_{21}
    \right),
    \ea
    \non
    \eeq
and this expression holds for all $k_z$, non-negative values of $\eps$, and $\beta \in (0,1)$.

\section{Singular perturbation analysis of\/ $\bH_{rj}$ with\/ $\{r = u$; $j = 1 \}$ and $\{ r = v,w$; $j = 2,3 \}$}
    \label{sec.Pi_a}

Here, we apply singular perturbation methods to examine how the $\We$-independent frequency response operators
    \[
    \ba{c}

    {\bH}_{u 1}
    \, = \,
    (1 \, + \, \mri \omega)
    \,
    \bG_u \bKsq \bF_1,
    ~~
    {\bH}_{r j}
    \, = \,
    (1 \, + \, \mri \omega)
    \,
    \bG_r \bKos \bF_j,
    ~~
    r = v,w; ~
    j = 2,3,
    \ea
    \]
depend on $\eps$ in flows with $\eps = 1/\mu \ll 1$. The aggregate steady-state variance of these operators is captured by the function $f$ in~\ref{eq.Ev} whose unfavorable scaling with $\eps$ was demonstrated in~\ref{sec.f}. Here, we show that the origin of this unfavorable scaling arises from the broadening of the temporal spectrum of these operators with a decrease in $\eps$. Furthermore, we demonstrate that while the inertialess model correctly predicts behavior of these operators at low temporal frequencies, it provides a poor approximation at high temporal frequencies. We also show that, from a physical point of view, no important viscoelastic effects take place in the contribution of the function $f$ to the spatial frequency responses of velocity fluctuations in elasticity-dominated flows.

As shown in~\ref{sec.f-ee}, the operators $\bH_{rj}$ with $\{r = u$; $j = 1 \}$ and $\{ r = v,w$; $j = 2,3 \}$ admit evolution equations given by
    \beq
    \ba{rcl}
    \tbo{\dot{x}_{r j}}{\eps \dot{z}_{r j}}
    & \!\!\! = \!\!\! &
    \tbt{-\bI}{\bI}{(1 - \beta) \bS_{\mrk}}{\beta \bS_{\mrk}}
    \tbo{x_{r j}}{z_{r j}}
    \, + \,
    \tbo{0}{\bF_j} d_j,
    \\[0.25cm]
    r
    & \!\!\! = \!\!\! &
    \obt{0}{\bG_r}
    \tbo{x_{r j}}{z_{r j}},
    \ea
    \label{eq.Pa-ss}
    \eeq
with $\{ \mrk = \mathrm{sq}$ for $r = u$; $\mrk = \mathrm{os}$ for $r = v,w \}$, homogeneous Dirichlet  boundary conditions on $x_{u 1}$ and $z_{u 1}$, and homogeneous Cauchy boundary conditions on $x_{r j}$ and $z_{r j}$ for $\{ r = v,w$; $j = 2,3 \}$. From a physical point of view, $z_{r j}$ and $x_{r j}$ with $\{ r = v,w$; $j = 2,3 \}$ are, respectively, determined by the streamfunction $\psi$ and the scalar field $\xi$ that represents a low-pass version of $\psi$, i.e.,
    \[
    z_{r j}
    \; = \;
    \psi,
    ~~
    x_{r j}
    \; = \;
    \xi
    \; = \;
    \psi/(\mri \omega + 1),
    ~~
    \{ r \, = \, v,w;
    ~
    j \, = \, 2,3 \}.
    \]
Note that, at any time instant, $\bphi_2$ is obtained from the wall-normal and spanwise gradients in $\xi$, $\bphi_2 = \bS_{21} \xi$ (for details, see~\ref{sec.f-ee}). On the other hand,
    \[
    z_{u 1} \; = \; u,
    ~~
    x_{u 1}
    \; = \;
    \zeta
    \; = \;
    u/(\mri \omega + 1),
    \]
where $\zeta$ is obtained by filtering high temporal frequencies in the streamwise velocity fluctuation. This scalar field determines $\bphi_4$ through a static-in-time relationship, $\bphi_4 = \bS_{43} \zeta$.

The approximate solutions for $x_{rj}$ and $z_{rj}$ in~(\ref{eq.Pa-ss}) can be found by performing a slow-fast decomposition of the system's dynamics. By setting $\eps = 0$ in the $z_{rj}$-equation of system~(\ref{eq.Pa-ss}), we obtain
    \beq
    \bar{z}_{rj}
    \, = \,
    - (1/\beta)
    \left(
    (1 - \beta) \, \bar{x}_{rj}
    \, + \,
    \bS_{\mrk}^{-1} \bF_j d_j
    \right).
    \label{eq.zbar}
    \eeq
As in~\ref{sec.g}, the white noise component $d_j$ in the expression for $\bar{z}_{rj}$ prevents it from being a valid approximation of $z_{rj}$.
Nevertheless, $\bar{z}_{rj}$ can still be employed as an approximation of an input $z_{rj}$ to the $x_{rj}$-subsystem in~(\ref{eq.Pa-ss}) as the slow system filters out the white noise component in $\bar{z}_{rj}$. On the other hand, an approximation of the fast subsystem is given by
    \[
    \eps \dot{z}_{rj,f}
    \, = \,
    \beta \bS_{\mrk} z_{rj,f}
    \, + \,
    \bF_j d_j,
    \]
with
    \[
    z_{r j} (t)
    \, = \,
    z_{rj,f} (t)
    \, - \,
    \left( (1-\beta)/\beta \right) x_{r j,s}(t)
    \, + \,
    \cO (\eps^{1/2}).
    \]
Thus, the slow component of $z_{rj}$ arises from a contribution of $(1 - \beta) \bS_{\mrk} x_{rj}$ and not from a contribution of the white noise input $d_j$, as would be common in singularly perturbed systems subject to slow inputs~\citep{kokkharei99}. To summarize, the slow-fast decomposition of system~(\ref{eq.Pa-ss}) is given by
    \beq
    \ba{rcl}
    \tbo{\dot{x}_{rj,s}}{\dot{z}_{rj,f}}
    & \! = \! &
    \tbt{\bA_{rj,s}}{0}{0}{\frac{1}{\eps} \bA_{rj,f}}
    \tbo{x_{rj,s}}{z_{rj,f}}
    \, + \,
    \tbo{\bB_{j,s}}{\frac{1}{\eps} \bB_{j,f}} d_j,
    \\[0.35cm]
    r
    & \! = \! &
    \obt{\bC_{r,s}}{\bC_{r,f}}
    \tbo{x_{rj,s}}{z_{rj,f}},
    \ea
    \non
    \eeq
where
    \[
    \ba{c}
    \bA_{rj,s} \, = \, - (1/\beta) \bI,
    ~~
    \bB_{j,s} \, = \, - (1/\beta) \bS_{\mrk}^{-1} \bF_j,
    ~~
    \bC_{r,s} \, = \, \left( (\beta - 1)/\beta \right) \bG_r,
    \\[0.1cm]
    \bA_{rj,f} \, = \, \beta \bS_{\mrk},
    ~~
    \bB_{j,f} \, = \, \bF_j,
    ~~
    \bC_{r,f} \, = \, \bG_r.
    \ea
    \]
Consequently, each velocity component can be decomposed into its slow and fast parts,
    \beq
    r \, = \,
    r_s \, + \, r_f,
    ~~
    r_s
    \, = \,
    \bH_{rj,s} d_j,
    ~~
    r_f
    \, = \,
    \bH_{rj,f} d_j,
    \non
    \eeq
where the slow frequency response is a function of $\omega$
    \beq
    \bH_{rj,s} (k_z,\omega;\beta)
    \, = \,
    \dfrac{1 \, - \, \beta}{\beta \left( \beta \, \mri \omega \, + \, 1  \right)}
    \,
    \bG_r \bS_{\mrk}^{-1} \bF_j,
    \non
    \eeq
and the fast frequency response is a function of $\bar{\omega} = \eps \, \omega$
    \beq
    \bH_{rj,f} (k_z,\bar{\omega};\beta)
    \, = \,
    \bG_r
    \left(
    \mri \bar{\omega} \bI \, - \, \beta \bS_{\mrk}
    \right)^{-1}
    \bF_j,
    ~~
    \bar{\omega} \, = \, \eps \, \omega.
    \non
    \eeq
In this scaling, the frequency $\bar{\omega}$ becomes important, i.e.\ $\cO (1)$, only for $\omega$ of $\cO (1/\eps)$ or higher~\citep{kokkharei99}. Hence, for low temporal frequencies $\omega = \cO (1)$, the fast frequency response can be approximated by its steady-state gain, $\bH_{rj,f} \approx - (1/\beta) \bG_r \bS_{\mrk}^{-1} \bF_j$. On the other hand, since $\bH_{rj,s}$ exhibits a low-pass property it becomes negligible at high temporal frequencies. Therefore, a low-frequency approximation of $\bH_{rj}$ is given by
    \beq
    \ba{rcl}
    \bH_{rj} (k_z,\omega; \beta,\eps)
    & \! = \! &
    \bH_{rj,s} (k_z,\omega;\beta)
    \, + \,
    \bH_{rj,f} (k_z,0;\beta)
    \, + \,
    \cO (\eps)
    \\[0.15cm]
    & \! = \! &
    - \,
    \dfrac{\mri \omega \, + \, 1}{\beta \, \mri \omega \, + \, 1}
    \,
    \bG_r \bS_{\mrk}^{-1} \bF_j
    \, + \,
    \cO (\eps),
    ~~
    | \omega | \, \leq \, \omega_1,
    \ea
    \label{eq.Hrj-low}
    \eeq
for some fixed positive $\omega_1$, and a high-frequency approximation of $\bH_{rj}$ is given by
    \beq
    \ba{rcl}
    \bH_{rj} (k_z,\bar{\omega}/\eps; \beta,\eps)
    & \! = \! &
    \bH_{rj,f} (k_z,\bar{\omega};\beta)
    \, + \,
    \cO (\eps)
    \\[0.15cm]
    & \! = \! &
    \bG_r
    \left(
    \mri \bar{\omega} \bI \, - \, \beta \bS_{\mrk}
    \right)^{-1}
    \bF_j
    \, + \,
    \cO (\eps),
    ~~
    | \bar{\omega} | \, \geq \, \omega_2,
    \ea
    \label{eq.Hrj-high}
    \eeq
for some fixed positive $\omega_2$.

\begin{figure}
    \centering
    {
    \subfloat[]{\includegraphics[width=0.45\textwidth]{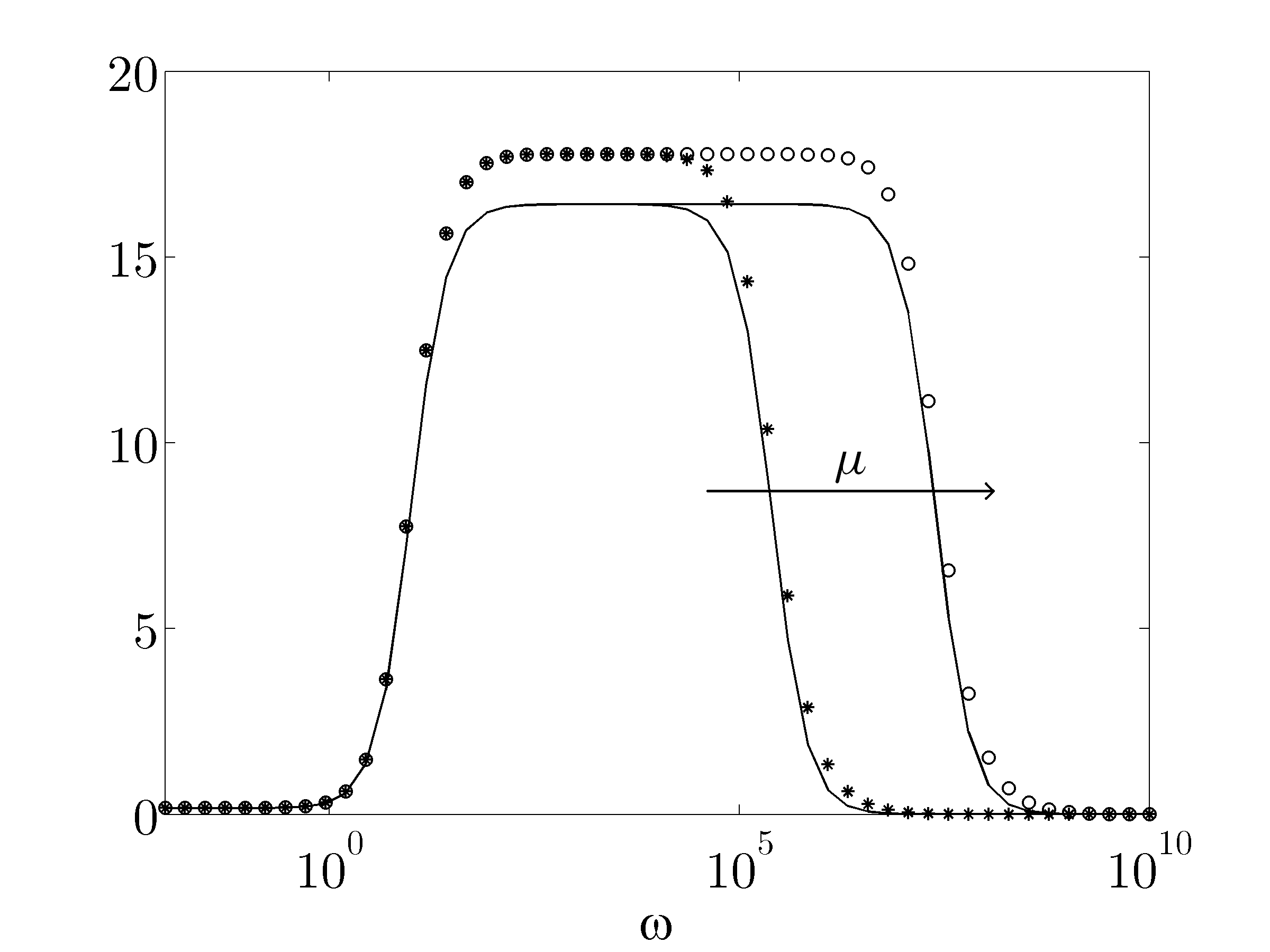}
    \label{fig.Hu1PSD}}
    \subfloat[]{\includegraphics[width=0.45\textwidth]{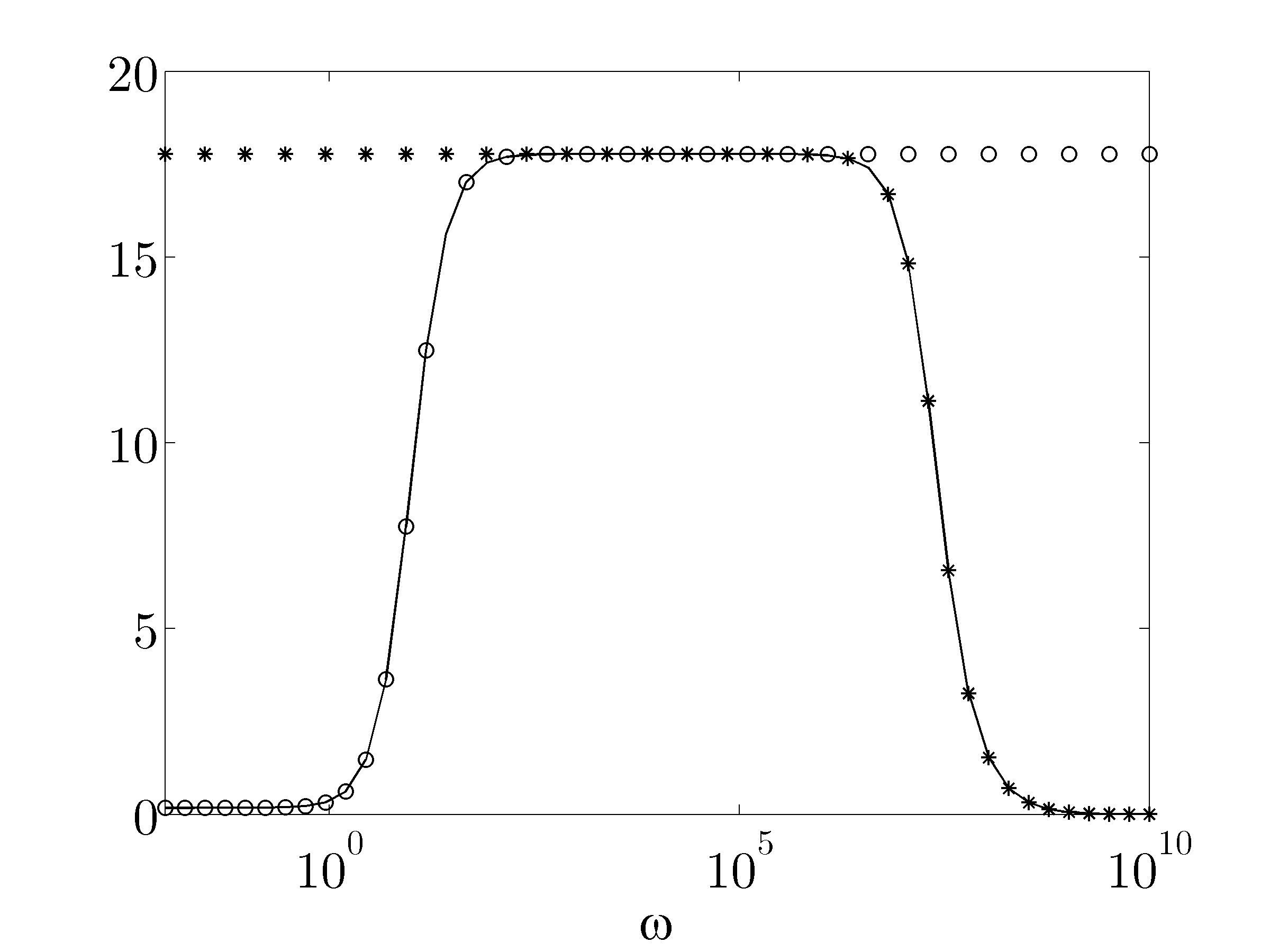}
    \label{fig.Hu1-low-high}}
    }
    \caption{The temporal frequency dependence of $\Pi_{u1}$ in flows with $k_z = 0$, $\beta = 0.1$, and $\eps = \{ 10^{-6}, 10^{-8} \}$. (a) The solid lines represent the results obtained by approximating $\bH_{u1}$ with~(\ref{eq.Hu1-approx}), and the symbols represent the results for the full operator $\bH_{u1}$. (b) The solid line denotes $\Pi_{u1} (0,\omega;0.1,10^{-8})$, the circles denote $\Pi_{u1}$ obtained using a low-frequency approximation~(\ref{eq.Hrj-low}), and the stars denote $\Pi_{u1}$ obtained using a high-frequency approximation~(\ref{eq.Hrj-high}).}
    \label{fig.Hu1}
    \end{figure}

Intuition about the temporal spectrum of the above frequency response operators in elasticity-dominated flows can be developed by analyzing properties of the operator $\bH_{u1}$. The spectral decomposition of the operator $\bSsq$ in the expressions for $\bH_{u1,s}$ and $\bH_{u1,f}$ can be used to represent these two operators as
    \beq
    \ba{rcl}
    \bH_{u1,s} (k_z,\omega;\beta)
    & \!\! = \!\! &
    \diag
    \left\{
    \dfrac{- (1 - \beta)}{\beta \, | \gamma_n (k_z) | \, (\beta \, \mri \omega \, + \, 1)}
    \right\}_{n \, \in \, \bbN},
    \\[0.25cm]
    \bH_{u1,f} (k_z,\bar{\omega};\beta)
    & \!\! = \!\! &
    \diag
    \left\{
    \dfrac{1}{\beta \, | \gamma_n (k_z) | \, (\frac{\mri \bar{\omega}}{\beta \, | \gamma_n (k_z) |} \, + \, 1)}
    \right\}_{n \, \in \, \bbN},
    \ea
    \non
    \eeq
where $\gamma_n (k_z) = - (k_z^2 + (n \pi/2)^2)$ are the eigenvalues of the Squire operator, and $\bbN$ denotes the set of natural numbers, $\bbN = \{ 1,2, \ldots \}$. By projecting $\bH_{u1,s}$ and $\bH_{u1,f}$ on the first eigenfunction of $\bSsq$, we obtain the following approximate expression for $\bH_{u1}$
    \beq
    \bH_{u1} (k_z,\omega; \beta,\eps)
    \, \approx \,
    \dfrac{\mri \omega \, + \, 1}
    {
    | \gamma_1 (k_z) |
    \left(
    \beta \, \mri \omega \, + \, 1
    \right)
    \left(
    \frac{\eps \, \mri \omega}{\beta \, | \gamma_1 (k_z) |} \, + \, 1
    \right)
    }
    \, + \,
    \cO (\eps).
    \label{eq.Hu1-approx}
    \eeq
The breakpoint frequencies in~(\ref{eq.Hu1-approx}) are determined by $\omega_1 = 1$, $\omega_2 = 1/\beta$, and $\omega_3 = \beta \, | \gamma_1 (k_z) | / \eps$, and the power spectral densities of $\bH_{u1}$ in flows with $\beta = 0.1$ and $\eps = \{ 10^{-6}, 10^{-8} \}$ are shown in Figure~\ref{fig.Hu1PSD}. For simplicity, the spanwise wavenumber is set to zero but similar trends are observed for other values of $k_z$. The solid lines represent the results obtained by approximating $\bH_{u1}$ with~(\ref{eq.Hu1-approx}), and the symbols represent the results for the full operator $\bH_{u1}$. We note that the projection of $\bH_{u1}$ on the first eigenfunction of $\bSsq$ captures well all essential trends, especially in the region of low temporal frequencies. Furthermore, we see that the peaks of the power spectral densities remain invariant under the change in $\eps$. Instead, increased elasticity spreads $\Pi_{u1}$ over a broader range of temporal frequencies. Since the cutoff frequency (i.e., the bandwidth) of $\bH_{u1}$ scales as $1/\eps$, it is not surprising that the variance maintained in $u$ by $d_1$ (which is obtained by integrating $\Pi_{u1}$ over all $\omega$) is also inversely proportional to $\eps$.

We note that, at $\eps = 0$, i.e.\ in creeping flow of an Oldroyd-B fluid, the operators $\bKos$ and $\bKsq$ simplify to
    \beq
    \bK_{\mrk}
    \, = \,
    - \,
    \dfrac{1}{\beta \, \mri \omega \, + \, 1}
    \,
    \bS_{\mrk}^{-1},
    ~~
    \mrk \, = \, \{ \mbox{os}, \, \mbox{sq} \},
    \non
    \eeq
which yields the expression for $\bH_{r j}$ that corresponds to the low-frequency approximation in~(\ref{eq.Hrj-low}). As illustrated in Figure~\ref{fig.Hu1-low-high}, this representation is characterized by the absence of a roll-off at high temporal frequencies and it is a poor approximation of the high-frequency dynamics~(\ref{eq.Hrj-high}). In particular, this implies that in inertialess flows stochastic forcing $d_1$ induces infinite variance in the  streamwise velocity component; similarly, stochastic forcing in either $d_2$ or $d_3$ yields wall-normal and spanwise velocities with unbounded variances. While the analysis conducted in~\ref{sec.f} confirms that in the limit of infinitely large elasticity number this is indeed the case, the analysis of this section shows that this is simply a consequence of the temporal spectrum of $\bH_{rj}$ becoming broader and broader with an increase in $\mu$ (cf.\ Figure~\ref{fig.Hu1PSD}). Furthermore, the increased elasticity does not change the value of the peaks of the frequency responses from $d_1$ to $u$ and from $d_2$ or $d_3$ to $v$ or $w$. Finally, we have shown that, from a physical point of view, no important viscoelastic effects take place in the variance amplification of operators $\bH_{r j}$ with $\{r = u$; $j = 1 \}$ and $\{ r = v,w$; $j = 2,3 \}$. Namely, in strongly elastic flows, the $(1/\eps)$-term in the expression for the function $f (k_z;\beta,\eps)$ in~(\ref{eq.f-high-mu}) only depends on the Orr-Sommerfeld and Squire operators in the streamwise-constant model of Newtonian fluids with $Re = 1$ and it is thus characterized by viscous dissipation effects.

As a consequence of the above analysis, we conclude that determination of the function $f$ in inertialess flows is an ill-posed problem. In the absence of inertia, white noise forcing -- which has contributions from arbitrarily large frequencies -- has a direct influence on certain velocity components ($d_1$ on $u$ and ($d_2,d_3$) on ($v,w$); cf.\ low-frequency approximation~(\ref{eq.Hrj-low}) and Figure~\ref{fig.Hu1-low-high}). Thus, at sufficiently high temporal frequencies, inertial effects become important and need to be retained in order to compute variance amplification from these forcing to these velocity components.
As we have shown in~\ref{sec.g} and~\ref{sec.Etau}, functions $g$ in~\ref{eq.Ev} and ($a,b,c$) in~\ref{eq.Etau} become independent of $\eps$ in the high-elasticity-number limit and consequently do not suffer from this problem.

%% References

\end{document}